\documentclass[aps,prd,preprint,eqsecnum,amsmath,amssymb,amsfonts,nobibnotes,nofootinbib]{revtex4}

\usepackage{bm}
\usepackage{epsf}

\DeclareMathOperator{\curl}{curl}

\newcommand{\pul}{{{\frac{1}{2}}}}
\newcommand{\tretina}{{{\frac{1}{3}}}}
\newcommand{\dtretiny}{{{\frac{2}{3}}}}
\newcommand{\sestina}{{{\frac{1}{6}}}}
\newcommand{\tripul}{{{\frac{3}{2}}}}
\newfont{\boldit}{cmbxti10}

\makeatletter
\newcommand{\spdot}[1][1]{^{\hbox{\raise#1\ex@\hbox{\normalfont .}}}}
\makeatother

\renewcommand{\tilde}{\widetilde}

\begin{document}
\centerline {Published in PHYSICAL REVIEW D{\bf 76}, 063501 (2007), 31 pages}
\title{Cosmological perturbation theory, instantaneous gauges and local inertial frames}

\author{Ji\v{r}\'{\i} Bi\v{c}\'ak}

\affiliation{
  Institute of Theoretical Physics, Faculty of Mathematics and Physics, Charles University,\\
  V Hole\v{s}ovi\v{c}k\'{a}ch 2, 180 00 Prague 8, Czech Republic\\ and Max Planck Institute for Gravitational Physics, Albert Einstein Institute,
  Am M\"{u}hlenberg 1, D--14476 Golm, Germany
}

\author{Joseph Katz}

\affiliation{
  Racah Institute of Physics, Hebrew University,\\
  Jerusalem 91904, Israel
}

\author{Donald Lynden-Bell}

\affiliation{
  Institute of Astronomy, The Observatories,\\
  Madingley Road, Cambridge CB30HA, United Kingdom\\
}


\begin{abstract}
Linear perturbations of Friedmann-Robertson-Walker universes with any curvature
and cosmological constant are studied in a general gauge without decomposition
into harmonics. Desirable gauges are selected as those which embody best
Mach's principle: in these gauges local inertial frames can be determined
instantaneously via the perturbed Einstein field equations from the
distributions of energy and momentum in the universe. The inertial frames are
identified by their `accelerations and rotations' with respect to the
cosmological frames associated with the `Machian gauges'.
In closed spherical universes, integral gauge conditions are imposed to
eliminate motions generated by the conformal Killing vectors. The meaning of
Traschen's integral constraint vectors is thus elucidated. For all three types
of FRW universes the Machian gauges admit much less residual freedom than the
synchronous or generalized harmonic gauge.
Mach's principle is best exhibited in the Machian gauges in closed spherical
universes. Independent of any Machian motivation the general perturbation
equations and discussion of gauges are useful for cosmological perturbation
theory.
\end{abstract}

\maketitle
\section{Introduction}

Einstein preferred a finite universe, bounded in space, over an infinite one
because he wanted to avoid posing boundary conditions. What Einstein really
disliked was that in open universes some of the motion of inertial frames is due to dragging by
matter while the rest is due to the boundary conditions at infinity.
Thirty-four years ago the authors
of acclaimed ``Gravitation'' \cite{MTW} commented on the Einstein view in a
footnote on p. 704:  ``Many workers in cosmology are skeptical of Einstein's
boundary condition of closure of the universe, and will remain so until
astronomical observations confirm it''. The Wilkinson Microwave Anisotropy
Probe has now provided data \cite{WM} which, among many other things, have
constrained the present value of total mass-energy density parameter of the
universe to be $\Omega_0 = 1.02 \pm 0.02$. With such a result, all three basic
sets of standard Friedmann-Robertson-Walker (FRW) cosmological models (see,
e.g., \cite{MTW}, \cite{weinb}) are compatible: the models with flat spatial
sections (with curvature index $k = 0, \Omega_0 = 1$), positive spatial
curvature models $(k = + 1, \Omega_0 > 1)$, as well as negative curvature $(k =
-1, \Omega_0 < 1)$. Nevertheless, the WMAP data `marginally prefer' $k=+1$
(see, in particular, \cite{Efst}), and, indeed, recently several authors studied
closed models again in detail (see, e.g., \cite{Sto}, \cite{Las}), after years
of preference of flat universes which have been considered as natural outcomes
of inflation. Even an idea going back to the Eddington-Lema\^{\i}tre cosmology
has now been revived: if our universe is closed today, it was always closed,
and perhaps inflation is `past-eternal' -- the universe, dominated at early
times by a single scalar field, could have started asymptotically from an
initial Einstein static universe which enters an inflationary expanding phase,
succeeded then by standard evolution (see \cite{EMT} and references therein).
The recent growing evidence for the existence of a cosmological constant
$\Lambda$ has been an inspiration for the reconsideration of spatially closed
universe of de Sitter  type \cite{Las}.

In the present work we do not, technically, bestow a privilege to any value of
spatial curvature. All three cases $k = 0, \pm 1$ are analyzed in equal detail,
and we even discuss, albeit briefly, closed hyperbolic 
 and closed flat 
 universes with multiconnected topologies. From the physical (to
some extent perhaps `philosophical') point of view, we adhere to the Einstein
preference, i.e., to the closed universes with standard (spherical) topology,
because our work on cosmological perturbation theory has been motivated by
Mach's principle.

\subsection{Mach's principle}

 Mach's principle has acquired certain unpopularity among some
relativity and cosmology circles. The primary reason is perhaps the fact that
under that name a range of meanings and interpretations, sometimes even mutually
contradictory, has gradually accumulated. During the Prague conference in 1988
to celebrate the 150th anniversary from Mach's birth \cite{UK} and, in particular,
at the T\"{u}bingen conference in 1993 devoted entirely to Mach's principle,
numerous interpretations have been given (see the excellent book \cite{Mach}).
More recently, Bondi and Samuel \cite{BoS} listed the `zeroth' plus ten other
versions of Mach's principle and described within which theoretical framework a
particular statement of the principle applies -- see also \cite{Dol}, where the
main formulations from \cite{Mach} and \cite{BoS} are summarized. A brief
history of Mach's principle and its meaning in general relativity and cosmology
is given in the Introduction to our first paper on the subject \cite{LKB}.

Despite a possible scepticism as regard the role of Mach's principle in
contemporary cosmology, most of the standard treatises on the subject do
include a discussion of the principle (see, e.g., \cite{MTW}, \cite{weinb},
\cite{Peeb}, \cite{Peacock}), and no one can deny Mach's ideas have been a
source of inspiration 
to many, not only Einstein. One of the purposes of the present work is to
demonstrate that a search for a framework in which Mach's principle can be best
embodied in the cosmological perturbation theory can lead to practical results,
such as the formulation of the perturbation theory in a completely general
gauge, followed by the selection of an advantageous `Machian gauge' for solving specific problems.

What then do we mean by Mach's principle? As in our previous work \cite{LKB},
as a starting point we adapt the original Bondi's formulation from his
classical `Cosmology' \cite{Bon}: ``Local inertial frames are determined
through the distributions of energy and momentum in the Universe by some
weighted averages of the apparent motions''. More specifically, we turn
primarily to those among Einstein's equations for linear perturbations of
the FRW models which represent the constraints, i.e., under suitable conditions
partial differential equations of elliptic type,  connecting the `initial
values' for matter perturbations with the perturbations of the metric. In
\cite{LKB} we went quite a way in realizing the Machian program. We  studied
the frame-dragging effects due to slowly, rigidly rotating, but collapsing or
expanding spheres in the (inhomogeneous) Lema\^{\i}tre-Tolman-Bondi universes,
and we analyzed the dragging effects of vector perturbations of the FRW
universes described in a special gauge such that three (momentum) constraint
equations enabled us to determine instantaneously metric perturbations
$h^0_k\;\; (k = 1,2,3)$ in terms of energy-momentum perturbations $\delta
T^0_k$. In the open universes, these are determined uniquely by requiring the
perturbations to vanish at infinity -- rotations are `absolute' in this sense.
In closed universes a linear combination of six Killing vectors (three rotations
plus three quasitranslations) may be added to the $h^0_k$. We still obtain the
solutions of the three constraint equations when angular momenta  corresponding
to the three rotations and quasimomenta corresponding to the three
quasitranslations of the sources (determined by $\delta T^0_k$) are given. In
this sense no absolute rotations exist in closed universe, only differences of
rotation rates are determinable -- in accord with Mach's ideas that `all
motions are relative'. If, however, the \textit{velocities} of the bodies, described by
perturbations of perfect fluid, are given, the metric perturbations $h^0_k$ are
determined uniquely.

The last result is related to the fundamental fact that six globally conserved
quantities, corresponding to the six Killing vectors in a FRW universe, must
all vanish if considered for the whole closed universe. The conserved
quantities, being the derivatives of  superpotentials, can be expressed as
surface integrals like an electric charge by using Gauss' theorem. As the
volume surrounded by the surface is expanded over all the universe at a given
time, the surface must shrink to zero. It was, among others,  an attempt to
understand Mach's principle in cosmological perturbation theory, which inspired
us  to formulate 
 conservation laws with respect to curved
backgrounds \cite{KBL}.  The resulting `KBL superpotential', using the
designation by Julia and Silva \cite{JSII}, was found, after applying certain
natural criteria, to be unambigous and most satisfactory in spacetimes with or
without a  cosmological constant, in any spacetime dimension $D\geq 3$ (see
\cite{JSII}, \cite{JSI}). It also found applications in the recent studies of the causal
generation of cosmological perturbations seeding large-scale structure
formation and of the back reactions in slow-roll inflation (see \cite{She}, \cite{LU},  
and references therein).

In the present paper we study \textit{general} linear perturbations of the FRW
universes from a `Machian perspective'. This leads us to investigate both
rotations and accelerations of local inertial frames in perturbed universes,
and to develop \textit{all} the perturbed Einstein equations in a general gauge
`ab initio'.

\subsection{Cosmological perturbation theory}

Observational evidence for isotropy and homogeneity of the Universe shows that
it is broadly well described by a FRW model, but the clustering, the galaxies
and the stars constitute local perturbations from the idealized substratums of
cosmological models. The other goal of cosmological perturbation theory is to
link the physical conditions in the early universe with structures observed
today. From the pioneering work of Lifshitz in 1946 (see review in \cite{LL}),
there appeared numerous papers studying linear perturbations of FRW universes;
for the more recent extensive reviews, see, for example, \cite{KodSas},
\cite{Stewart}, \cite{MFB}, \cite{Dur}, \cite{Bertschinger}. Recently,
several authors have even found the impetus and skill to write down the
complicated system of equations for the second-order perturbations of the FRW
models -- see \cite{HW} and references therein -- but applications of these are
yet to appear.

In any cosmological perturbation theory, two problems at once confront us: (i)
What smooth cosmological model is best suited to our Universe, and (ii) how do we
map the points of our inhomogeneous Universe onto a chosen cosmological model.
Both these problems are, in relativity, connected with the gauge freedom that
changes the apparent form of the perturbations by which our Universe differs
from the smooth cosmological model adopted. Although the first problem is
primarily connected with a difficult question of an appropriate averaging of an
inhomogeneous universe, or so-called `fitting problem' (see, e.g.,
\cite{ElBru}), it is also related to the gauge problem because two different FRW
universes can be close to one another, at least for some time, so one can be
considered as a perturbation of another. When we have chosen an appropriate
background, then we are confronted with the freedom to choose coordinates in
the real Universe differently and so to remap the Universe onto the background
model.  This gives rise to the commonly discussed gauge freedom. Of course,
what happens in the real world is independent of what background is used and
how we map onto it. This is the motive behind gauge invariant perturbation
theory, and this made the work of Bardeen \cite{Bard} in 1980, in which gauge
invariant quantities combined from cosmological perturbations were first
introduced, so influential. The gauge problem is explained in technical terms
from different (equivalent) perspectives in depth in the literature: one can
consider a one-parameter family of 4-dimensional manifolds, with $M_0$ a
background and $M_\varepsilon$ a perturbed universe, embedded in a
5-dimensional manifold $N$ and connected by a ``point identification map'' which
is specified by a vector field $X$ on $N$ transverse to the manifolds $M$;
the gauge transformation is then a change of $X$ (see, e.g., \cite{Stewart}, \cite{SW}).
Alternatively, in a more physical vein, in any chosen
coordinate system in the real Universe one assigns to all physical quantities
$Q(x^\alpha)$ also their background values $\overline{Q}(x^\alpha)$. These, in
contrast to $Q(x^\alpha)$, do not change their functional dependence on
coordinates under an infinitesimal coordinate transformation (see \cite{MFB}).
Mathematically, any of these approaches lead to the changes of physical
quantities as they appear in the following (see, in particular, Section IV).

There exists a well-known lemma 
\cite{SW} stating that the
linear perturbation of a quantity is gauge invariant only if the quantity
vanishes on the background or is a constant (scalar or linear combination of
products of Kronecker delta). 
 The density perturbation
$\delta \rho$, for example, is not gauge invariant since
$\overline{\rho}$ is time-dependent function in the FRW backgrounds. That is
why to obtain a gauge invariant quantity, one has to consider, e.g., the
gradients of density perturbation or combine $\delta \rho$ with some other
quantities. However, as in the black-hole perturbation theory, solving for
gauge invariant quantities may not mean finding all quantities of interest. For
example, in the problem of the motion of a charged black hole in a weak
asymptotically uniform electric field, there is only one gauge invariant
quantity. We need to fix the gauge at the end in order to find all
perturbations of the metric and electromagnetic field to see how the hole
accelerates \cite{Bi}. It is advantageous at the start to have the possibility
of a gauge choice according to  the problem in hand. 
 Selecting a gauge
which implies a physically preferable coordinate system may eventually give
both a better physical understanding and an easier mathematical procedure.
After all, motions in the solar system can be described as seen from a frame
that rotates to keep a planet `at rest', but are much more readily
comprehended in Newton's inertial axes.

Last but not least, physical effects associated with Mach's ideas like the
dragging of inertial frames are of a global nature and they do  require the
introduction of suitable coordinate frames (the `gauges'). A true
understanding of inertia and inertial frames must involve specific frames or
coordinates. To borrow Dieter Brill's comment from 
Ref. \cite{Mach}: ``Mach's principle may point the way toward giving physical
meaning to  quantities usually considered frame-dependent''.We return to this
issue in the concluding remarks 
 where some of our other work on
Machian effects \cite{KLB}, \cite{LBK} is summarized in the context of the
present paper.

\subsection{The goal}

There exist many frameworks for treating linearized perturbations of FRW
universes. The one which has been used most frequently involves the synchronous
gauge, with all quantities decomposed into suitable harmonics in accordance
with the spatial curvature. In what follows we make a general study of
advantageous gauges without imposing \textit{a priori} conditions on them and without
decomposition into harmonics. \textit{We identify desirable gauges as those
which embody best Mach's principle}. We find that these gauges are also
motivated by the gauge choices used in full nonlinear general relativity. Most
importantly, however, they are distinguished by the simplifications they bring
both to the perturbed Einstein equations and to their physical interpretation.

What do we
 mean by Mach's principle within this broader framework? 
  We again start from Bondi's formulation 
 that ``local inertial frames are determined
through the distributions of energy and momentum in the Universe by some
weighted averages''. However, to determine a local inertial frame in a general
situation means to \textit{find} both its `\textit{rotation and acceleration}'
from the 
 distributions of energy and momentum, represented 
 by 
$\delta T^\mu_\nu$. In a 
 general situation 
we need to know the full spacetime metric in a neighborhood of a point in order
to determine completely local inertial frames at that point; see, e.g.
\cite{MTW}, \cite{weinb}. In Wheeler's conception of Mach's principle (e.g.
\cite{MTW}, \cite{Ciu}), we have to specify the initial data on a Cauchy
spacelike surface like
 the conformal three-geometry and the mass-energy
currents, solve for the spacetime geometry $g_{\mu\nu}$, and thus determine
 local inertial frames. In general, gravitational waves will globally also
contribute to the dragging of the inertial frames but only when the waves are
nonlinear perturbations of a FWR universe. However, limiting ourselves
to the linear perturbations of the FRW universes, it is interesting to see
 what data are needed to determine the `accelerations' and
`rotations' of local inertial frames with respect to what we call the
\textit{cosmological (observers') frames}. In a perturbed FRW universe, can a
gauge be found such that the distribution of $\delta T^\mu_\nu$ determines
uniquely and instantaneously the rotations and accelerations of local inertial
frames via Einstein's field equations?

\subsection{The outline}

After first reviewing the properties of a general congruence of timelike
worldlines in a general spacetime (see, e.g., \cite{Ellis}), we consider the
congruence of `cosmological observers' in a perturbed FRW universe with
coordinates $\{x^\mu\}$  as a `perturbation' of the congruence of fundamental
observers in the FRW background. We assume that in a `cosmological gauge'
$\{x^\mu\}$ the cosmological observers move along $x^i =$ constants,  but we
describe the properties of their congruence by covariant expressions which can
be calculated in any coordinates. The cosmological observers are equipped with
their local frame vectors; the timelike ones are their 4-velocities, and the
spacelike ones lie along their connecting vectors. Now a cosmological observer
is, in general, \textit{accelerated} with respect to a local freely falling
inertial frame, in particular, the one which at  a given spacetime point moves
with the same 4-velocity. Expressing this acceleration in terms of the metric
perturbations, we find that only certain components of the metric perturbations
are needed. Next, we determine the \textit{rotation} of the axes of the
cosmological observer with respect to the nonrotating rigid orthogonal axes
(gyroscopes held in their centres of mass) of the local inertial frame. 
 Having the acceleration  $\boldsymbol\alpha$
and the angular velocity $\boldsymbol\omega$
expressed, we have \textit{determined the local inertial frame:} it accelerates
and rotates with respect to the corresponding cosmological frame with just the
opposite vectors, $-\boldsymbol\alpha$ and $-\boldsymbol\omega$.
 All these issues are analyzed in
Section II.

Assuming a general congruence of cosmological observers, i.e., equivalently, a
general gauge, we find that in order to determine the accelerations and
(avaraged) rotations of local inertial frames in the sense just described,  we
need to know the metric perturbations $\delta g_{00}$ and $\delta g_{0i}$ and
their first derivatives. The main issue in Sections III and IV is
to find and study the gauges in which these quantities can be determined
instantaneously from the knowledge of energy-momentum distributions $\delta
T^\mu_\nu$.
We give the perturbed Einstein equations for all three types of
FRW universes with any value of $\Lambda$, in an arbitrary gauge. We first
adopt the `relativists' attitude' and 
 start 
from the perturbed FRW metric in the form

\begin{equation}\label{1.1}
ds^2 = \left( \bar g_{\mu \nu} + h_{\mu \nu} \right) dx^\mu
dx^\nu = dt^2 - a^2(t) f_{kl}dx^k dx^l + h_{\mu \nu}
dx^\mu dx^\nu,
\end{equation}
where the spatial background metric is $f_{kl}(x^m), k,l,m... = 1, 2, 3; $ 
 $t$ is the `cosmic time', so $\delta g_{\mu\nu} \equiv h_{\mu\nu}$. Perturbations $h_{\mu\nu}$ are small so that quadratic terms can be neglected.
In one of the standard coordinate systems the background FRW metric $\overline{g}_{\mu\nu}$ reads
\begin{equation}\label{1.2}
ds^2 = dt^2 - a^2 (t) \left[ \frac{dr^2}{1-kr^2} + r^2(d\theta ^2
+\sin ^2 \theta d \varphi ^2) \right],
\end{equation}
where in a positive curvature (closed) universe $(k = +1)\; r \in <0,
1 >$, in flat $(k = 0)$ and negative curvature $(k = - 1)$ open
universes $r \in <0, \infty ), \;\theta \in <0, \pi>, \varphi\in <0, 2
\pi)$. We shall also employ other common alternatives as, e.g., hyperspherical coordinates,
\begin{equation}\label{1.3}
ds^2 = dt^2 - a^2 \left[ d\chi^2 + \Sigma ^2 _k (d \theta ^2 +
\sin^2 \theta d \varphi ^2) \right],
\end{equation}
with $\Sigma _k = \sin \chi ,\chi , \sin$h  $\chi$ for,
respectively, $k = +1, 0, -1$. 
 The perturbations 
  $\delta T^\mu_\nu$ are left general,  but 
   a perturbed perfect fluid is considered 
    as an 
     example. 
In Appendix \ref{App.A} we give all the perturbed Einstein's equations and the Bianchi identities starting off from \eqref{1.1}; in Section III we give them using conformal time $\eta$ and metric perturbations defined as is usual in  the cosmological literature, e.g., in \cite{KodSas}, \cite{MFB}, \cite{Bertschinger} -- again in a completely general gauge.

We do not decompose the perturbations in 
 harmonics nor do 
 we first separate them into the scalar, vector and tensor parts (used, e.g., in \cite{Stewart}).  Although both methods are very useful in cosmology, they involve nonlocal operations. In order to make Fourier-type analyses in the space variables, 
 one needs to know quantities in the whole space, which is not `typical' in cosmology. The splitting of a local perturbation into some scalar, vector and tensor perturbations is also nonlocal. Imagine a trivial (zero) perturbation in a given domain ${\cal{O}}$, and extend it to an annulus ${\cal{A}}$ so that it is nonvanishing there. Hence, in ${\cal{O}}$ the trivial perturbation will split into nontrivial (scalar, vector, tensor) pieces which depend on the extension into ${\cal{A}}$. Therefore,  a perturbation which is the sum of  scalar, vector, and tensor parts cannot be uniquely expressed in terms of the Bardeen gauge invariant variables \cite{Bard} which are defined separately for each part. Without using harmonics or splittings, the perturbed Einstein field equations are in a form suitable for searching for solutions in terms of Green's functions. How the Green's function approach can reveal new aspects of cosmological perturbation theory has been recently indicated by Bashinsky and Bertschinger~\cite{Bash}. 


In Section IV, the main purpose is to motivate and describe geometrically several gauges in which the accelerations and rotations of the local inertial frames follow instantaneously from the field equations. We call these \textit{Machian gauges}. We also clarify 
 the residual gauge freedom that these gauges admit, and make a comparison with two typically non-Machian gauges -- the synchronous gauge and  the generalized Lorenz-de Donder (`harmonic') gauge. The Machian gauges turn out to admit much less residual freedom.    
The freedom represented by the gauge transformations generated by the conformal Killing vectors in closed (spherical) universes is 
 removed by the \textit{integral gauge conditions} which we impose. In closed hyperbolic 
  universes 
   our Machian local gauge conditions fix coordinates uniquely. 

Finally, in Section V 
 we give the field equations in the Mach 1 gauge 
 and show how they can be solved to give the local inertial frames when the distribution of the matter energy-momentum is given. We also discuss Traschen's integral-constraint vectors \cite{Tr1}, \cite{Tr2.} restricting possible $\delta T^\mu_\nu$. 
 According to Traschen and others \cite{Tr1}, \cite{TrErd.}, their existence has implications for the Sachs-Wolfe effect and for microwave background anisotropies. 
 Traschen considered these vectors in the synchronous gauge. By contrast, in the Mach 1 gauge, 
 these constraints become a straightforward consequence of the constraint equations and acquire a simple, lucid meaning. 
 We 
  find integral constraints also on quantities not considered by Traschen. 
  In closed universes these integral constraints are satisfied automatically as a consequence of our integral gauge conditions by which motions generated by the spatial conformal Killing vectors are eliminated.
In Section V we also list all Green's functions known in literature which solve the constraint equations needed for the determination of the local inertial frames; some are still unknown. We then review
 our recent work \cite{BLK1}, \cite{BLK2} on vorticity perturbations of FRW universes and study their effect on local inertial frames. As a second example, we consider perturbations of potential type for which the vorticity vanishes. At the end 
 we analyze the `Machian' question of how uniquely local inertial frames are determined in perturbed universes.

In Concluding Remarks (Sec. VI)
 we briefly summarize the results and discuss global aspects of Mach's principle. 
In Appendix \ref{App.C} the Killing and conformal Killing vectors in FRW universes 
 are listed, 
  and those harmonics which are needed in the main text are given. In Appendix \ref{App.D} we discuss briefly the field equations in the other 
 gauges considered in Section~IV.

\section{The acceleration and rotation of local inertial frames}

\subsection{The congruence of cosmological observers}

Consider a {\em general\/} spacetime with coordinates $\{x^\mu\}$ in which a congruence of timelike,
non-intersecting worldlines of ''cosmological observers`` is given
by
\begin{equation}\label{2.1}
x^\mu = x^\mu(y^i ; p)\;\;\;\;\;\;\;\;\; i = 1,2,3.
\end{equation}
The choice of fixed $y^i$ determines the worldline of a particular
observer; $p$ is a parameter along the worldline, commonly chosen
as either the cosmological time $x^0 = t$ or the observer's proper
time $\tau$.
The cosmological observers use their 4-velocity as their normalized timelike frame vector,
\begin{equation}\label{2.3}
\tau ^\mu = (\partial x ^\mu /\partial \tau)_{y^i} = t^\mu/(g_{\alpha
\beta}t^\alpha
t^\beta)^{\frac{1}{2}},\,\,t^\mu = (\partial x ^\mu /\partial p)_{y^i}.
\end{equation}
For 
spatial frame vectors 
a cosmological observer naturally takes three independent 
vectors specified by $ \delta y^i$ pointing from him to three other
observers of the congruence, orthogonal to $\tau^{\mu}$:
\begin{equation}\label{2.6}
\delta x^\mu _{\bot} = P^\mu _{\nu} \delta x ^\nu =(\delta ^\mu _\nu - \tau ^\mu \tau _\nu)(\partial x ^\nu /\partial y ^i)_p \;\; \delta y^i.
\end{equation}
As a triad of spatial vectors $e^\mu _{(i)}$, any three linearly
independent vectors proportional to $\delta \underset{(i)}{x}{}_\bot^\mu $ can be taken.
A triad based on the connecting vectors is given at a fixed
spacetime point and can be extended along the observer's
worldline because connecting vectors are Lie-propagated (see, e.g.,
(\cite{Br-Sum}, \cite{Wald}) along the congruence. 
This gives
\begin{equation}\label{2.12}
P^\lambda _\mu \delta x_{\bot ; \nu} ^\mu  \tau ^\nu = \tau
^\lambda_
{; \nu} \delta x ^\nu _\bot.
\end{equation}
Three independent connecting vectors define the triad of
unit spacelike vectors $m^\mu _{(i)}$:
\begin{equation}\label{2.13}
\delta \underset{(i)}{x}{}_\bot ^\mu = \delta l_{(i)} m^\mu _{(i)},
\;\;\;\; m_{(i)\mu} m^\mu _{(i)} = -1,
\end{equation}
 with no summation over index $i$. Equation \eqref{2.12} implies the
propagation equations for scalar distances $\delta
\textit{l}_{(i)}$ -- the 'generalized Hubble's law` (admitting a possibly
anisotropic expansion) \cite{Ellis} --
and the propagation equations for triad $m^\mu _{(i)}$.
Decomposing the 
derivative of a 4-velocity 
in the
standard manner 
 (e.g. \cite{MTW}, \cite{Ellis}),
\begin{equation}\label{2.16}
\tau _{\mu ;\nu} = \tau _\nu \alpha_{\mu} + \omega_{\mu \nu} +
\sigma _{\mu \nu} + \begin{matrix}{\frac{1}{3}}\end{matrix}\theta P_{\mu \nu},
\end{equation}
the acceleration $\alpha _\mu$, vorticity $\omega
_{\mu\nu}$  (antisymmetric), shear $\sigma _{\mu\nu}$ (symmetric) and expansion $\theta$ are
given respectively by
\begin{equation}\label{2.17}
\alpha _\mu = \tau _{\mu ;\nu}\tau^\nu,
\end{equation}
\begin{equation}\label{2.18}
\omega _{\mu\nu} =  \begin{matrix}{\frac{1}{2}}\end{matrix}P_\mu ^\kappa P_\nu ^\lambda (\tau_{\kappa ;\lambda} - \tau _{\lambda
;\kappa}),
\end{equation}
\begin{equation}\label{2.19}
\sigma_{\mu\nu} = \begin{matrix}{\frac{1}{2}}\end{matrix}P^\kappa_\mu P_\nu^\lambda (\tau_{\kappa ;\lambda} + \tau _{\lambda
;\kappa})- \begin{matrix}{\frac{1}{3}}\end{matrix} \theta P_{\mu\nu},
\end{equation}
\begin{equation}\label{2.20}
\theta = \tau ^\nu _{;\nu}.
\end{equation}
We obtain, successively,
\begin{equation}\label{2.21}
P^\lambda _\mu \delta x ^\mu _{\bot ;\nu}\tau ^\nu = (\omega ^\lambda {}_\nu + \sigma ^\lambda _\nu + \begin{matrix}{\frac{1}{3}}\end{matrix} \theta P^\lambda _\nu)
\delta x^\nu _\bot.
\end{equation}
\begin{equation}\label{2.22}
\left.\frac{d}{d\tau}(\delta \textit{l}_{(i)})\right/\delta \textit{l}_{(i)} =
(\sigma _{\mu\nu} + \begin{matrix}{\frac{1}{3}}\end{matrix} \theta P_{\mu\nu})m^\mu _{(i)}
m_{(i)}^\nu,
\end{equation}
\begin{equation}\label{2.23}
P^\lambda _\mu m^\mu _{(i);\nu}\tau ^\nu = \left[ \omega  ^\lambda{}_\nu + \sigma ^\lambda_\nu +
(\sigma _{\alpha \beta}
m ^\alpha _{(i)} m^\beta _{(i)}) P^\lambda_\nu \right] m^\nu
_{(i)}.
\end{equation}


\subsection{Cosmological observers in a perturbed FRW universe: The frames}
Consider first an unperturbed FRW model described by metric
(\ref{1.1}) with $h_{\mu\nu} = 0$. Fundamental (cosmological) observers move
along the worldlines $x^i$ = constants with 4-velocity $\overline{\tau}^{\mu} = (1,0,0,0)$. 
As the spatial triad, they take three independent vectors
$\overline{e}^{\mu}_{(i)}$ perpendicular to $\overline{\tau}^\mu$.
These need not be chosen to be necessarily mutually orthogonal if,
for example, coordinates are used in which $f_{k l}$ in (1.1) is not
diagonal [as, e.g., in (\ref{6.3})]. In standard coordinates in FWR backgrounds like in Eqs. (\ref{1.2}) and (\ref{1.3}),
$f_{k l}$
is diagonal and the vectors
\begin{equation}\label{2.28}
\overline{e}^{\mu}_{(i)} = (0, \delta ^m _i),\quad
\overline{e}_{(i)\mu} = (0, \overline{g}_{im})
\end{equation}
are orthogonal. It is easy to normalize them:
\begin{equation}\label{2.29}
\overline{m}^\mu _{(i)} = (-\overline{g}_{ii})^{-\frac{1}{2}}
\left[ 0, \delta _i^m \right],\quad \overline{m}_{(i)\mu} =
(-\overline{g}_{ii})^{-\frac{1}{2}} \left[ 0, \overline{g}_{im} \right]\,,
\end{equation}
again with no summation over $i$, index $m=1,2,3$. 
The quantities (\ref{2.17})--(\ref{2.20}) characterizing the congruence of the
fundamental observers are well known: $\overline{\alpha}  _\mu = \overline{\omega}_{\mu \nu} =
\overline{\sigma}_{\mu \nu} = 0$, 
 $\overline{\theta} = 3\dot{a}/a$, 
the dot is $d/dt$.

In a linearly perturbed FRW universe the metric is given by Eq.~(\ref{1.1});
the indices of the first-order quantities (including $h_{\mu\nu}$) are
shifted by the background metrics $\overline g_{\mu\nu}$,
respectively $\overline g^{\mu\nu}$.
The congruence of cosmological observers will, in general coordinates,
be given by (2.1). The frame vectors can be written in the form
$\tau ^\mu = \overline{\tau} ^\mu + \delta\tau ^\mu, \; e^\mu
_{(i)} = \overline{e}^\mu _{(i)} + \delta e^\mu _{(i)}$, 
similarly for covariant components, and for $\alpha^\mu, \omega_{\mu\nu}, \sigma
_{\mu\nu}$, and $\theta$. In general coordinates these quantities
can be found easily from the expressions given in Section~II~A. In
the following we shall assume 
 that
coordinates $\{x^\mu \}$ represent a ``cosmological gauge'', in which
the congruence of cosmological observers is given by $x^i = y^i =$
constants. 
We find $\tau ^\mu$ to be given by
\begin{equation}\label{2.35}
\tau ^\mu = \overline{\tau}^\mu + \delta\tau ^\mu = ( 1 -
\begin{matrix}{\frac{1}{2}}\end{matrix}h_{00}, 0,0,0 ).
\end{equation}
The spatial triad, determined by connecting vectors orthogonal to
$\tau ^\mu$ and lying along coordinate lines, 
 is
\begin{equation}\label{2.37}
e^\mu_{(i)} = {\overline{e}}^{\mu}_{(i)} + \delta e ^\mu_{(i)} = (- h_{i0}, \delta
^{m}_i),
\end{equation}
from which the corresponding unit spacelike vectors 
$m^\mu_{(i)}$ can be found:
\begin{equation}\label{2.39}
m^\mu_{(i)} = \overline{m}^\mu_{(i)} + \delta m^\mu_{(i)} =
(-\overline{g}_{ii})^{-\frac{1}{2}} \left[ -h_{i0}, \delta^m_i
(1-\begin{matrix}{\frac{1}{2}}\end{matrix}h_{ii}/\overline{g}_{ii}) \right].
\end{equation}

We gave here both the background and perturbed frames for completeness. In the following we shall often use just the background frames because these are only needed when a small, first-order quantity is projected.

\subsection{The acceleration of local inertial frames}

We shall designate the local frame of a cosmological observer (CO)
given by tetrad $\tau ^\mu$, $e^\mu_{(i)}$, respectively $m^\mu_{(i)}$,
by the COF -- cosmological observer frame. This frame moving along $x^i =$ constants is, in general,
{\em accelerated\/} with respect to local freely falling \textit{inertial}
frames. Among the inertial frames there is a frame which, moving
at a given spacetime point with 4-velocity $\tau^\mu$, is
momentarily at rest with respect to the COF; such a frame is called
the LIF -- local inertial frame%
\footnote{There are of course infinitely many LIFs moving
with the 4-velocity $\tau^\mu$ at a given point. However, they differ just by purely spatial
transformations or constant shifts of time. Among them, there is
also such a LIF that its origin coincides with that of a corresponding
COF and its acceleration 
 is 
  $-\alpha^l$.}.
The 4-acceleration of the COF with respect
to the LIF is given by
Eq. (\ref{2.17}). Using $\tau^\mu$ given by Eq. (\ref{2.35}) and
perturbed
metric (\ref{1.1}), we find
\begin{equation}\label{2.49}
\alpha ^\lambda = (0, \alpha ^l),
\end{equation}
where
\begin{equation}\label{2.50}
\alpha ^\textit{l} = \delta \Gamma ^l_{00} = \overline{g}^{lm}
(-\begin{matrix}{\frac{1}{2}}\end{matrix}h_{00,m} + \dot{h}_{0m}).
\end{equation}
We see that {\em only $h_{00}$ and $\dot{h}_{0m}$ are needed in determining the
acceleration of the COF with respect to the LIF\/} {\it or, equivalently, the
acceleration of the LIF with respect to the COF} (which is  $-\alpha ^l$).
Spatial metric perturbations do not even enter in the frame
components of the acceleration because the unperturbed spatial
triad is needed to the zeroth order only:
\begin{equation}\label{2.51}
\alpha_{(i)} = e_{(i)\lambda} \alpha^\lambda \cong
\overline{e}_{(i)\lambda}\alpha^\lambda
\end{equation}
(similarly with projections on unit vectors $m^\lambda_{(i)}$).
Although we calculated the acceleration in coordinates adapted to
 cosmological observers, it is given by a
covariant expression (\ref{2.17}) which can be expressed in {\em any\/}
coordinates. The result is also invariant under gauge
transformations (see Section IV) since in 
 the
background this acceleration vanishes.

\subsection{The rotation of local inertial frames}

Next we wish to determine the rotation of the axes of the COF with respect
to the non-rotating rigid orthogonal axes (gyroscopes held in their
centres of mass) of the LIF at a given point and thus, vice versa,
the rotation of the LIF with respect to the COF.

First consider a cosmological observer carrying a gyroscope
described by a spacelike vector $W^\mu$, perpendicular to $\tau
^\mu$. The gyroscope is transported along observer's worldline by
Fermi-Walker transport. 
 Another gyroscope,
carried by an inertial observer moving with the same $\tau ^\mu$
at a given point, does not rotate relative to $W^\mu$. However, a
vector $S^\mu$, perpendicular to $\tau^\mu$, which is transported
along the worldline of CO in a \textit{general} manner, will rotate
relative to $W^\mu$ by $(D_F S^\mu)\Delta \tau$, where $D_F S^\mu$
is the Fermi-Walker time derivative defined by (see Fig.~1, and e.g.~\cite{Tho-Mac})
\begin{equation}\label{2.52}
\frac{D_F S^\mu}{d \tau} \equiv P^\mu _\rho S^\rho_{;\nu} \tau^{\nu}
= S^\mu_{;\nu} \tau^\nu + (\alpha_\nu S^\nu) \tau^\mu,
\end{equation}
where $\alpha_\nu$ is the acceleration (\ref{2.17}) and
 $S^\mu \tau_\mu = 0$ was used. 
For the gyroscope, $D_F W^\mu/d\tau = 0$.

\begin{figure}[h]
\epsfbox{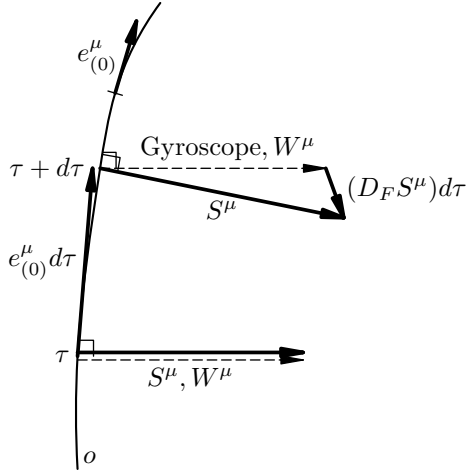}
\caption{The Fermi-Walker time derivative $D_F S^\mu$ (based in part
on Fig.~2 in \cite{Tho-Mac}).
The cosmological observer $o$ with four-velocity $e^\mu_{(0)}$
carries with himself a gyroscope, represented by the spatial
vector $W^\mu$ (dashed arrow), and a spatial vector $S^\mu$ which
are both perpendicular to $e^\mu_{(0)}$ and identical at
observer's proper time $\tau$. After $d\tau$, both $S^\mu$ and
$W^\mu$ remain perpendicular to $e^{\mu} _{(0)}$ but the generally
propagating vector $S^\mu$ will differ from the nonrotating,
Fermi-Walker transported gyroscope by the Fermi-Walker time
derivative $(D_F S^\mu)d\tau$.}
\end{figure}

Now regarding Eq. (\ref{2.21}) we see that the left-hand side (l.h.s.) 
 is just equal to the Fermi-Walker derivative of the connecting vector
so that
\begin{equation}\label{2.54}
\frac{D_F \delta x^\mu_\perp}{d\tau} = (\omega^\mu{}_\nu +
\sigma^\mu_\nu + \begin{matrix}{\frac{1}{3}}\end{matrix}\theta P^\mu_\nu) \delta x^\nu_\perp.
\end{equation}
Therefore, since the congruence of cosmological observers has, in
general, a non-vanishing vorticity and shear, the connecting
vectors rotate with respect to gyroscopes. The last term in Eq. (\ref{2.54}) is proportional
to $\delta
x^\mu_\perp$and represents only a dilation of the connecting
vector due to the (isotropic) expansion of the congruence.
Similarly, unit vectors $m^\mu_{(i)}$ of the COFs rotate with
respect to gyroscopes according to Eq. (\ref{2.23}):
\begin{equation}\label{2.55}
\frac{D_Fm^\mu_{(i)}}{d\tau} = \left[ \omega^\mu{}_\nu +
\sigma_\nu^\mu+ (\sigma_{\alpha\beta} m^\alpha_{(i)}m^\beta_{(i)})
P_\nu^\mu \right] m^\nu_{(i)}.
\end{equation}

Turning now to the perturbed FRW universes we find, using $\tau^\mu$ from
Eq.~(\ref{2.35}) and the perturbed metric (\ref{1.1}),
the vorticity (\ref{2.18}) to have a simple form
\begin{equation}\label{2.56}
\omega_{k l} = \delta\omega_{k l} = \frac{1}{2} (h_{0 k
,l}- h_{0 l,k}), \;\omega_{0\alpha} = \delta \omega _{0\alpha} = 0,
\end{equation} 
the shear (\ref{2.19}) turns out to be
\begin{equation}\label{2.58}
\sigma_{k l} = \delta\sigma _{k l} =
\frac{1}{2}\dot{h}_{k l} - \frac{1}{6}\dot{h}^m_m \overline{g}_{k l}- \frac{\dot{a}}{a}h_{k l},\;
\sigma_{0\alpha} = \delta \sigma _{0\alpha} = 0,
\end{equation} 
and the expansion (\ref{2.20}) reads
\begin{equation}\label{2.60}
\theta = \overline{\theta} + \delta\theta = \frac{3\dot{a}}{a} +
\frac{1}{2}(\dot{h}^m_m - \frac{3\dot{a}}{a} h^0_0).
\end{equation}
Since both $\omega^\mu{}_\nu$ and $\sigma^\mu_\nu$ are of the first
order in $h_{\mu\nu}$, on the right--hand side (r.h.s.)of Eq. (\ref{2.55}) only
$\overline{m}^\mu_{(i)}$ enters and the equation takes the form
\begin{equation}\label{2.61}
\frac{D_F m^l_{(i)}}{d\tau} = \left[ \omega^l{}_k +
\sigma_k ^l + (\sigma _{ab}
\overline{m}^a_{(i)}\overline{m}^b_{(i)}) \delta_ k ^l \right]
\overline{m}^k_{(i)},
\end{equation}
where $\overline{m}^k_{(i)} = (-\overline{g}_{(ii)})^{-\frac{1}{2}} 
\delta^k_i$, and $\omega^l{}_k$, $\sigma_k^l$ are
given by Eqs. (\ref{2.56}) and (\ref{2.58}). 

Clearly, the vector $m^l_{(i)}$ rotates relative to the gyroscopes
and, hence, a gyroscope will rotate relative to COF not only due
to a nonvanishing vorticity but also due to the presence of a
shear. A gyroscope will precess in a gravitational wave described
by $\dot{h}_{k l}$ (cf. the discussion in \cite{Bini}); it is
{\it not} true, as sometimes stated (\cite{Bertschinger},
p.~334) that a spin (a gyroscope) precesses relative to the
cosmological frame at a rate given just by the vorticity~$\omega_{k l}$.

The axes of a LIF are determined by three orthogonal
gyroscopes, while those of a COF are determined by three approximately orthogonal
vectors $e^\mu_{(i)}$ or, after their normalization, by unit
vectors $m^\mu_{(i)}$. 
 As a consequence, \textit{on average} the
rotation of the COF relative to the LIF (moving with the same
$\tau^\mu$ at a given point) is determined just by vorticity
$\omega_{k l}$. Indeed, there is a significant difference
between the terms on the r.h.s. of Eq. (\ref{2.61}): $\omega_{k l}$ is
antisymmetric while $\sigma _{k l}$ is symmetric and
traceless. If at a given instant a vector $m_{(i)}^\mu$ lies along
a principial direction of $\sigma_{k l}$, its direction will
be changed only by the vorticity. As in fluid kinematics \cite{Ba}, it is
just the vorticity which describes the ``effective angular
velocity'' of the fluid~(see~Fig.~2).

\begin{figure}[h]
\epsfbox{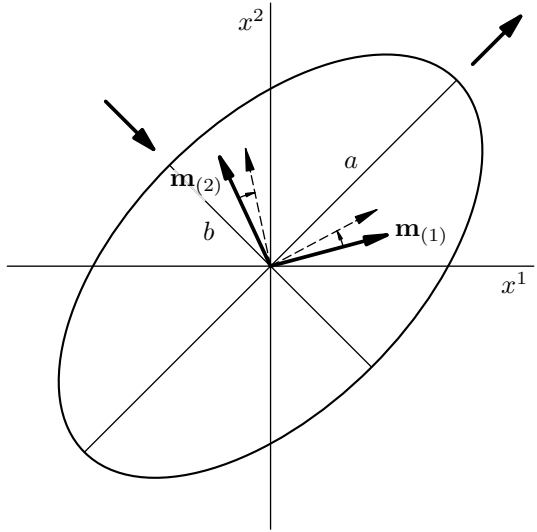}
\caption{Because of the shear with the principal axes $a,b$, almost orthogonal
unit vectors $\textbf{m}_{(1)}$,
$\textbf{m}_{(2)}$ change their directions but they do not,
on average, rotate with respect to the fixed axes $x^1$, $x^2$.}
\end{figure}

Hence, we conclude that in order {\em to determine the averaged
rotations of local inertial frames with respect to the
cosmological frames in the perturbed FRW universes, it is
sufficient to determine the vorticity tensor (\ref{2.56}), i.e., spatial
gradients of $h_{0 k}$.\/}


Rather than by $\omega_ {k l}$ the rotation is usually
represented by the ''cosmologist's vorticity vector"
\begin{equation}\label{2.66}
\omega^\lambda = \frac{1}{2}
\overline{\varepsilon}^{\lambda\sigma\mu\nu}\overline{\tau}_{\sigma}\omega_{\mu\nu},
\end{equation}
where 
\begin{equation}\overline{\varepsilon}^{\alpha\beta\gamma\delta} =
(-\overline{g})^{-{\frac{1}{2}}} \left[ \alpha\beta\gamma\delta
\right],\; \overline{g} = \text{det} (\overline{g}_{\mu\nu}), \label{2.67}
\end{equation}
and  $[\alpha\beta\gamma\delta]$ is the permutation symbol. In our case
we get
\begin{equation}\label{2.68}
\omega^\lambda = (0, \omega^l), \;\;\;\; \omega^l = \frac{1}{2}
\varepsilon^{lmn} h_{0n,m},
\end{equation}
$\varepsilon^{l m n} = (\overline{\gamma})^{-\frac{1}{2}}[lmn], \overline{\gamma} = $ det
$(-\overline{g}_{ik})$. Considering $h_{0k}$ as a
3-dimensional velocity vector, the cosmologist's vorticity (\ref{2.66}) yields
$\frac{1}{2}$ of the standard vorticity,
$\curl{\bm{v}}$,
in fluid dynamics. However,
$\curl{\bm{v}}$ represents
twice the effective rigid local angular velocity of the fluid in
an inertial frame \cite{Ba}. Therefore, the averaged
rigid angular velocity of COFs with respect to LIFs is determined
exactly by $\omega^l$ given by Eq. (\ref{2.68}). Equivalently, LIFs
rotate with respect to COFs with angular velocity $-\omega^l$.

\subsection{A note on generalized backgrounds}

When the background is not FRW but, say, a  Lema\^{\i}tre-Tolman-Bondi or Bianchi model,
the accelerations and the averaged rotations of the LIFs with respect to the COFs can still be determined from just 
$h_{00}$ or $h_{i0}$.
For example, if the coordinates can be
chosen such that $g_{\mu\nu} = \overline{g}_{\mu\nu} + h_{\mu\nu}$, where 
$\overline{g}_{0i} = 0$,
$\overline{g}_{00}$ is an arbitrary function of time $x^0$, and
$\overline{g}_{ik}$ are arbitrary functions of $x^\lambda$, then
 cosmological observers given by $x^i =$
constant have their accelerations with respect to LIFs equal to $\alpha^l = \overline{g}^{lm} (-\frac{1}{2} h_{00,m} + h_{0m,0} -
\frac{1}{2}\overline{g}^{00} \overline{g}_{00,0}h_{m0})$,
and their vorticity is $\overline{\omega}_{\mu\nu} = 0$, $\omega_{kl} = \delta\omega_{k
l}= \frac{1}{2}(\overline{g}_{00})^{-\frac{1}{2}}(h_{0k,l} -
h_{0l,k})$, $\omega_{0\alpha} = 0$.


\subsection{Sources and their description in the cosmological frame}

In the FRW universes the background energy-momentum tensor is
commonly taken to be the perfect fluid stress tensor, $\overline{T}_\mu^\nu = (\overline{\rho} + \overline{p})
\overline{U}_\mu \overline{U}^\nu - \overline{p}\delta^\nu_\mu$, 
so that in the comoving coordinates 
\begin{equation}\label{2.75}
\overline{T}^0_0 = \overline{\rho},\;\;\; \overline{T}^i_j = -\overline{p}\delta ^i_j, \;\;\;
\overline{T}^0_j = \overline{T}^j_0 = 0.
\end{equation}
The energy density $\overline{\rho}$ and the pressure
$\overline{p}$
of the matter can describe a standard perfect fluid with a given
equation of state. 
Alternatively,
one may regard these expressions as the stress tensor components of a
homogeneous time-dependent scalar field $\Phi$ of an inflationary
model with the energy density $\overline{\rho} = \rho_{\Phi}$ and
effective pressure $\overline{p}=p_\Phi$ (see, e.g.,~\cite{LiddtecLyth}).
The special case with $\overline{\rho} + \overline{p} = 0$, $\overline{\rho} =
-\overline{p}= \text{constant}$,
corresponds to the de Sitter vacuum spacetime with a cosmological
constant $\Lambda = -\overline{p} = \overline{\rho}$, commonly
interpreted as a vacuum energy. 
Any of these background matter contents can be considered in the
present work. We shall thus not, in general, specify the form of the
perturbations $\delta T_{\mu}^{\nu}$ of the energy-momentum tensor. 
Employing the frame vectors $e^\mu_{(\alpha)}$ given by \eqref{2.37}, we find
the frame components of perturbations (indicated by [e] and [m])for a general energy-momentum tensor to be given
by
\begin{equation}\label{2.81}
\delta {T}{}_{[e](0)}^{(0)} = \delta {T}{}_{[m](0)}^{(0)} = \delta T ^0_0,
\;
\delta {T}{}_{[e](i)}^{(0)} = (-\overline{g}_{ii})^\frac{1}{2} \delta {T}{}_{[m](i)}^{(0)} = \delta T_i^0 - (\overline{\rho} +
\overline{p})h_{0i},
\end{equation}
\begin{equation}\label{2.83}
\delta {T}{}_{[e](0)}^{(i)} = (-\overline{g}_{ii})^{-\frac{1}{2}} \delta
{T}{}_{[m](0)}^{(i)} = \delta T_0^i,
\;\delta {T}{}_{[e](i)}^{(k)} =
(\overline{g}_{ii}/\overline{g}_{kk})
^\frac{{}{1}}2\delta {T}{}_{[m](i)}^{(k)} = \delta T_i^k,
\end{equation}
with no summation over $i,k$. By employing the ``mixed''
tensorial coordinate components of perturbations, we see that their
values, except for $\delta T^0_i$, coincide -- up to the background
``normalization'' factors $\sim (-
\overline{g}_{ii})^{\pm\frac{1}{2}}$ -- with their frame scalar
components.

In the case of a perfect fluid the coordinate components read 
\begin{equation}\label{2.85}
\delta T^0_0 = \delta\rho,\; \delta T^0_i = (\overline{\rho} +
\overline{p})(h_{i0} + V_{i}), \;\delta T^i_0 = (\overline{\rho} + \overline{p})V^i, \;
\delta T^k_i = -\delta p\; \delta^k_i,
\end{equation}
where $\delta\rho$ and $\delta p$ are perturbations of the matter
density and pressure. The velocity 
\begin{equation}\label{2.86}
V^i = \frac{dx^i}{dt}\,,
\end{equation}
is the spatial part of the perturbation of the fluid's 4-velocity
\begin{equation}\label{2.87}
U^\mu = \overline{U}^\mu + \delta U^\mu = (1 - \begin{matrix}{\frac{1}{2}}\end{matrix}h_{00},
V^i).
\end{equation}
It is easy to see that the 4-velocity is approximately the unit
timelike vector since we assume $V^i\ll 1$, and terms proportional
to $V^2$ and $Vh$ can thus be neglected. 
The 4-acceleration of the fluid is defined by $A^\mu = U^\mu_{;\nu} U^\nu$.
Since the background value $\overline{A}^\mu = 0, A^\mu$ is of the
first order. The standard condition $A^\mu U_\mu = 0$ thus implies
$A^\mu \overline{U}_\mu = 0$, and hence $A^0 = 0$.
The calculation of the spatial components yields
\begin{equation}\label{2.92}
A^i = \dot{V}^i + 2 \frac{\dot{a}}{a}V^i -
\frac{1}{2}\overline{g}^{is} h_{00,s} +
\overline{g}^{is}\dot{h}_{s0} = \dot{V}^i + 2\frac{\dot{a}}{a} V^i
+ \alpha^i,
\end{equation}
where $\alpha^i$ is the acceleration (\ref{2.50}) of the cosmological
frame with respect to the local inertial frame, or $-\alpha^i$ is the
acceleration of the LIF with respect to the COF. The acceleration (\ref{2.92})
is the fluid's acceleration with respect to the LIF, whereas $\dot{V}^i$
characterizes its acceleration with respect to the COF. If the fluid
is momentarily at rest in the COF, $V^i = 0$, and the fluid has the same acceleration
with respect to the COF as the LIF has, $\dot{V}^i = -\alpha ^i$, then $A^i =
0$, as it should. Since the fluid's acceleration vanishes for the
background, its frame components are just
 $\!{A}{}_{[e]}^{(i)} = (-\overline{g}_{ii})^{-\frac{1}{2}}{A}{}_{[m]}^{(i)} =
A^i$. 
It is not difficult to check that the acceleration (\ref{2.92})
satisfies the perturbed relativistic Euler's equations,
\begin{equation}\label{2.94}
(\overline{\rho} +\overline{p})A^i = \delta (\Pi ^{i\mu}p_{,\mu})\,,\quad
\Pi^{i\mu} = g^{i\mu} - U^i U^\mu,
\end{equation}
where $\Pi^{i\mu}$ is the projection tensor
into the 3-space orthogonal to $U^\mu$. As we shall notice in
Section III B, these are just the spatial parts of the perturbed Bianchi
identities.

The vorticity of the fluid is defined by
\begin{equation}\label{2.95}
\Omega^\alpha = \begin{matrix}{\pul}\end{matrix}\varepsilon^{\alpha\beta\gamma\delta}
U_\beta\Omega_{\gamma\delta}, \;
\Omega_{\gamma\delta} = \pul(U _{\gamma;\mu}\Pi^\mu_\delta
- U_{\delta;\mu} \Pi^\mu_\gamma ).
\end{equation}
 Since $\Omega^\alpha U_\alpha = U^\gamma \Omega_{\gamma\delta} =
0$ and $\Omega$'s are of the first order, we again get $\Omega^0 = 0 = \Omega_{0\alpha}$.
The nonvanishing spatial parts turn out to be
\begin{eqnarray}\label{2.98}
\Omega_{kl} = \begin{matrix}{\pul}\end{matrix} \left[ (V_k + h_{0k})_{,l} - (V_l +
h_{0l})_{,k} \right], \;\;
\Omega ^i = \begin{matrix}{\pul}\end{matrix} \varepsilon^{ikl}(V_l + h_{0l})_{,k}
= \begin{matrix}{\pul}\end{matrix} \varepsilon^{ikl} V_{l,k} + \omega^i;
\end{eqnarray}
 $\omega^i$ is the vorticity vector (\ref{2.68}) of the
cosmological frame. Again, this result 
 is plausible
in the following sense: Since the LIF rotates with respect to the COF with
$-\omega^i$, then if the fluid rotates with respect to the COF with $\frac{1}{2}\varepsilon^{ikl}V_{l,k} =
-\omega^i$, it does not rotate with respect to the LIF, $\Omega^i =
0$. As with the acceleration, the frame components are simply given by
$\Omega{}_{[e]}^{(i)} = (-\overline{g}_{ii})^{-\frac{1}{2}}\Omega{}_{[m]}^{(i)} =
\Omega^i$. 

From Eqs.~\eqref{2.81}--\eqref{2.85} it is evident that to give
the {\em frame\/} components of the source we need
to know only the perturbations $\delta\rho$, $\delta p$, and $V^i$
of the fluid. No metric perturbations are needed ---
in contrast to the coordinate components $\delta T_i^0$
in which $h_{0i}$ enters. This is important for our understanding
of Mach's principle.

\section{Field equations}

We have seen that the accelerations and rotations of 
 LIFs with respect to the 
COF are determined in a general gauge by $h_{00}$ and $h_{0l}$
components of the perturbations. We shall now write the perturbed
Einstein's equations for the FRW backgrounds in a general gauge 
\footnote {Some of the equations presented here have been derived independently by
Langlois (1994) in his PhD thesis and by P. Uzan and N.
Deruelle (private communication). The perturbed Ricci tensor components and the equations
of motion have been written down by Bardeen (1980) \cite{Bard} after a
decomposition of the~metric into scalar, vector, and tensor parts using
his specific amplitudes.}.
We then shall see later which gauges will enable us to determine
instantaneously perturbations $h_{00}$ and $h_{0l}$ (separately
from $h_{kl}$) in terms of matter perturbations.

A straightforward way to express the perturbations of Einstein's
equations is in terms of a physical cosmic time $t$ and some
convenient spatial coordinates $x^l$ of the FRW background.
However, there are 
 advantages in using conformal time $\eta$, given by $a(\eta)\,d\eta=dt$.
Both $t$ and $\eta$ are common in the literature and we shall thus
give explicitly the perturbation equations in two forms --- with
$t$ in Appendix \ref{App.A} and with $\eta$ in this section. 

\subsection{{{Perturbed field equations with a conformal time $\eta$}}}
In terms  of coordinates $\tilde{x}^\mu =(\eta , x^k)$ the~metric
of the~background is
\begin{equation}\label{3.23}
d\bar{s}^2=\tilde{\bar{g}}_{\mu\nu}d\tilde{x}^\mu d\tilde{x}^\nu=
    a^2e_{\mu\nu}d\tilde{x}^\mu d\tilde{x}^\nu=
    a^2\left[ d\eta^2-f_{kl}dx^kdx^l\right]\ ,
\end{equation}
where we introduced the conformally related {\em static\/} background
metric $e_{\mu\nu}$ by
\begin{equation}\label{3.24}
e_{00}=1\ ,\ \ e_{0l}=0\ ,\ \ e_{kl}=-f_{kl}(x^i)\ .
\end{equation}
The~components
of a tensor, say $\widetilde{W}^\nu_\mu$, are related to those of
the tensor $W_\mu^\nu$ in $x^\mu = (t, x^k)$ coordinates as
follows:
\begin{equation}\label{3.25}
\widetilde{W}^0_0=W^0_0\ ,\ \ \widetilde{W}^0_l=a^{-1}W^0_l\ ,\ \
\widetilde{W}^l_0=aW^l_0\ ,\ \ \widetilde{W}^l_k=W^l_k\ .
\end{equation}
Defining the~dimensionless ``relative Hubble constant'' by ${\cal{H}}=\frac{1}{a}\frac{da}{d\eta}=\frac{a'}{a} = \dot a = aH$,
we can write  the~non-vanishing background Christoffel symbols as
\begin{equation}\label{3.27}
{\tilde{\bar{\Gamma}}}{}_{00}^0={\cal{H}}\ ,\ \
{\tilde{\bar{\Gamma}}}{}_{kl}^0={\cal{H}} f_{kl}\ ,\ \
{\tilde{\bar{\Gamma}}}{}_{0l}^m={\cal{H}}\delta^m_l\ ,\ \
{\tilde{\bar{\Gamma}}}{}_{kl}^m=\bar{\Gamma}_{kl}^m\ ,
\end{equation}
where $\bar{\Gamma}_{kl}^m$ is given in Appendix \ref{App.A}.
The~prime 
 hereafter denotes
the~derivative with respect to 
$\eta$. (Later, it will also be used to denote a coordinate change, but no confusion should arise.)
The~nonzero components of the background Einstein equations 
 become 
\begin{equation}\label{3.28}
{\tilde{\bar{G}}}{}_{0\ }^0=\bar{G}^0_0
            =\frac{3}{a^2}\left( k+{\cal{H}}^2\right)=\kappa\bar{\rho} + \Lambda\ ,\ \
{\tilde{\bar{G}}}{}_k^m=\bar{G}^m_k
          =\frac{1}{a^2}\delta^m_k \left( k+{\cal{H}}^2+2{\cal{H}}'\right)
          =-(\kappa\bar{p} - \Lambda)\delta^m_k\ .
\end{equation}

The~linearly perturbed Einstein equations will 
be written in terms of the~dimensionless perturbations $\tilde{h}_{\mu\nu}$
of $e_{\mu\nu}$,
\begin{eqnarray}\label{3.29}
&\;&ds^2=
(\widetilde{\overline{g}}_{\mu\nu} + \delta
\widetilde{g}_{\mu\nu})d\widetilde x^\mu d\widetilde x^\nu =
a^2\left( e_{\mu\nu}+\tilde{h}_{\mu\nu}\right)
                d\tilde{x}^\mu d\tilde{x}^\nu\ = \nonumber \\
                &=&a^2 \left[ (1+\widetilde{h}_{00}) d\eta ^2 + 2 \widetilde{h}_{0k}\;d\eta dx^k - (f_{kl} - \widetilde{h}
                _{kl}) dx^k dx^l  \right],
\end{eqnarray}
which means that $\delta \widetilde{g}_{\mu\nu}=a^2\tilde{h}_{\mu\nu}$.
It is important to emphasize that, in contrast to tensors like in
Eq. (\ref{3.25}), $\widetilde{h}_{\mu\nu}$'s are \textit{not}
components (in coordinates $\widetilde x^\mu$) of $h_{\mu\nu}$ used in \eqref{1.1} and Appendix \ref{App.A}; as seen from
Eq. (\ref{3.29}), they represent the perturbations of the conformal
static metric $e_{\mu\nu}$, whereas $h_{\mu\nu}$'s 
represent perturbations of the physical background metric $\bar g_{\mu\nu}$.
In a (1+3)-decomposition, i.e., ~in quantities $\tilde{h}_{00}$, $\tilde{h}_{0l}$,
$\tilde{h}_{kl}$, we do not raise the~index $0$ and we
raise the~spatial
indices only with $f^{kl}$; thus $\tilde{h}^m_0=f^{ml}\tilde{h}_{0l}\ ,\ \tilde{h}^{mn}=f^{mk}f^{nl}\tilde{h}_{kl}\ ,  \text{etc.}$
The~explicit relations between $h_{\mu\nu}$, $h^\nu_\mu$, or
$h^{\mu\nu}$ 
and
$\tilde{h}_{\mu\nu}$, $\tilde{h}^\nu_\mu$, or $\tilde{h}^{\mu\nu}$
are given in Appendix \ref{App.A}.


The~perturbations of Einstein's equations in terms of
${\tilde{h}}_{\mu\nu}$ can be obtained from equations in Appendix \ref{App.A}.
We shall introduce two special symbols which not only
simplify the~equations but are also helpful in suggesting
particularly useful gauge conditions. 
 We set
\begin{equation}\label{3.35}
\mathcal{K} =\tripul {\mathcal{H}}{\tilde h}_{00}\
         +\pul \left({\tilde h}^n_n\right)^{'}
         -\nabla_n {\tilde h}^n_0\ .
\end{equation}
 The~second quantity we shall
employ is defined with the~traceless part of ${\tilde{h}}^l_k$,
\begin{equation}\label{3.36}
{\tilde {h}}{}^l_{T k}={\tilde{h}}^l_k
     -\tretina\delta^l_k{\tilde{h}}^n_n\ .
\end{equation}
We set
\begin{equation}\label{3.37}
{\cal T}_k=\nabla_l {\tilde {h}}{}^l_{T k}\ .
\end{equation}
Now we give Einstein's perturbation equations, separating
$\delta \tilde {G}{}^l_{T k}$, the~traceless part of
$\delta {\tilde G}^l_k$,  from the~trace $\delta {\tilde G}^n_n$  which
we combine with $\delta {\tilde G}^0_0$ for a reason to be seen below.
Thus, recalling that $\nabla^2=f^{kl}\nabla_{kl}$, we have the~following dimensionless equations
\begin{eqnarray}
a^2\kappa\delta {\tilde T}^0_0&=&a^2\delta {\tilde G}^0_0=
     \tretina\nabla^2 {\tilde h}^n_n
     +k{\tilde{h}}^n_n-2{\cal{H}}\mathcal{K}
     -\pul\nabla_k {\cal T}^k\ ,\label{3.38}\\
a^2\kappa\delta {\tilde T}^0_k&=&a^2\delta {\tilde G}^0_k=
     \pul\nabla^2{\tilde{h}}_{k0}
    +k{\tilde{h}}_{k0}
    +\sestina\nabla_{kl}{\tilde{h}}^l_0
   \nonumber\\
   &&\ \
   {}+\dtretiny\nabla_k \mathcal{K}
    -\pul \left({{\cal T}_k}\right)^{'}\ ,\label{3.39}\\
a^2\kappa\left(\delta {\tilde T}^0_0-\delta {\tilde T}^n_n\right)&=&
     a^2\left(\delta{\tilde G}^0_0-\delta {\tilde G}^n_n\right)=
     \nabla^2{\tilde h}_{00}
    \nonumber\\
    &&\ \
     {}+3a\left(\frac{1}{a}{\cal{H}}\right)^{\!'}{\tilde h}_{00}
     +\frac{2}{a}\left( a\mathcal{K}\right)^{'}\ ,\label{3.40}
\end{eqnarray}
and
\begin{eqnarray}
a^2\kappa\left(\delta{\tilde T}^l_k-
     \tretina\delta^l_k\delta {\tilde T}^n_n\right)&=&
     a^2\delta\tilde {G}{}^l_{T k}=-\pul\nabla^2
        \tilde {h}{}^l_{T k}
          +k\tilde {h}{}^l_{T k}+\frac{1}{2a^2}\left[ a^2
           \left(\tilde {h}{}_{T k}^{l}\right)^{'}\right]^{'}
\nonumber\\
&&\ \
      {}+f^{lm}\left(\nabla_{(m}{\cal T}_{k)}
      -\tretina f_{mk}\nabla_n {\cal T}^n\right)
\nonumber\\
&&\ \
     {}-\frac{1}{a^2}f^{lm}\left[ a^2\left(\nabla_{(m}{\tilde h}_{k)0}
       -\tretina f_{mk}\nabla_n{\tilde h}^n_0\right)\right]^{'}
\nonumber\\
&&\ \ {}+\pul f^{lm}
      \left(\nabla_{mk}-\tretina f_{mk}\nabla^2\right)
        \left({\tilde h}_{00}-\tretina {\tilde h}^n_n\right)\ .
\label{3.41}
\end{eqnarray}
For completeness we also write down the equation which can be derived from
Eq.~(\ref{3.39}): 
\begin{equation}\label{3.41aa}
a^2\kappa \;\delta \widetilde{T}_0^k
= -\frac{1}{2} \nabla ^2 \widetilde{h}^k_0 + \left[ k -
2a(\frac{1}{a}\mathcal{H})'\right] \widetilde{h}^k_0 -
\frac{1}{6}\nabla ^k \nabla _l \widetilde{h}^l_0 -\frac{2}{3}
\nabla ^k \mathcal{K}+ \frac{1}{2} (\mathcal{T}^k)'.
\end{equation}
This equation follows from Eq. (\ref{3.39}) by using the relation 
\begin{equation}\label{3.41bb}
\delta \widetilde{T}^0_k = -f_{kl}\;\delta \widetilde{T}^l_0 - \frac{2}{\kappa a^2} \left[ -k + a \left( \frac{{\cal{H}}}{a}\right)' \right] \widetilde{h}_{0k} .
\end{equation}

In the~case of perfect-fluid perturbations we define the local {\em coordinate\/}
velocity by
\begin{equation}\label{3.42}
{\tilde V}^k=\frac{dx^k(\eta)}{d\eta}\,.
\end{equation}
Notice that $\tilde V^k$ is not equal to $V^k$ [defined
in Eq.~(\ref{2.86})] 
 expressed in
coordinates $\tilde x^\mu$ because $\tilde V^k$ is defined with respect
to the conformal time. Since $\tilde x^k = x^k$, in both
coordinates we have simple relations,
\begin{eqnarray}
\tilde V^k &=& a V^k\,,\nonumber\\
\tilde V_n &=& f_{nm}\tilde V^m = f_{nm} a V^m = -a^{-1}\bar g_{nm} V^m = -a^{-1} V_n\,.\label{3.41a}
\end{eqnarray}
Nevertheless, the fluid's 4-velocity components (\ref{2.87})
transform as a general tensor like (\ref{3.25}).

In the case of perfect fluid the energy-momentum tensor perturbations become
\begin{eqnarray}
a^2\kappa\,\delta{\tilde T}^0_0 &=& a^2\kappa\,\delta\rho\,,\quad
  a^2\kappa\,\delta{\tilde T}^0_k=2\left( k+{\cal H}^2-{\cal H}^{'} \right)
    \left( -{\tilde V}_k+{\tilde h}_{k0} \right)\,,\nonumber\\
a^2\kappa\,\delta \widetilde{T} ^k_0 &=& 2(k+\mathcal{H}^2 - \mathcal{H}') \widetilde{V}^k\,,\quad
  a^2\kappa\,\delta\tilde T_k^l = -a^2\kappa\,\delta_k^l\delta p\,,\label{3.43}
\end{eqnarray}
so that the left-hand sides of equations (\ref{3.40}) and (\ref{3.41}) are
\begin{equation}\label{3.44}
a^2\kappa\left( \delta {\tilde T}^0_0 -\delta {\tilde T}^n_n\right)=
a^2\kappa\left(  \delta\rho+3\delta p\right)\ ,\ \
a^2\kappa\left( \delta {\tilde T}^l_k -
\tretina\delta^l_k\delta {\tilde T}^n_n\right)=0\ .
\end{equation}

We combined Einstein's
equations in such a way that 
 Eqs.~(\ref{3.38}) and (\ref{3.40}) contain 
scalars
under the transformation of spatial coordinates, 
whereas Eq.~(\ref{3.41}) involves tensorial quantities only.
In the perfect-fluid case, the `source' in Eq.~(\ref{3.41})
vanishes so that this equation represents propagation
of a free gravitational field, i.e.,~of gravitational waves described
by traceless quantities ${\tilde h}{}^l_{T k}$.
The first and the third term on the
r.h.s. of Eq.~(\ref{3.41}) combine into a d'Alembert
wave operator modified by the time dependence
of the expansion factor $a(\eta)$. More generally,
however, the perturbed fluid could
be an imperfect fluid which includes shear viscosity.
This can be described by an additional term
in $\delta T_\mu^\nu$, given by a symmetric shear
tensor $\delta\Sigma_\mu^\nu$ which is traceless, $\delta\Sigma_\mu^\mu = 0$, and
purely spatial in the fluid rest frame,
$U^\nu\delta\Sigma_\mu^\nu = 0$ (see, e.g.,~\cite{Bertschinger}).
Then the shear 
 would appear as a source in Eq.~(\ref{3.41}).

\subsection{Bianchi identities and conservation laws}

The perturbed contracted Bianchi identities, $\nabla _\nu G^\nu_\mu =
0$ for $\mu = 0$ and $\mu = k$, imply 
\begin{eqnarray}
&\!&\frac{1}{a^2}(a^2\delta\widetilde{G}^0_0)^{\displaystyle\cdot} + \frac{\dot{a}}{a}
(\delta\widetilde{G}^0_0 - \delta \widetilde{G}^n_n) -
\frac{1}{a}\nabla^k \delta\widetilde{G}^0_k \nonumber \\
&+&\frac{3}{2a}\kappa (\overline{\rho} + \overline{p})
\left(\dot{a}\widetilde{h}_{00}-\frac{2}{3}\mathcal{K}\right) = 0, \label{3.45} \\
&\;&\frac{1}{a^3}(a^4 \delta\widetilde{G}^0_k)^\cdot + \nabla_m
\delta\widetilde{G}^m_k - \frac{1}{2}\kappa (\overline{\rho} +
\overline{p})\nabla_k \widetilde{h}_{00} = 0.\label{3.46}
\end{eqnarray}
Replacing $\delta
\widetilde{G}^\mu_\nu$ 
 by $\delta\widetilde{T}^\mu_\nu$
from the field equations we get the conservation laws
for $\delta\widetilde{T}^\mu_\nu$:
\begin{eqnarray}
&\!&(\delta\widetilde{T}^0_0)^{\displaystyle\cdot} +
\frac{\dot{a}}{a}(3\delta\widetilde{T}^0_0 -
\delta\widetilde{T}^k_k) - \frac{1}{a}\nabla^k\delta
\widetilde{T}^0_k \nonumber \\
&+&\frac{3}{2a}( \overline{\rho} + \overline{p}) \left(\dot{a}\widetilde{h}_{00} -\frac{2}{3}
{\cal K}\right) = 0,\label{3.47}
\end{eqnarray}
\begin{equation}\label{3.48}
\frac{1}{a^3}(a^4 \delta\widetilde{T}^0_k)^{\displaystyle\cdot} + \nabla_m
\delta\widetilde{T}^m_k - \frac{1}{2}(
\bar{\rho}+\bar{p})\nabla_k\widetilde{h}_{00} = 0.
\end{equation}
In the case of a perfect fluid the conservation laws become
\begin{eqnarray}\label{3.49}
&\!&(\delta\rho)^{\displaystyle\cdot} +\frac{3\dot{a}}{a}(\delta\rho + \delta p) +
\frac{1}{a}( \overline{\rho} + \overline{p}) \nabla_k(\widetilde{V}^k - \widetilde{h}^k_0) \nonumber \\
&+&\frac{3}{2a}( \overline{\rho} + \overline{p}) \left(\dot{a}\widetilde{h}_{00}-
\frac{2}{3}{\cal K}\right) = 0,
\end{eqnarray}
\begin{equation}\label{3.50}
\frac{1}{a^3}\left[ a^4 ( \overline{\rho} + \overline{p})(
\widetilde{V}_k - \widetilde{h}_{k0})\right]^{\displaystyle\cdot} + \nabla_k \delta p
+ \frac{1}{2} ( \overline{\rho} + \overline{p})\nabla_k
\widetilde{h}_{00} = 0.
\end{equation}
The first equation expresses the conservation of the mass-energy
$\delta\rho$. The second is the 
 equation of motion; when the
time-derivative term is negligible it 
 represents the
equilibrium condition between the gradients of pressure and
gravitational potential $\frac{1}{2}\widetilde{h}_{00}$, which
would be much harder to see in the synchronous gauge with $\widetilde{h}_{00} =
0$. Until now, all equations have been in an arbitrary gauge. The next section  is devoted 
to the choice of `appropriate gauges'.

\section{Gauges}

A change of the gauge can be regarded as an
infinitesimal coordinate transformation $x^\mu \rightarrow x'^{\mu} = x^\mu + \zeta ^\mu
(x)$. Under the gauge transformations, 
the metric changes by the Lie derivative (e.g.~\cite{Stewart}) as  $\Delta g_{\mu\nu} = \pounds_{\zeta}g_{\mu\nu}\equiv
\lim_{\zeta\rightarrow 0} \left[g_{\mu\nu}(x') - g'_{\mu\nu} (x')
\right] = \zeta_{\mu;\nu} + \zeta_{\nu;\mu}.$
The explicit formulas are given in Appendix \ref{App.B}.

Since gauge transformations contain four arbitrary functions, we
can impose four gauge conditions. Regarding the field equations
(\ref{3.38}) - (\ref{3.41}) we see instantly
that four gauge conditions ${\cal K} = 0 = \mathcal{T}_k$ decouple the
first three equations from the rest. Equations (\ref{3.38}),
(\ref{3.39}), and (\ref{3.40}) determine directly the metric components
$\widetilde{h}_{00}, \widetilde{h}_{k0}$, and $\widetilde{h}^n_n$
from the \textit{instantaneous} distribution of sources given by $\delta\widetilde{T}^0_0,
\delta\widetilde{T}^0_k$ and $\delta\widetilde{T}^0_0 -
\delta\widetilde{T}^n_n$; no time integration is needed.
Accelerations and rotations of local inertial frames follow then
from (\ref{2.50}) and (\ref{2.56}). Such an instantaneous determination of local
inertial frames is also possible by employing other
gauges. We call these gauges \textit{Machian}.

The purpose of this section is to motivate and describe geometrically
several Machian gauges, and to clarify what is the
residual gauge freedom these gauges admit. For a comparison we
shall also consider two typically {\em non\/}-Machian gauges --- the
synchronous gauge and the generalized Lorenz-de Donder, or ``harmonic'' gauge. In
the next section these gauges will be used to analyze the field
equations and the way they can be solved to determine local
inertial frames. In the Machian gauges we shall always
restrict the spatial part of the metric by requiring the three
gauge conditions $\mathcal{T}_k = 0$, where $\mathcal{T}_k$ is
given by Eq. (\ref{3.37}).
 These conditions will be motivated first.

\subsection{Gauge conditions on the spatial metric}

We start beyond the linear
perturbation theory. Smarr and York (1978) \cite{SY}, in treating full
general relativity as an evolution from given initial Cauchy data
on a spacelike slice 
 to the next slice,
studied the kinematics of the observers threading the slices. The
evolution is represented in terms of coordinates attached to
these ``coordinate observers''. 
Kinematical and dynamical effects can be suitably separated if a
relative velocity of the coordinate observers, with respect to the
(Eulerian) observers whose worldlines are perpendicular to the
given slicing, is such that the shear of coordinates arising if
one goes from one slice to the next is 
minimized. For a given slicing the relative
velocity is determined by the shift vector 
 and,
therefore, Smarr and York require ``the minimal-distortion'' shift
vector. This 
 condition is equivalent to the
equation
\begin{equation}\label{4.10}
              D^j \dot{\tilde{\gamma}}_{ij} = 0,
\end{equation}
where $\tilde \gamma_{ij} = {(\text{det} \gamma)}^{- \frac {1}{3}}
\gamma_{ij}$ is the conformal 3-metric on a given slice, the
dot denotes the time derivative, and $D^j$ denotes the 
 covariant
derivative with respect to the spatial metric $\gamma_{ij}$
induced on given slicing by four-dimensional metric $g_{\mu
\nu}$. 
The condition (\ref{4.10}) is a natural choice from a number of points of view.  
We refer the reader to the original paper \cite{SY}
for the details; 
here we wish to make just a few comments. 
In the weak-field
limit in the wave zone, 
condition (\ref{4.10}) generalizes
and includes the well-known ``transverse-traceless''  gauges of Arnowitt,
Deser and Misner (1962) and of Dirac (1959) (see, e.g.,~\cite{MTW}).
In the linearized gravity in generally curved coordinates with $g_{\mu \nu} = \bar g_{\mu \nu} + h_{\mu \nu}$,
where $\bar g_{00} = 1, \bar g_{0i} = 0, \partial_t \bar g_{ij}
= 0$, the condition (\ref{4.10}) implies $\partial_t \bar D^j {h}{}_{T ij} = 0$,
where ${h}{}_{T ij} = h_{ij} - \frac{1}{3}\, h \bar g_{ij}$, $h \equiv
\bar g^{ij} h_{ij}$, and $\bar D^j$ is the spatial covariant
derivative with respect to $\bar g_{ij}$. 
This is analogous to the radiation (or Coulomb) gauge
condition in electrodynamics. 
In stationary spacetimes with a timelike Killing vector
$\xi ^\alpha$ the gauge condition (\ref{4.10}) is satisfied if the slicing
is carried into itself by the $\xi ^\alpha$ isometry and $\xi
^\alpha$ is tangent to coordinate observers. 
It is interesting to consider 
more general slicings. Choosing in
Schwarzschild spacetime the 
 slices orthogonal to the
geodesics of particles freely falling from rest at infinity,
one finds that the condition (\ref{4.10})
implies 
\begin{equation}\label{4.13}
ds^2 = (1 - 2M/r) d \tau ^2 -2(2M/r)^{\frac {1}{2}} dr d \tau -
dr^2 - r^2 (d \theta^2 + \sin^2 \theta \varphi ^2),
\end{equation}
i.e., one obtains the time-independent form (\ref{4.13}) 
with the
spatially flat metric on the slices.%
\footnote{More recently, (\ref{4.13}) was rediscovered as a
technically suitable form of the Schwarzschild metric for
describing the Hamiltonian dynamics for spherically symmetric
gravitating shells by Kraus and Wilczek (1995) \cite{KW} without any
geometrical argumentation.}

Now in our case of the perturbed FRW metric 
 it is easy to
see that the conformal 3-metric is $\tilde \gamma_{ij} = f^{- \frac {1}{3}} [f_{ij} - {\tilde
{h}}{}_{T ij}]$, 
 $f \equiv \text{det}(f_{ij})$, and ${\tilde {h}{}_{T ij}}$ is given by
(\ref{3.36}).
Therefore, the 
 condition (\ref{4.10}) implies
$\dot {\cal{T}}_k = 0$,
with ${\cal{T}}_k$ given by (\ref{3.37}). 
It is 
the last
condition which converts 
Eq. (\ref{3.39}) into the equation
for $\tilde {h}_{k0}$ without the terms depending on the traceless part of $\tilde {h}_{kl}$.

Motivated by the analysis 
 above, we shall assume that, 
  in fact, a slightly stronger condition,
\begin{equation}\label{4.17}
{\cal T}_k  = \nabla_l{\tilde{h}}{}^l_{T k} = 0,
\end{equation}
is satisfied. This just means that the spatial
coordinates are restricted on an initial slice and this
restriction is then maintained by the original condition. 
Notice that \eqref{4.17} is
covariant under 3-dimensional coordinate transformations within
chosen slices. 

\subsection{Gauge conditions on the time slicing}

The three gauge conditions $\mathcal{T}_k = 0,\,k=1,2,3$, do not restrict the time
coordinate, i.e., the slicing by spatial hypersurfaces $x^0 =
\text{constant}$. We thus supplement them with the fourth gauge condition
fixing the slices.
In order to understand its geometrical meaning, 
we first calculate the geometrical quantities characterizing the
slices. Using the perturbed FRW metrics 
in a general gauge,
we find that the unit timelike vector field $n^\mu$
orthogonal to each slice 
 is given by
\begin{equation}\label{4.18}
\widetilde{n}^\mu = \widetilde{\bar {n}}^\mu + \delta \widetilde{n}^\mu,
\end{equation}
where $\widetilde{\bar{n}}^\mu = \left( a^{-1},0,0,0\right),\,
\delta \widetilde{n}^\mu = a^{-1}\left(-\frac{1}{2} \tilde {h}_{00},
\tilde {h}^j_0\right)$.
Calculating the expansion $\theta = \tilde n^\mu_{;\mu}$
of the congruence of timelike curves that meet the slices
orthogonally, we find
\begin{equation}\label{4.20}
\theta = \bar {\theta} + \delta \theta = 3 \frac{a'}{a^2} -
\frac {3}{2a} \left(\frac {a'}{a} \tilde {h}_{00}+\frac{1}{3}\bigl(\widetilde{h}_n^n\bigr)' - \frac
{2}{3} \mathcal{P}\right) \nonumber = \bar \theta - \mathcal{K},
\end{equation}
where $\mathcal{K}$ is given by Eq.~(\ref{3.35}), and
\begin{eqnarray}\label{P}
\mathcal{P} =\nabla_l\widetilde{h}^l_0.
\end{eqnarray}
For the shear of the congruence we obtain $
\tilde\sigma_{\alpha \beta} = \delta \tilde\sigma_{\alpha \beta},\;\delta
\tilde\sigma_{00} = 0, \; \delta \tilde\sigma_{0i} = 0$,
\begin{eqnarray}\label{4.21}
\delta \tilde\sigma_{ij} & = & - a \nabla_{(i } \tilde h_{j)0} +
\frac {1}{3} a f_{ij}
\nabla_m \tilde h^m_0 + \frac{1}{2} a \tilde {h}'{}_{T ij} \nonumber \\
& = & - a \nabla_{(i } \tilde h_{j)0} + \frac{1}{3} a f_{ij} \mathcal{P}
+ \frac{1}{2} a \tilde {h}'{}_{T ij}.
\end{eqnarray}

The {\em uniform-Hubble-expansion gauge\/}, introduced by
Bardeen (1980) \cite{Bard}, but apparently not used much later (though see \cite{BardTurn}),
requires $\delta \theta = 0$, i.e.
\begin{equation}\label{4.22}
\mathcal{K} =\frac{3}{2} \frac {a'}{a} \tilde {h}_{00} + \frac{1}{2}(\widetilde{h}^n_n)' - \mathcal{P} = 0.
\end{equation}
This gauge condition is again motivated by 
the popular choice of the `constant
mean curvature slices' in the full theory (the trace of
the extrinsic curvature tensor -- the mean curvature -- of a
spacelike hypersurface  with normal $n^{\mu}$ 
is $K = -n^{\mu}_{; \mu}$). 
The
condition (\ref{4.22}) thus means that we choose such a time coordinate in
the perturbed FRW universes that the extrinsic curvature of the
$\eta =$ constant hypersurfaces is the same as in the unperturbed
universe, i.e., it is constant along each hypersurface. Much work
has been done on the existence and properties of such 
 foliations 
(see \cite{Ren} for
the recent review and 
 references).

The gauge condition (\ref{4.22}) for time slicing combined with the
gauge conditions (\ref{4.17}) for spatial part of the metric will be
called the \textit{Mach 1 gauge}. We have not found it in the
literature, although the gauge conditions (\ref{4.17}) and (\ref{4.22}) were used separately.

Another basic geometrical object associated with a spacelike slice
is its intrinsic curvature, the simplest measure of which is the
intrinsic (3-dimensional) scalar curvature. In the perturbed FRW
universes ${\cal R} = {\cal\overline{R}} + \delta {\cal R},\;
{\cal\overline{R}} = -\frac{6k}{a^2}$, 
 and
\begin{equation}\label{4.24}
\delta {\cal R} = -\frac{2}{3a^2}(\nabla^2 \widetilde{h}^n_n +
3k\widetilde{h}^n_n) + \frac{1}{a^2}\nabla_n{\cal T}^n.
\end{equation}
When the gauge condition (\ref{4.17}) is combined with ``the uniform-intrinsic-scalar
curvature'' condition $\delta{\cal R} = 0$,
i.e.
\begin{equation}\label{4.25}
\nabla^2 \widetilde{h}^n_n + 3 k\widetilde{h}^n_n = 0,
\end{equation}
we speak about the \textit{Mach 2 gauge}. A stronger version 
 -- the Mach $2^*$ gauge -- requires
\begin{equation}\label{4.26}
\widetilde{h}^n_n = 0.
\end{equation}
Another possible 
condition 
 for the choice of slicing is $\nabla_i\nabla_j
K^{ij} = 0$, where $K^{ij}$ is the extrinsic curvature tensor.
Nothing appears to be known about this choice in the nonlinear
context. In our formalism 
this condition
reads [using Eq. (\ref{4.21}) for the shear]
\begin{equation}\label{4.27}
0 = \nabla^i\nabla^j \delta\tilde\sigma_{ij} = -\frac{2}{3}a (\nabla^2 + 3k) \nabla_l \widetilde{h}^l_0 +
\frac{1}{2}a \nabla^l {\cal{T}}_l',
\end{equation}
which justifies the name ``minimal-shear hypersurface condition''
suggested by Bardeen \cite{Bard}. Combined with the conditions
(\ref{4.17}) the last equation implies the gauge condition
\begin{equation}\label{4.28}
(\nabla^2 + 3k) \nabla_l\widetilde{h}^l_0 = 0.
\end{equation}
The \textit{Mach 3 gauge} is defined by the gauge conditions (\ref{4.17}) and (\ref{4.28}).
Its stronger version, the \textit{Mach 3$\;^*$ gauge},

\begin{equation}\label{4.29}
{\cal P} = \nabla_l \widetilde{h}^l_0 = 0,
\end{equation}
combined with (\ref{4.17}), has been called  the \textit{Poisson
gauge} by Bertschinger \cite{Bertschinger} in 1995. He analyzed 
 its
advantages for physical interpretation of cosmological
perturbations, in particular, as compared with the
synchronous gauge. The same gauge 
 has already been proposed in 1994 by Bombelli,
Couch, and Torrence \cite{Bom} who called it the ``cosmological
gauge''.

We also
mention the standard \textit{synchronous gauge}, still used most commonly
in cosmology,
\begin{equation}\label{4.30}
\widetilde{h}_{00} = \widetilde{h}_{0i} = 0,
\end{equation}
and, in more detail, the generalized \textit{Lorenz-de Donder gauge} (frequently also called the
harmonic gauge -- cf., e.g., \cite{weinb}, recently \cite{BiKa}),
\begin{equation}\label{4.31}
\overline{\nabla}_\mu\delta ( \sqrt{-g}g^{\mu\nu}) = 0.
\end{equation}
This has been extensively used in a number of problems, 
in particular, in weak-field approximations dealing with equations of motion
and gravitational radiation (see, e.g.,
\cite{Bla}), but not in cosmology. The
gauge conditions (\ref{4.31}) for $\mu = 0$ imply
\begin{equation}\label{4.32}
-\nabla_l \widetilde{h}^l_0 + \begin{matrix}{\frac{1}{2}}\end{matrix}\bigl(\widetilde{h}_{00}+ \widetilde{h}_n^n\bigr)' + {\cal{H}}\bigl(3\widetilde{h}_{00} + \widetilde{h}^n_n\bigr) = 0,
\end{equation}
for $\mu = k$ we get
\begin{equation}\label{4.33}
-\nabla_l\,\underset{T}{\widetilde{h}}{\,}^{\,l}_k + \begin{matrix}{\frac{1}{6}}\end{matrix}\nabla _k (\widetilde{h}^n_n - 3 \widetilde{h}_{00})
+\widetilde{h}{\,}\raisebox{0.8ex}{$'$}{\!\!}_{0k} + 4 {\cal{H}}\widetilde{h}_{0k} = 0.
\end{equation}
We shall now analyze the residual gauge freedom which the gauges
introduced above admit.


\subsection{Gauge-fixing and residual gauge freedom}

From relations (\ref{4.4}) -- (\ref{4.7}) in Appendix \ref{App.B} we readily obtain the changes
of the geometrical quantities defined in (\ref{3.35}), (\ref{3.37}), 
(\ref{P}), and (\ref{4.21}) under gauge transformations:
\begin{equation}\label{4.34}
\Delta\mathcal{T}_k = \Delta(\nabla_l {\widetilde{h}}{}_{T k}^l) =
-(\nabla^2\zeta_k + 2k\zeta_k + \frac{1}{3}\nabla_k
\nabla_l\zeta^l),
\end{equation}
\begin{equation}\label{4.35}
\Delta{\cal P} = \Delta(\nabla_l\widetilde{h}^l_0) = \frac{1}{a}
\nabla^2\zeta^0 - a \nabla_k\dot{\zeta}^k,
\end{equation}
\begin{equation}\label{4.36}
\Delta{\cal K} = -\frac{1}{a} \left[ \nabla^2 \zeta^0 + 3a^2 \left (
\frac{\dot{a}}{a}\right )\spdot \zeta^0 \right],
\end{equation}
\begin{equation}\label{4.37}
\Delta{\cal R} = \frac{4}{3a^2} (\nabla^2 + 3k) (\nabla_l\zeta^l +
3\frac{\dot{a}}{a}\zeta^0),
\end{equation}
\begin{equation}\label{4.38}
\Delta (\nabla^k\nabla^l\delta\widetilde{\sigma}_{kl}) = -\frac{2}{3}(\nabla^2
+ 3k) \nabla^2\zeta^0.
\end{equation}

We shall 
discuss
first \textit{the minimal-distortion spatial gauge condition: $\mathcal{T}_k =
0$}.

Starting from a general gauge, we reach the required condition by
purely spatial gauge transformations given by $\zeta_k$, which
satisfy the inhomogeneous equation with given l.h.s. $\Delta
\mathcal{T}_k$. The residual gauge freedom is determined by $\zeta_k$
solving the homogeneous equation
\begin{equation}\label{4.39}
\nabla^2 \zeta_k + 2k\zeta_k + \frac{1}{3}\nabla_k\nabla_l\zeta^l
= 0.
\end{equation}
There are solutions of this equation given by linear combinations
(with time-dependent coefficients) of the conformal Killing
vectors in the constant-curvature spaces $S^3, R^3, H^3$. To see
this, recall that in 3-dimensional space a conformal Killing
vector satisfies
\begin{equation}\label{4.40}
\nabla_l\zeta_k + \nabla_k\zeta_l = \frac{2}{3}f_{kl}
\nabla_n\zeta^n.
\end{equation}
Since spaces of constant curvature are conformally flat, they
admit ten linearly independent conformal Killing vectors as $E^3$
(see, e.g., \cite{Eisenh}). Their explicit forms 
 are given in Appendix~\ref{App.C}, where their relationship to
the scalar and vector hyperspherical harmonics is also elucidated. Among
the ten conformal Killing vectors, six are pure Killing vectors,
$\xi^{(A)}_i, A = 1,2,...,6$; the remaining four $\psi^{(B)}_i,
B=
1,...,4$ do not reduce to the Killing vectors. In spaces of
nonvanishing constant curvature, $\psi^{(B)}_i$ can be written as
gradients of scalars:
\begin{equation}\label{4.41}
k = \pm 1:\;\;\;\;\psi^{(B)}_i = \partial_i Q^B = \nabla_iQ^B,
\;\;\; B = 1,...4.
\end{equation}
The four scalar fields, $Q^B$, are equal, up to a multiplicative
constant, to the following four scalar harmonics (see Appendix \ref{App.C} for details):
\begin{align}
k=+1&:\;Q_{(L=1,\; l=0,\; m=0)} \simeq \cos \chi,\;Q_{(L=1,\; l=1,\; m=-1,\;0,+1)} \simeq  \sin \chi Y_{1m}(\theta,\varphi),\label{4.42}\\
k=-1&:\;Q_{(\lambda = 2i,\; l = 0,\; m = 0)} \simeq \cosh \chi,\; Q_{(\lambda = 2i,\; l = 1,\, m = -1,0,+1)} \simeq \sinh \chi\; Y_{1m}(\theta,\varphi).\label{4.43}
\end{align}
Taking the divergence of 
Eq. \eqref{4.40} and using \eqref{3.2} and \eqref{3.3} in Appendix \ref{App.A} to commute the derivatives,
we obtain Eq. (\ref{4.39}). Hence, any conformal Killing vector $\zeta_k$
solves Eq. (\ref{4.39}). In open universes, all such solutions $\zeta_k$
diverge at infinity $(\chi\rightarrow \infty)$, except for
translations in flat $(k = 0)$ universe when $\zeta_k$ are constant
in Cartesian-like coordinates. We now prove that \textit{there exist no
bounded solutions of Eq.} (\ref{4.39}) \textit{other than conformal Killing
vectors in $S^3$ and translations in} $E^3$.
To prove this we decompose $\zeta_k$ into a gradient of a scalar
and a transverse vector:
\begin{equation}\label{4.44}
\zeta_k = \nabla_k Z + {\zeta}{}_{T k}, \;\;\;\nabla^k
{\zeta}{}_{T k}= 0.
\end{equation}
This decomposition is unique up to $ Z \rightarrow  Z+$constant, if for open universes
we require $\zeta_{k}$ to decay asymptotically so that $\int \nabla^k \zeta_k
dV$ converges \cite{Stewart}. Substituting then (\ref{4.44}) into
(\ref{4.40}) and 
commuting the derivatives, we find
\begin{equation}\label{4.45}
\nabla^2\nabla^2 Z + 3 k\nabla^2Z = 0.
\end{equation}
In $S^3$ the only smooth solutions of the equation
\begin{equation}\label{4.46}
\nabla^2 Q + 3 Q = 0
\end{equation}
are given by the linear combination of the four scalar harmonics
(\ref{4.42}), the gradients of which give the conformal Killing
vectors. In closed space the solution of (\ref{4.45}) is thus, with $\beta_B(t)$ arbitrary, 
\begin{equation}\label{4.47}
Z = \sum^{4}_{B=1} \beta_B(t)Q^B + Z_0(t).
\end{equation}
In $H^3$ equation (\ref{4.45}) becomes
\begin{equation}\label{4.48}
\nabla^2Q - 3Q = 0.
\end{equation}
The solutions are four scalar harmonics given in (\ref{4.43}) -- these,
however, diverge at infinity. In open universes the only solution
of (\ref{4.48}) leading to asymptotically well-behaved $\zeta_k$ is $Z =
Z_0(t)$, the gradient of which does not contribute to $\zeta_k$.
Substituting now for $\zeta_k$ in Eq. (\ref{4.39}) the decomposition
(\ref{4.44}), and regarding the above results for $Z$, we find that
 Eq. (\ref{4.39}) reduces to the equation $
\nabla^2 {\zeta}{}_{T k} + 2k{\zeta}{}_{T k}= 0$.
In open universes this equation does not admit any asymptotically
well-behaved solutions, except for $\zeta{}_{T k}$ = constant for $k = 0$. In a closed universe the equation is
equivalent to the Killing equation (Appendix \ref{App.C}).
Hence, 
under the assumption that the
vector $\zeta^i$  
 is bounded,
the 
 condition (\ref{4.17}) 
  fixes the spatial
coordinates uniquely in $H^3$; and in $E^3$ the remaining gauge
freedom is just $\zeta^i = \sum_{A=1}^3 f_A (t)
\zeta^{(A)i}_{\text{tr}}$, corresponding to a time-dependent linear
combination of translations. In $S^3$ the residual gauge freedom
is given by a linear combination of ten conformal Killing vectors
(six Killing and four conformal Killing):
\begin{equation}\label{4.50}
\zeta^i = \sum^{6}_{A=1} \alpha_A (t) \xi^{(A)i} +
\sum^{4}_{B=1} \beta_B (t) \psi^{(B)i}.
\end{equation}
We now discuss 
the three Machian gauges 
successively.\\

{\boldit 1. Mach 1: Uniform-Hubble-expansion gauge}\\
From Eq. (\ref{4.36}) we see that the residual freedom in
$\zeta^0$ is given by the solutions of 
\begin{equation}\label{4.51}
\nabla^2 \zeta^0 + 3a^2 \left( \frac{\dot{a}}{a}\right )\spdot\zeta^0 = 0.
\end{equation}
Multiplying Eq.~(\ref{4.51}) by $\zeta^0$ and integrating by parts over a
domain ${\cal D}$, we find
\begin{eqnarray}\label{4.52}
&\!&\int_{{\cal D}}\zeta^0\nabla^2\zeta^0 d^{(3)} V = \int_{{\partial{\cal
D}}}\zeta^0\nabla_k\zeta^0 dS^k - \int_{{\cal D}}f^{kl}
\nabla_k\zeta^0\nabla_l\zeta^0 d^{(3)} V \nonumber \\
&=&- \int_{{\cal D}}
3a^2 \left ( \frac{\dot{a}}{a}\right )\spdot(\zeta^0)^2 d^{(3)} V,
\end{eqnarray}
where $d^{(3)} V=\sqrt{f}\,d^{3}x,\,f=det(f_{kl});\,a_{0}^{3}\,d^{(3)}V,\,a_{0}=a(x^{0})$ is the proper volume in a slice $x^0 =$ constant.
Taking ${\cal D}$ to be all space, then the integral over the
boundary vanishes in open spaces because of the boundary condition
on $\zeta^0$, and it is zero in closed spaces because there is no
boundary. Therefore, 
\begin{equation}\label{4.53}
\int_{{\cal D}} f^{kl} \nabla_k \zeta^0 \nabla_l \zeta^0 d^{(3)} V = 3a^2
\left ( \frac{\dot{a}}{a}\right )\spdot \int_{\cal D}(\zeta^0)^2 d^{(3)} V.
\end{equation}
The 
factor on the r.h.s. can be
rewritten using 
 the FRW background equations:
\begin{equation}\label{4.54}
\!{\cal A} (t) \equiv 3a^2 \left ( \frac{\dot{a}}{a}\right )\spdot = 3a^2 \dot{H} =
3k - \textstyle{\frac{3}{2}} a^2\kappa ( \overline{\rho} + \overline{p})
=- 3a^2 (H^2-\textstyle{\frac{1}{3}\Lambda)} - \textstyle{\frac{1}{2}} a^2 \kappa ( \overline{\rho} + 3
\overline{p}).
\end{equation}
In all standard 
 models the strong energy condition $\overline{\rho} + 3\overline{p} >
0$ is valid so that ${\cal A} < 0$ ($H^2 - \frac{1}{3}\Lambda > 0$ is satisfied in 
 realistic models). In inflationary universe
models with $\overline{\rho} + \overline{p} = 0 \;(\Lambda = 0)$, the function ${\cal A}(t) <
0$ for open universes. In all these cases the r.h.s. of Eq.
(\ref{4.54}) is nonpositive, whereas the l.h.s. is non-negative.
Therefore, the only solution of Eq. (\ref{4.51}) is $\zeta^0 = 0$. In the
standard inflationary model with $k = 0, \;\overline{\rho} + \overline{p} =
0$, we have ${\cal A} = 0$, and $\zeta^0 = \zeta^0(t)$ is an
admissible solution of Eq. (\ref{4.51}) which is bounded and has a
vanishing gradient (reflecting the higher symmetries of de Sitter
space to which the FRW models reduce). If $k = +1$ \textit{and} $\overline{\rho} + \overline{p} =
0$, the relation (\ref{4.54}) turns Eq. (\ref{4.51}) (for any~$\Lambda$) into
$\nabla^2 \zeta^0 + 3\zeta^0 = 0$,
which is Eq. (\ref{4.46}), the solutions thus being
\begin{equation}\label{4.56}
\zeta^0 = \sum^{4}_{B=1} \sigma_B (t) Q^B,
\end{equation}
where 4 scalar harmonics $Q^B$ are given in (\ref{4.42}), $\sigma_B$
 are arbitrary.
 
 Let us summarize. Assuming $\zeta^\mu$ bounded, the Mach 1
 gauge 
fixes the coordinates uniquely in the open universes with $k =
 -1$, and for  $k = 0$ 
  it determines the
 spatial coordinates up to time-dependent translations, $\zeta^k
 (t)$, whereas the time slicing is unique if the background matter
 satisfies the strong energy condition; 
 $x^0$ can be shifted by $\zeta^0 (t)$ in the inflationary
 universe. In closed universes the spatial coordinates
 are determined up to the time-dependent motions (\ref{4.50}) given
 by the Killing and conformal Killing vectors; the time slicing is
 unique in the standard backgrounds with the strong energy
 condition satisfied. In the inflationary backgrounds the time can
 be shifted by $\zeta^0$ determined by Eq. (\ref{4.56}). 
\\

 {\boldit 2. Mach 2: Uniform-scalar-curvature gauge}\\
 Requiring the scalar 3-curvature of the time slices to be equal
 to the background values fixes the gauge 
  up to the
 transformations satisfying 
  [see Eq. (\ref{4.37})]
\begin{equation}\label{4.57}
(\nabla^2 + 3k) (\nabla_l\zeta^l + 3\frac{\dot{a}}{a}\zeta^0) = 0.
\end{equation}
Assuming again the 
  condition (\ref{4.17}), we restricted
$\zeta^l$ already by Eq. (\ref{4.39}), which implies the divergence
$\nabla_l\zeta^l$ to satisfy $(\nabla^2 + 3k)\nabla_l\zeta^l = 0$.  
 Equation (\ref{4.57}) thus reduces to
(assuming $\dot{a}\neq 0$) $(\nabla^2 + 3k) \zeta^0 = 0$.
As discussed above [cf. (\ref{4.46}) or (\ref{4.48})], the only bounded
solutions are $\zeta^0 = 0$ if $k = -1$, $\zeta^0 = \zeta^0 (t)$
if $k = 0$, and $\zeta^0$ is given in terms of 
$Q^B$ for $k=+1$. 

{\it Mach $2^*$: The traceless gauge}. -- The gauge condition $\widetilde{h}^n_n = 0$ implies the previous
one, and is stronger. Indeed, regarding Eq. (\ref{4.7}), we see that the
residual gauge freedom is given by
$\nabla_k \zeta^k + 3\frac{\dot{a}}{a} \zeta^0 = 0$,
which determines $\zeta^0$ in terms of $\zeta^k$ (assuming
$\dot{a} \neq 0$). With the gauge conditions (\ref{4.17}), $\zeta^0 =
0$ in open spaces, in $S^3$ the residual freedom in $\zeta^k$ is
given by Eq. (\ref{4.50}) which implies $\nabla_k\zeta^k = {\sum_{B=1}^{4}} \beta_B (t)
\nabla^2Q^B$ [see Eq. (\ref{4.41})], and thus leads to [using (\ref{4.46})]
\begin{equation}\label{4.60}
\zeta^0 = (a/\dot{a}) \sum^{4}_{B=1} \beta_B(t)Q^B,
\end{equation}
$\beta_B (t)$ are arbitrary.\\

{\boldit 3. Mach 3: The minimal-shear gauge}\\
As seen from Eq. (\ref{4.38}) this gauge condition allows the 
transformations restricted by 
\begin{equation}\label{4.60a}
(\nabla^2 + 3k) (\nabla^2 \zeta^0 - a^2 \nabla_k \dot{\zeta}^k) = 0,
\end{equation}
which, using $\zeta^k$ that satisfy
(\ref{4.17}), reduces to 
$(\nabla^2 + 3k) \nabla^2 \zeta^0 = 0$.
This is the same as~(\ref{4.45}). In open spaces the only bounded
solutions are $\zeta^0 = \zeta^0(t)$. In closed spaces
\begin{equation}\label{4.62}
\zeta^0 = \sum^{4}_{B=1} \sigma_B (t) Q^B + \sigma^0(t),
\end{equation}
where $\sigma$'s are arbitrary, and $Q^B$ given by Eq. (\ref{4.42}).

\textit{Mach 3*: The Poisson gauge}.--
The condition $\nabla_l h^l_0 = 0$ 
 admits a smaller 
 freedom. Equation (\ref{4.60a}) becomes 
\begin{equation}\label{4.63}
\nabla^2\zeta^0 - a^2 \nabla_k\dot{\zeta}^k = 0,
\end{equation}
which, after substituting for $\zeta^k$ from Eq. (\ref{4.50}),  for
$k = +1$ gives
\begin{equation}\label{4.64}
\zeta^0 = \sum^{4}_{B=1} a^2 \dot{\beta}_B(t) Q^B + \sigma^0(t),
\end{equation}
where $\sigma^0$ is arbitrary but $\beta_B(t)$ are the same
functions as those in $\zeta^k$ in Eq. (\ref{4.50}) -- in contrast to
Eq. (\ref{4.62}) where $\sigma$'s are independent of $\beta$'s. In open
universes the only residual freedom in the choice of time is given
by arbitrary $\zeta^0(t)$.

Regarding the gauge freedom in ${\cal T}_k = 0$, we see that the
Poisson gauge in the case $k = -1$ fixes the spatial coordinates
uniquely; the time coordinate is fixed up to $\zeta^0(t)$. In the case
$k = 0$ the spatial coordinates are fixed up to translations
$\zeta^i(t)$ and time shifts $\zeta^0(t)$. In the closed case the
freedom in spatial coordinates is determined by linear
combinations of the Killing and conformal Killing vectors
(\ref{4.50}), whereas the time coordinate by the combination
(\ref{4.64}) of scalar harmonics $Q^B$. Hence, in the closed case
there are 11 arbitrary functions of time which represent the gauge
freedom.

These results are at variance with Bertschinger's statement \cite{Bertschinger} that
there is ``an almost unique transformation from an arbitrary gauge
to the Poisson gauge''. 
 Clearly, Bertschinger does not consider
the possibility that his $\beta$ solves equation $(\nabla^2 + 3k)\nabla^2\beta =
0$ [i.e.~our Eq.~(\ref{4.45})] which preserves the gauge condition
$\nabla_l {h}_{T k}^l = 0$. Solutions for $\beta$ for $k =\pm 1$
are as in Eq. (\ref{4.47}), where $Z_0(t)$ indeed has no effect but the
terms containing $Q^B$ do have an effect -- not only on $\zeta_k$
but also on $\zeta^0$ as described above.

Let us now mention the gauge freedom 
in two typical `non-Machian' gauges.\\

{\boldit 4. Synchronous gauge}\\
From Eqs. (\ref{4.4}) and (\ref{4.5}) it is immediately seen that the synchronous
gauge admits the well-known residual freedom given by
transformations satisfying
\begin{equation}\label{4.65}
\dot{\zeta}^0 = 0,\;\;\;\;\; \nabla_l\zeta^0 = a^2\dot{\zeta}_l,
\end{equation}
which imply
\begin{equation}\label{4.66}
\zeta^0 = \zeta^0(x^i), \;\;\; \zeta_l = \left[\int\frac{dt}{a^2(t)}
\right] \nabla_l\zeta^0 (x^i) + Z_l(x^i).
\end{equation}
Functions $\zeta^0(x^i)$ and Z$_l(x^i)$ are arbitrary. The
gauge freedom is the same for both open and closed universes.\\

{\boldit 5. The generalized Lorenz-de Donder gauge}\\
Requiring the gauge conditions (\ref{4.32}) and (\ref{4.33})
to be satisfied, we can use relations (\ref{4.4}) - (\ref{4.7}) 
 to find
out the residual freedom in this gauge. It turns out to be
restricted by 
\begin{equation}\label{4.67}
\nabla^2 \zeta^0 - a^2 \ddot{\zeta}^0 - 3
a\dot{a}\dot{\zeta}^0 +3(a\ddot{a} + \dot{a}^2)\zeta^0 - 2
a\dot{a}\nabla_l\zeta^l= 0,
\end{equation}
\begin{equation}\label{4.68}
\nabla^2 \zeta_k - a^2 \ddot{\zeta}_k + 2k\zeta_k - 5 \dot{a}a
\dot{\zeta}_k + 2( \dot{a}/a)\nabla_k\zeta^0 = 0.
\end{equation}
The only feasible way to solve this coupled system appears to
be the use of harmonics, but here we shall just restrict ourselves
to noticing that, for a slowly changing expansion factor
($\dot{a}$, $\ddot{a}$ small), the system turns just into two
decoupled wave equations,
\begin{equation}\label{4.69}
\frac{1}{a^2}\nabla^2 \zeta^0 - \ddot{\zeta}^0 = 0,\;\;\;
\frac{1}{a^2}\nabla^2\zeta_k - \ddot{\zeta}_k +
\frac{2k}{a^2}\zeta_k = 0.
\end{equation}

In the flat case these are just wave equations in flat space with
coordinates $ax^i$ (which give the proper lengths in $k = 0$
universes). The gauge freedom is thus analogous to the freedom of the
Lorenz gauge in electrodynamics. Any solution of a wave equation
can be characterized by its Cauchy values -- here $\zeta^0(x^i)$,
$\dot{\zeta}^0(x^i), \zeta_k(x^i)$, and $\dot{\zeta}_k(x^i)$, i.e., 
by
eight
functions of spatial coordinates.

Summarizing, 
 we find that the \textit{Machian gauges are substantially more
restrictive than
the synchronous gauge and the  generalized Lorenz-de Donder gauge}. The last two
gauges admit transformations characterized by several (two and~eight)
\textit{arbitrary functions of three variables} -- of the spatial coordinates
$x^i$. All the Machian gauges admit only several arbitrary
functions of time. In some cases they fix the coordinates
uniquely.
An arbitrary additive function of time, $\zeta^0(t)$, 
 represents just the changes of the units of
time: $dt' = (1+\dot{\zeta}^0)dt$. The spatially homogeneous changes of
$x^i$ by $\zeta^i (t)$
describe just the shifts of the origin of spatial coordinates. 
Otherwise, \textit{all three Machian gauges fix the coordinates
uniquely in the hyperbolic universes $H^3$} as a consequence of
 boundary conditions at infinity. The remaining functions
of time in spherical universes $S^3$ will be interpreted in the
following.

\subsection{Integral gauge conditions in closed universes with standard spherical topology}

 In the closed spherical spaces
the spatial coordinates $x^i$ are fixed up to the transformations
$x^i \rightarrow x'^{i} = x^i + \zeta^i$, where $\zeta^i$ is given
by a linear combination of six Killing and four (proper) conformal
Killing vectors of $S^3$, in which the coefficients 
 are 
  arbitrary
functions of time.

In order to acquire an insight into the effects such coordinate
changes can produce, consider an unperturbed FRW universe with
standard spherical topology $(k = +1)$. Transform the metric in the
hyperspherical coordinates (\ref{1.3}) by a gauge
transformation generated by one translational, one rotational, and
one conformal Killing vector which have the simplest forms in the
hyperspherical coordinates: $\zeta^i_{\text{tr}} =
 (\cos \theta, - \cot \chi \sin\theta, 0), \zeta^i_{\text{rot}} = (0,0,1),
 \zeta^i _{\text{conf}} = (-\sin \chi, 0,0)$
  (cf. Appendix \ref{App.C}). Admitting the time-dependent
  coefficients, the transformation has the form
\begin{eqnarray}\label{4.87}
&\!&\chi' = \chi + \alpha (t) \cos\theta - \gamma(t) \sin \chi,
\nonumber \\
&\;&\theta' = \theta - \alpha(t) \cot \chi \sin \theta, \;\;\;
\varphi' = \varphi + \beta(t),
\end{eqnarray}
which can easily be inverted since $\alpha,\beta,\gamma$ are
small. In addition to the transformation (\ref{4.87}) we consider a
change of the time coordinate (time slicing) allowed by our Machian
gauge conditions in closed universes.

Hence, we take $\zeta^0$ of the form (\ref{4.62}) because other
possibilities (\ref{4.56}), (\ref{4.60}), and (\ref{4.64}) are included in
(\ref{4.62}). 
 However, since in (\ref{4.87}) only the
simplest conformal Killing vector 
enters, it is sufficient to take only those time transformations
which are associated with this vector and with the shift of the
time origin which is also allowed by Eq.~(\ref{4.60a}):
\begin{equation}\label{4.88}
t' = t + a^2 \left[ \delta (t) \cos \chi + \sigma (t) \right],
\end{equation}
where for convenience the expansion factor is pulled out. Under
the transformations (\ref{4.87}) and (\ref{4.88}) the standard FRW
metric with $k = +1$ becomes
\begin{eqnarray}\label{4.89}
&\!&ds^2 = \left[ 1 - (2a^2 \delta)\spdot \cos \chi ' - (2a^2
\sigma)\spdot \right] dt'^2 \nonumber \\
&-&a^2 \left[ 1 + 2(\gamma + a\dot{a}\delta) \cos \chi ' + 2
a\dot{a}\sigma\right]\left[ d\chi '^2 + \sin ^2 \chi ' (d\theta '^2 + \sin ^2 \theta
' d\varphi '^2) \right] \nonumber \\
&+&2a^2 \{\left[ \dot{\alpha}\cos \theta ' + (\delta -
\dot{\gamma})\sin \chi ' \right] d\chi ' \nonumber \\
&-&\dot{\alpha}\sin \chi ' \cos \chi ' \sin \theta ' d\theta ' +
\dot{\beta}\sin^2 \chi ' \sin^2 \theta ' d\varphi ' \} dt'.
\end{eqnarray}
Since $\alpha,...,\sigma$ are, in general, time-dependent and 
$h'_{00}, h'_{0i}$ nonvanishing, the
frames associated with $\chi ', \theta ',\varphi '$ fixed are noninertial  in
general,
 and the inertial frames, momentarily at rest with respect
to them, are seen to have the acceleration [cf.
Eq. (\ref{2.50})]
\begin{eqnarray}\label{4.90}
&\!&\alpha'^{\chi} = - (1/a^2) \left\{ \left[ 2a^2 (\delta -
\dot{\gamma}) \right]\spdot \sin \chi ' + (a^2\dot{\alpha})\spdot \cos
\theta'\right\}, \nonumber \\
&\;&\alpha'^{\theta } = -(1/a^2) (a^2\dot{\alpha})\spdot \cot \chi '
\sin \theta ', \;\;\; \alpha'^{\varphi} = -(1/a^2)
(a^2\dot{\beta})\spdot,
\end{eqnarray}
and to rotate with the vorticity [cf. Eq. (\ref{2.56})]
\begin{eqnarray}\label{4.91}
&\!&\omega'_{\chi\theta} = a^2 \dot{\alpha}\sin ^2 \chi ' \sin
\theta ', \;\;\; \omega'_{\chi\varphi} = a^2 \dot{\beta} \sin
\chi' \cos \chi ' \sin^2 \theta ', \nonumber \\
&\;&\omega'_{\theta\varphi} = -a^2 \dot{\beta} \sin ^2 \chi '
\sin \theta ' \cos \theta ',
\end{eqnarray}
i.e., with the vorticity 3-vector (\ref{2.68}) given by
\begin{equation}\label{4.92}
\omega'^{\chi} = (1/a) \dot{\beta}\cos \theta ', \;\; \omega'^{\theta}= (1/a) \dot{\beta} \cot \chi ' \sin \theta ',\;\;
\omega'^{\varphi} = (1/a) \dot{\alpha}.
\end{equation}
The above results are easily understood: time-dependent
rotations in the $\varphi$ direction, with $\chi = 0$, and the $\varphi$ axis
fixed, imply nonvanishing accelerations in this direction only, whereas
the corresponding vorticity vector has no $\varphi$ component. The translational Killing vector which for, say, $\theta =
0$ represents rotations in the $\chi$ direction (with  $\chi =
\pi/2$ fixed), leads to accelerations only in
the $\chi$ direction, and the vorticity vector in the $\varphi$ direction.

None of these acceleration or vorticity vectors can be compensated
by an allowed change (\ref{4.88}) of time slicing. As expected, the
shift of the time origin, $\sigma(t)$, does not enter Eqs.
(\ref{4.90})-(\ref{4.92}). However, the effect of the transformation
generated by the conformal Killing vector, which appears only in the
$\chi$ component of the acceleration, can be annulled by
choosing $\delta = \dot{\gamma}$.
Nevertheless, this condition does not remove the effect of
both $\delta$ and $\gamma$ in the conformal factor of the
spatial background metric. Metric (\ref{4.89}) implies a
nonvanishing trace of the form (omitting ``the time shift''
$\sigma)$
\begin{equation}\label{4.94}
h'^{n}_n = - \widetilde{h}'^{n}_n = 6 (\gamma +
a\dot{a}\delta)\cos \chi '.
\end{equation}
Therefore, the spatial metric differs from the canonical metric of
a homogeneous and isotropic 3-sphere. This metric is preserved only by
transformations representing real symmetries, i.e., those
generated by the Killing vectors.\footnote{As in flat space, the canonical
flat-space metric in Cartesian coordinates is preserved only by rigid
translations and rotations, but not by dilatations.}

We have described the particular effects of the gauge freedom
corresponding to the Killing and (proper) conformal Killing
vectors 
 in order to show 
 their different
character. 
In a general, linearly perturbed, FRW universe, the metric will be
much more complicated than that of Eq. (\ref{4.89}). If it contains terms
appearing in (\ref{4.89}) (and those corresponding to other Killing
and conformal Killing vectors), they can, of course, be removed by
gauge transformations of the form (\ref{4.87}) and (\ref{4.88}).
Since a natural goal in a relativistic perturbation theory is to
fix the gauge at the end as uniquely as possible, we shall now
require, in all Machian gauges, additional gauge conditions which
exclude the freedom corresponding to the gauge transformations
generated by (proper) conformal Killing vectors. However, we leave
the freedom corresponding to the 
Killing vectors since
these exhibit the symmetry of the background universe at any given
time.

The trace (\ref{4.94}) is, at given time, proportional to $\cos \chi$
[we omit primes in metric (\ref{4.89})], i.e., just to the first of the
scalar harmonics in Eq. (\ref{4.42}). 
 Neglecting
the time shift $\sim\sigma(t)$, the perturbation $h_{00} =
\widetilde{h}_{00}$ in (\ref{4.89}) is also proportional to this
harmonic. The term in $h_{0i} (= a\widetilde{h}_{0i})$ in
(\ref{4.89}) corresponding to the same conformal transformation of
the spatial coordinates and of the time slicing is proportional to the
gradient of $\cos \chi$ but the scalar ${\cal P} =
\nabla_l\widetilde{h}^l_0$  is again proportional to
this harmonic.\footnote{We do not consider the scalar
$\nabla_l\nabla_k{\widetilde{h}}^{lk}_{T}$since the gauge condition
(\ref{4.17}) guarantees that it vanishes in all Machian gauges.}
The metric perturbations which arise or may be removed by
gauge transformations generated by the conformal Killing vectors (i.e., the vectors of the form $\zeta^i = \sum^{4}_{B=1}\beta_B \psi^{(B)i} =
\sum^{4}_{B=1}\beta_B\nabla_iQ^B$)
will be eliminated by the following integral gauge conditions. These will be imposed at all times:
\begin{equation}\label{4.96}
\!\int_{S^3}\widetilde{h}^n_n Q^B d^{(3)} V =
\int_{S^3}\widetilde{h}_{00} Q^B d^{(3)}V
=\int_{S^3} {\cal P}Q^B d^{(3)} V = 0,
\end{equation}
where $\widetilde{h}^n_n, \widetilde{h}_{00}, {\cal
P} = \nabla_l\widetilde{h}^l_0$ are functions of all spacetime
coordinates, harmonics $Q^B (\chi, \theta, \varphi)$ are given in
Eq. (\ref{4.42}). 
 The integral gauge conditions
(\ref{4.96}) require that spatial scalars $\widetilde{h}^n_n,
\widetilde{h}_{00}$, and $\mathcal{P}$ are orthogonal to the
4-dimensional functional space spanned by $Q^B$, i.e., by harmonics
which are eigenfunctions with zero eigenvalues of the operator $(\nabla^2 +
3)$ in $S^3$. In Section V we shall notice that 
 conditions
(\ref{4.96}) are closely related to Traschen's integral constraints
\cite{Tr1}, \cite{Tr2.} which restrict 
 perturbations of
energy-momentum tensors representing sources. In this way we 
make sure that 
 conditions (\ref{4.96}) do not restrict
physics.

\subsection{Machian gauges in closed spherical universes: Summary}

After adopting the integral gauge conditions (\ref{4.96}), the gauge
freedom in all three Machian gauges becomes transparent and
simple: It 
 reflects the proper (Killing) symmetry of the
background universe at any fixed time. The minimal-distortion shift
gauge condition (\ref{4.17}), together with integral gauge conditions
(\ref{4.96}), which we assume in all Machian gauges, fix the spatial
coordinates uniquely up to transformations $x^i\rightarrow x'^{i} = x^i +
\zeta^i$ with
\begin{equation}\label{4.97}
\zeta^i = \sum^{6}_{A=1}\alpha_A(t) \xi^{(A)i},
\end{equation}
where $\xi^{(A)i}$ are six spatial Killing vectors 
-- three (quasi)rotations and three (quasi)translations.

In the 
 Mach 1 gauge  the time slicing is
unique even without requiring integral gauge conditions (\ref{4.96})
if the background matter satisfies the strong energy condition.
After requiring (\ref{4.96}), it is unique also in the inflationary
backgrounds. In the uniform-scalar-curvature gauge (Mach 2), with
(\ref{4.96}) satisfied, the time coordinate is unique, the same being
true for the special case of the traceless gauge (Mach $2^*$).
Finally, in the minimal-shear gauge (Mach 3) and its special case
of the Poisson gauge (Mach $3^*$), the adoption of the integral
conditions (\ref{4.96}) leaves the only freedom in Eqs. (\ref{4.62}) and
(\ref{4.64}) to be $\zeta^0 = \sigma^0(t)$, i.e., the time coordinate
is fixed up to trivial, `universal' shifts depending just on an
arbitrary 
 function of time. This, as noticed below
Eqs. (\ref{4.90})-- (\ref{4.92}), does not influence accelerations and rotations
of local inertial frames. Therefore, all our Machian gauges fix
coordinates uniquely up to the `time-dependent' Killing motions
(\ref{4.97}) of spatial coordinates. The Machian gauges with the
integral gauge conditions (\ref{4.96}) are thus determining
coordinates both more restrictively and more
plausibly than the synchronous and generalized Lorenz-de
Donder gauge. 

\subsection{On closed hyperbolic and flat universes}

Although in this work we generally assume the cosmological
backgrounds with standard topologies only, and thus with
geometries which are homogeneous and isotropic also globally, in
this intermezzo we consider 3-dimensional backgrounds represented by
closed flat $(k = 0)$ and hyperbolic $(k = -1)$
3-manifolds. Finite universes with multiconnected topologies have
become popular in recent years in the light of new theories
extending general relativity, and in the view of a possibility (in
principle) to determine the topology of our universe by means of
cosmic microwave background observations, or from the distribution of distant
sources. A comprehensive, nice review containing many references
appeared  recently \cite{Lev}. 

Globally, these universes are different; in
particular, they admit smaller families of continuous symmetries.
Closed hyperbolic 3-manifolds do not have smooth Killing vectors
at all \cite{Lev} and do not possess nontrivial solutions of
 the equation $\nabla^2 \phi + 3 k\phi = 0$. Indeed,
multiplying this equation by $\phi$, integrating by parts over a
3-dimensional domain $\mathcal{D}$ [cf. Eq. (\ref{4.52})], we get
\begin{equation}\label{4.98}
\int_{\partial{\cal D}}\phi\nabla^j \phi dS_j - \int_{\cal
D}\nabla_j\phi\nabla^j\phi \; d ^{(3)} V + 3 k \int_{\cal D} \phi^2 d ^{(3)}V = 0.
\end{equation}
Taking ${\cal D}$ to be whole closed space, the first integral
vanishes, because there is no boundary, and since both the second
and the third integrals are non-negative, Eq. (\ref{4.98}) for $k = -1$
can be satisfied only with $\phi = 0$. Hence, Eq. (\ref{4.48})
has only solutions $Q = 0$, so $Z$ in Eq. (\ref{4.44}) does not
contribute to $\zeta_k$. Analogously, equation 
 for the
transverse part of $\zeta_k$, after being multiplied by ${\zeta}_{T m}\gamma^{km}$,
and integrated by parts, becomes
\begin{equation}\label{4.99}
\!\int_{\partial{\cal D}}\gamma^{mk}{\zeta}_{T m} \nabla^j{\zeta}_{Tk}
dS_j - \int_{\cal D}\gamma^{mk} \nabla_j\zeta_
{T m}\nabla^j{\zeta}_{T k} d^{(3)}V 
+ 2k  \int_{\cal D}\gamma^{mk}{\zeta}_{T m}{\zeta}_{T k} d^{(3)}V = 0.
\end{equation}
Again, taking ${\cal D}$ to be all closed space, the first term is
zero, and as the other integrals are non-negative, 
(\ref{4.99}) for $k = -1$ is solved only by ${\zeta}_{T k} = 0$.
It is easily seen that, except for 
 trivial shifts $\sigma(t)$ in the 
  Mach 3
gauge, 
 \textit{in closed hyperbolic
 universes} our instantaneous \textit{Machian local gauge conditions
fix coordinates uniquely}, without integral gauge conditions being
imposed.

For closed flat $(k = 0)$ universes, Eq.
(\ref{4.98})
implies $\phi = \phi(t)$ and Eq. (\ref{4.99}) gives $\nabla_j {\zeta}_{T m} =
0$ so that $\zeta^{i} = \sum^{3}_{A=1}f_A(t)\xi^{(A)i}_{\text{tr}}$, where $ \xi^{(A)i}_{\text{tr}}, A = 1,2,3$
are translation Killing vectors. Hence the 
condition (\ref{4.17}) determines the spatial coordinates up to
arbitrary time-dependent linear combinations of translations. That
the rotational Killing vectors are globally ruled out can be well
understood in the simplest example of a compact flat 3-manifold --
a 3-torus $T^3$. 
All closed flat 3-manifolds are given in Fig. 26 in
\cite{Lev}. In fact, only the 3-torus admits globally a 3-parameter
family of translational symmetries given in Cartesian coordinates
by the 3~independent constant Killing vectors. 
Considering just
$T^3$, 
 we find that the gauge condition (\ref{4.17})
fixes spatial 
 coordinates 
  up to
\begin{equation}\label{4.100}
x^i\rightarrow x'^{i} = x^i + \zeta^i(t) = x^i + \sum^3_{A=1}
f_A(t)\xi^{(A)i}_{\text{tr}},
\end{equation}
where $\xi^{(A)i}_{\text{tr}}, A = 1,2,3,$ are 3 translation Killing
vectors. In the Cartesian-type coordinates these can be chosen as
$\zeta^i_{(1)} = (1,0,0)$, etc., so that the transformation
generated by them is 
\begin{equation}\label{4.101}
x = x' - f_{(1)} (t), \;\;\; y = y' - f_{(2)}(t), \;\;\; z = z' -
f_{(3)}(t).
\end{equation}
This brings the FRW background metric with $k=0$ into the form
\begin{equation}\label{4.102}
\!ds^2 = dt^2 - a^2 (dx'^2 + dy'^2 + dz'^2)
+ 2a^2 ( \dot{f}_{(1)} dt dx' + \dot{f}_{(2)} dt dy' +
\dot{f}_{(3)} dt dz').
\end{equation}
The acceleration of the local inertial frames with respect to the
frame with $x', y', z'$ fixed is thus given by [cf. Eq. (\ref{2.50})]
\begin{eqnarray}\label{4.103}
\alpha'^{x}  = -(1/a^2) (a^2\dot{f}_{(1)})\spdot[4], \;\;\;  \alpha'^{y} =
-(1/a^2) (a^2\dot{f_{(2)}})\spdot[4],
\alpha'^{z} = -(1/a^2)(a^2\dot{f}_{(3)})\spdot[4].
\end{eqnarray}
Since the gradients $h'_{0k,l}$ are vanishing, local
inertial frames do not rotate. 

\section{Field equations, integral constraints, solutions\\ 
and inertial frames.}

We now turn to the 
 equations for 
perturbations 
in the Machian gauges.
 Here we pay attention to the Mach 1 gauge. Its choice of the constant mean curvature slices is most natural 
 from the perspective of the full nonlinear theory. Moreover, the structure of the field equations for linear perturbations and their solutions do not differ significantly for the Mach gauges considered. The equations in Mach 2 and 3 gauges and in the generalized Lorenz-de Donder gauge are briefly discussed in Appendix \ref{App.D}.  Whenever solutions are known in terms of
Green's functions, we write them down. They can be used to
determine the accelerations and rotations of local inertial frames.
 Alternatively,
solutions in terms of 
 harmonics \cite{KodSas}, \cite{To} can be obtained by direct calculations, but they will not be studied in the present work.

\subsection{Field equations in Mach 1 gauge}

The minimal-distortion shift 
 condition
(\ref{4.17}) is combined with the constant mean external curvature
condition, i.e., with 
 [c.f. Eq. (\ref{4.22})]
\begin{equation}\label{5.1}
\frac{3}{2}{\dot{a}}\tilde{h}_{00}+\frac{1}{2}a\dot{\tilde{h}}{}^n_n-{\cal{P}}=0,\;\;
{\cal{P}} = \nabla _l\tilde{h}^l_0.
\end{equation}
In addition, we impose integral gauge conditions (\ref{4.96}). As a
consequence of the differential gauge conditions (\ref{4.17}) and (\ref{5.1}),
the field equations \eqref{3.38}, \eqref{3.39}, \eqref{3.40}, \eqref{3.41}, and \eqref{3.41aa} simplify considerably:
\begin{equation}\label{5.2}
\nabla ^2 \tilde{h}^n_n + 3 k\tilde{h}^n_n = 3a^2\kappa
\delta\tilde{T}^0_0,
\end{equation}
\begin{equation}\label{5.3}
\nabla ^2\tilde{h}_{k0} + 2k\tilde{h}_{k0} +
\frac{1}{3}\nabla_k
{\cal{P}}=2a^2\kappa\delta\tilde{T}^0_k,
\end{equation}
\begin{equation}\label{5.4}
\nabla ^2\tilde{h}_{00}+3a^2\left(
\frac{\dot{a}}{a} \right)\spdot \tilde{h}_{00} =
a^2\kappa(\delta\tilde{T}^0_0 - \delta \tilde{T}^n_n ).
\end{equation}
Instead of Eq. (\ref{5.3}) we may, equivalently, consider the equation 
\begin{equation}\label{5.3a}
\nabla^2 \tilde{h}_0^k - 2 \left[ k - 2a \left(\frac{1}{a}{\cal{H}}\right)'{} \right] \tilde{h}^k_0 + \frac{1}{3}\nabla ^k {\cal {P}} = - 2a^2 \kappa\, \delta \tilde{T}^k_0,
\end{equation}
in which $\delta \tilde{T}^k_0$ plays the role of a source.
The field equation (\ref{3.41}) can be written in the form
\begin{eqnarray}\label{5.5}
\nabla^2{\tilde{h}}{}^l_{T k} - 2k{\tilde{h}}^l_{T k} -
\frac{1}{a}\left( a^3\dot{\tilde{h}}{}_{T k}^{l}\right)\spdot +\,{\cal{F}}\{ \tilde{h}_{00}, \tilde{h}^n_n, \tilde{h}_{0k}, \dot{\tilde{h}}_{0k}
\} = -2a^2\kappa\delta{\tilde{T}}^l_{T k},
\end{eqnarray}
where ${\cal{F}}\{...\}$ denotes terms linear in the quantities in
the brackets and in their spatial derivatives. If $\mathcal{F}$
and $\delta{\tilde{T}}^l_{T k}$ are known, the last equation is a
wave-type equation for ${\tilde{h}}^l_{T k}$.

It is remarkable that in this gauge neither of equations
(\ref{5.2})-(\ref{5.3a}) contain any time derivative. All four of these equations
are elliptic
equations for $\tilde{h}^n_n, \tilde{h}_{k0},$ and $\tilde{h}_{00}$
when the right-hand sides are given. The first two are standard
constraint equations, and the third became elliptic equation for
$\tilde{h}_{00}$ as a consequence of the gauge conditions.  
 Another
remarkable feature of (\ref{5.2})-(\ref{5.3a}) is that, with $\delta \tilde{T}^0_0, \delta
\tilde{T}^0_k, \delta\tilde{T}^n_n,\delta\tilde{T}^k_0$ given, they represent a
completely separated system of four equations for, subsequently, $\widetilde{h}^n_n,
\tilde{h}_{k0}$, and $\tilde{h}_{00}$. 
 $\mathcal{P}=\nabla _l
\tilde{h}^l_0$ in Eq. (\ref{5.3}) is governed by a separate equation.
Applying $\nabla ^k$ on Eq. (\ref{3.39}) 
and
commuting derivatives, 
 we obtain 
\begin{equation}\label{5.6}
\nabla ^2 \mathcal{P}+3k\mathcal{P}+\nabla
^2\mathcal{K}=\frac{3}{2}a^2\kappa \nabla ^k\delta
\tilde{T}^0_k,
\end{equation}
which, in gauges for which $\mathcal{K}=0$, turns into
\begin{equation}\label{5.7}
\nabla ^2\mathcal{P} + 3k\mathcal{P} = \frac{3}{2}a^2 \kappa \nabla
^k \delta \tilde{T}^0_k.
\end{equation}
This has exactly the same form as (\ref{5.2}) for
$\tilde{h}^n_n$. With $\delta\tilde{T}^0_k$ given, we can
solve (\ref{5.7}) for $\mathcal{P}$ and substitute into Eq.~(\ref{5.3}), which can
then be written as
\begin{equation}\label{5.8}
\nabla ^2 \tilde{h}_{k0} +2k\tilde{h}_{k0} = 2a^2
\kappa\delta\tilde{T}^0_k
- \frac{1}{3}\nabla
_k \mathcal{P},
 \end{equation}
where the 'source` term on the r.h.s. is known. Considering, alternatively, Eq.(\ref{5.3a}), we get 
\begin{equation}\label{5.7a}
\nabla ^2\mathcal{P} + 3a \Bigl( \frac{1}{a} {\cal{H}}\Bigr)' {\cal{P}} = -\frac{3}{2} a^2\, \kappa\nabla_k\, \delta \tilde{T}^k_0, 
\end{equation}
\begin{equation}\label{5.8a}
\nabla^2 \tilde{h}^k_0 - 2 \left[ k - 2a \Bigl( {\frac{1}{a}} {\cal{H}}\Bigr)' \right] \tilde{h}^k_0 = -2a^2\kappa \,\delta \tilde{T}^k_0 - \frac{1}{3}\nabla_k {\cal{P}}.
\end{equation}

\subsection{Global gauge conditions and integral-constraint vectors for spherical universes}

Let us now consider the
integral gauge conditions (\ref{4.96}). 
We wish to elucidate their relation to
Traschen's constraint vectors \cite{Tr1}, \cite{Tr2.}. An integral-constraint vector
$V^\mu$ is defined by the relation 
\begin{equation}\label{5.9}
\int_{\mathcal{D}}   \delta T^\alpha_\mu V^\mu n_\alpha d^{(3)}V = \int
_{\partial \mathcal{D}} d\Sigma _l B^l,
\end{equation}
in which $\mathcal{D}$ is (possibly a part of) a spacelike
hypersurface, $n^\alpha $ its normal, $\partial\mathcal{D}$
its 2-dimensional boundary; $B^l$ depends on $h_{\mu\nu}$ and its
derivatives and it vanishes if these are zero on
$\partial\mathcal{D}$; $V^n$ is gauge independent. Since $V^\mu $
does not depend on $\delta T^\alpha_\mu$, Eq. (\ref{5.9}) represents
simple constraints on 
 source perturbations.

There exist 10 integral-constraint vectors in each of the FRW
universes but 6 of them are just spatial Killing vectors. The
other 4 are more interesting -- Traschen and others considered
their 
 implications for microwave background anisotropies
(see, e.g., \cite{Tr1}, \cite{TrErd.}). 
 The time components
of the 4 Traschen vectors are proportional to the 
scalars $Q^B$ 
 [Eqs. (\ref{4.42}), (\ref{4.43})], the spatial parts -- to 
$\nabla ^iQ^B$.
In a closed spherical universe
\begin{equation}\label{5.10}
V^\mu _{(B)} = (Q^B, a^{-1}\dot{a}\nabla ^i Q^B).
\end{equation}
Applying 
(\ref{5.9}) to the whole closed universe, 
it takes the form
\begin{equation}\label{5.11}
\int_{S^3} \left[ Q^{B}\delta T^0_0 + \frac{\dot{a}}{a} \nabla
^i Q^{B}\delta T_i^0 \right] d^{(3)}V = 0.
\end{equation}
Integrating by parts in the second term, we obtain
\begin{equation}\label{5.12}
\int_{S^3} \left[ Q^{B} \delta T^0_0 -
\frac{\dot{a}}{a}Q^{B}\nabla^i\delta T^0_i \right] d^{(3)} V = 0.
\end{equation}
In order to deduce the simplest constraints on the matter
perturbations, Traschen \textit{et al}. \cite{Tr1}, \cite{Tr2.}, \cite{TrErd.} consider the synchronous
gauge and, in addition, restrict ''physics`` in assuming vanishing
pressure so that the synchronous coordinates can be chosen to be
comoving with the fluid (in fact dust) since the flow is
irrotational. Then 
 the fluid velocity
$V_k = \bar{g}_{kl}V^l = \bar{g}_{kl}\delta U^l = 0$ and 
 $\delta T^0_i = (\bar{\rho}+\bar{p})(h_{i0}+V_i)=0$
in the synchronous gauge. Since the second integral in \eqref{5.11}
vanishes in this case, the constraints 
 imply just 
 [see, e.g., (14) in \cite{TrErd.}]
\begin{equation}\label{5.13}
\int_{S^3}  Q^{B}\delta \rho\; d^{(3)}V = 0.
\end{equation}

In the Mach 1 gauge the constraints \eqref{5.11} and \eqref{5.12} have clear,
simple consequences without any necessity to restrict physics.
Since both the constraint equations \eqref{5.2} and \eqref{5.7} 
have on the l.h.s.
the operator $\nabla ^2 + 3$ which has eigenfunctions $Q^{B}$
with zero eigenvalues (see Appendix \ref{App.C}), it is evident that the sources on the
r.h.s. must be orthogonal to the 4-dimensional function
space spanned by 4 harmonics $Q^{B}$. Therefore, in 
closed universes the perturbations $\delta T^\mu _\nu$ of any type
of matter 
 have to satisfy
separately the constraints
\begin{equation}\label{5.14}
\int _{S^3}  Q^{B} \delta T^0_0 d^{(3)} V = 0,
\end{equation}
and
\begin{equation}\label{5.15}
\int_{S^3}  Q^{B}\nabla ^i \delta T^0_i d^{(3)}V = 0,
\end{equation}
the last being equivalent to
\begin{equation}\label{5.16}
\int_{S^3}  \nabla ^i Q^{B}\delta T^0_i d^{(3)}V = 0.
\end{equation}
The same 
is true for $\delta \tilde{T}^0_i = a^{-1}\delta T^0_i$ and  $\delta
\tilde{T}^0_0=\delta T^0_0$. 
Hence, Traschen's constraints \eqref{5.9}, resp. \eqref{5.11}, are indeed
satisfied -- in such a way that, in fact, both integrals in the
constraints have to vanish separately. The 
 constraints now 
 become a straightforward consequence of the
Einstein equations. This is not the case in the synchronous gauge
where the constraint equations are coupled and there are more complicated
equations for $\tilde{h}^n_n, \mathcal{T}_k = \nabla
_l{\tilde{h}}{}^l_{T k}$ and their 
 derivatives,  as can be seen 
 from Eqs. (\ref{3.37}) and (\ref{3.38}) with $\widetilde{h}_{00} = \widetilde{h}_{k0} = 0$.

The constraints (\ref{5.14}) and (\ref{5.15}) and the constraint field equations
(\ref{5.2}) and (\ref{5.7}) also 
 demonstrate, why our global gauge
conditions (\ref{4.96}) do not restrict physics. They just eliminate
solutions of the homogeneous equations 
 which, in any case, can be removed by gauge
transformations generated by conformal Killing vectors. The
 gauge condition (5.1) implies that the
same integral gauge constraint, satisfied for $\widetilde{h}^n_n$
and $\mathcal{P}$, is valid also for the spatial scalar
$\widetilde{h}_{00}$, as is also required in (\ref{4.96}). As a
consequence, from the field equation (\ref{5.4}) another constraint,
which has not been discussed by Traschen \textit{et al.}, follows:
\begin{equation}\label{5.17}
\int_{S_3}  Q^{B} \delta T^n_n  d^{(3)} V = 0,
\end{equation}
the same for $\delta \tilde{T}^n_n(=\delta T^n_n)$. Hence, in
the Mach 1 gauge the whole picture of the 
 Traschen-type
constraints and our global integral gauge conditions is nicely
symmetrical: all scalar perturbations in both the metric and the
energy-momentum tensor,
$\tilde{h}^n_n, \tilde{h}^0_0, \mathcal{P} =
\nabla _k\tilde{h}^k_0, \delta\tilde{T}^n_n,
\delta\tilde{T}^0_0,$ and $\nabla^k\delta \tilde{T}^0_k$,
are orthogonal to the 4-dimensional space spanned by harmonics
$Q^{B}$.


\subsection{Solutions of the field equations and local inertial frames}

We are interested 
 in solutions for
$\widetilde{h}_{00},\widetilde{h}_{0i}$, and $\widetilde{h}^n_n$
when the sources are given in terms of $\delta \widetilde{T}^\mu_\nu$. These
quantities determine local inertial frames. Hence, we wish to
solve elliptic equations (\ref{5.2})-(\ref{5.4}). 
Solutions can be given in terms of harmonics but these will not be considered here.
However, several of these equations 
have been solved in literature in terms of Green's functions.
The Green's functions for the equation
\begin{equation}\label{5.18}
\nabla ^2 \Phi (x^i) + 3k\Phi (x^i) = - 2P(x^i),
\end{equation}
where $x^i = \{ \chi , \theta , \varphi\}$ are the hyperspherical coordinates, 
 are 
  \cite{TrErd.}: 
\begin{align}
G_{S^3} (x,x') &= -(1/4\pi ) \left[ \frac{\cos 2\psi}{\sin
\psi}\left(1-\frac{\psi}{\pi}\right) - \frac{1}{2\pi}\cos \psi \right],\label{5.19}\\
G_{E^3}(x,x') &= - (1/4\pi) \frac{1}{\tilde{l}},\label{5.20}\\
G_{H^3} (x,x') &= -(1/4\pi) \left[ \frac{\cosh 2\alpha}{\sinh\alpha}-2\cosh\alpha \right], \label{5.21}
\end{align}
where $\cos \psi = \cos \chi \cos \chi ' + \sin \chi \sin \chi ' \cos \gamma \;(k=+1),
\widetilde{l} = l^2 + l'^2 - 2ll' \cos \gamma \;(k=0),\; \cosh \alpha = \cosh \chi \cosh \chi '
- \sinh \chi \sinh \chi' \cos\gamma \;(k =-1)$, and 
$\cos\gamma = \cos \theta \cos \theta' + \sin \theta \sin\theta' \cos
(\varphi-\varphi')$. Here $\alpha$ is the geodesic distance under the metric $f_{ij}$ between the 'source point' $x'^{i} = \{ \chi ', \theta ', \varphi ' \}$ and the 'field point' $x^i = \{\chi, \theta, \varphi \}$. 
 The Green's functions satisfy the equations
\begin{equation}\label{5.25a}
(\nabla ^2 + 3k) G (x, x') = [det(f_{ij}]^{-\frac{1}{2}} \delta(x,x'),
\end{equation}
where $\nabla^2$ refers to the point $x^i$, $\delta (x',x')$ is the Dirac distribution.
In terms of the Green's functions (\ref{5.19})--(\ref{5.21})
the solution to Eq. \eqref{5.18} is given by 
\begin{equation}\label{5.22}
\Phi (x) = -2 \int  G (x, x') P(x')d^{(3)} V'.
\end{equation}
Hence, given $\delta\tilde{T}^0_0$ and $\nabla
^k\delta\tilde{T}^0_k$ we can determine 
$\tilde{h}^n_n$ and $\mathcal{P}$ from Eqs. (\ref{5.2}) and (\ref{5.7}). 
 $\tilde{h}_{00}$ can 
  be determined from the
gauge condition \eqref{5.1}, or by solving \eqref{5.4} if $\delta
\tilde{T}^n_n$ is known.
Notice, however, that to determine $\widetilde{h}_{00}$ from 
 \eqref{5.1} we need to know 
  also the time derivative $\dot{\widetilde{h}}{}_n^n$. This can
 be found by taking the 
  derivative of Eq. (\ref{5.2})
and assuming that $\delta \dot{\tilde{T}}{}_0^0$ is known. $[\delta \dot{\tilde{T}}{}_0^0$ 
can be expressed from the Bianchi identity (\ref{3.50}) in terms of $\delta\tilde{T} ^\mu_\nu$ 
and $\tilde{h}_{00}.]$ The solutions for $\dot{\tilde{h}}{}_n^n$ can then be constructed.

Knowing $\mathcal{P}$ from 
 Eq. (\ref{5.7}), we can determine
   $\tilde{h}_{k0}$ from Eq. \eqref{5.8} or \eqref{5.8a}. In general, we need to find a Green's function bitensor $G_a{}^{b'}(x, x')$ satisfying 
\begin{equation}\label{5.23}
f^{lm}\nabla_l \nabla_m G_a{}^{b'}(x, x') + 2k G_a{}^{b'}(x, x') = f^{-1/2}(x) \delta_a^{b'}\delta(x, x')
\end{equation}
in the case of Eq.(\ref{5.8}) [analogously for Eq. (\ref{5.8a})]. 
 Then the solution for $\tilde{h}_{0k}$ can be written as
\begin{equation}\label{5.24}
\tilde{h}_{0k} = \int G_k{}^{b'}(x, x') \mathcal{S}^{(A)}_{b'} (x') d^{(3)} V';
\end{equation}
 by the source $\mathcal{S}_b{}^{(A)}$, $A = I, II$, the r.h.s. of Eq. \eqref{5.8},  respectively \eqref{5.8a}, is denoted. 

In 
 a spatially flat universe, 
  the easiest way is to write Eqs. \eqref{5.8} and \eqref{5.8a} in Cartesian coordinates. Then \eqref{5.8} decouples into three Poisson equations for each $\tilde{h}_{k0}$, the Green's functions are standard, 
 and the solutions 
  are given by Poisson integrals over the source:
\begin{equation}\label{5.25}
\tilde{h}_{0k} (x^i, t) = \int \frac{\mathcal{S}^{(I)}_k(x')}{|x - x'|} d^3 x', \;
\mathcal{S}^{(I)}_k = 2a^2 \kappa \delta \tilde{T}^0_k - \frac{1}{3}\partial _k \mathcal{P.}
\end{equation}
Equation (\ref{5.8a}) turns into three equations of the Yukawa-type, as noticed recently by Schmid \cite{Schm}. Indeed, the l.h.s. of eq. (\ref{5.8a}) is of  the form $\nabla^2 \widetilde{h}^k_0 - \lambda^2(\eta)\widetilde{h}^k_0$ , so two Green's functions are given by
\begin{equation}\label{5.27}
G(x,x') = -\frac{1}{4\pi} \frac{e^{\mp \lambda |x-x'|}}{|x - x'|},
\end{equation}
where
\begin{equation}\label{5.28}
\lambda^{2}(\eta) = -4a({\cal{H}}/a)', \;\;\lambda^{2} (t) = -4a^2\dot{H}.
\end{equation}
Usually 
 $\dot{H}<0$, so $\lambda$ is real. The well-behaved solution of Eq. (\ref{5.8a}) is thus
\begin{equation}\label{5.30}
\tilde{h}^k_0 = - \frac{1}{2\pi}\int \mathcal{S} ^{(II)}_k (x') \frac{e^{-\lambda|x-x'|}}{|x-x'|} d^3x', \;
\mathcal{S}^{(II)}_k = -2 a^2\kappa \delta \tilde{T}^k_0 + \frac{1}{3}\partial_k\mathcal{P.}
\end{equation}

For open universes, the properties of the Green's bitensor $G_a{}^{b'}$ solving Eq. (\ref{5.23}) with $k = -1$ have been studied by d'Eath \cite{Dea}. In particular, it can be shown that such $G_a{}^{b'}$ exists which satisfies the boundary conditions at the source points and decays as $\exp [-3d(x,x')]$ as $d(x,x') \rightarrow \infty$, with $d(x,x')$ being the geodesic distance between the points $x, x'$ under the metric of an open universe. In hyperspherical coordinates, $d = \alpha$, where $\alpha$ is given below Eq. (\ref{5.21}). In fact, it was d'Eath \cite{Dea} who found the explicit form of the (scalar) Green's function (\ref{5.21}), but the explicit form of the Green's bitensor for solving the equations for vector perturbations for $k=-1$ does not seem to be known. The same is the case with spherical universes where only the Green's function (\ref{5.19}) for the scalar equation (\ref{5.18}) with $k = +1$ is known. Nevertheless, we can find explicit solutions for quite general classes of the vector perturbations also in case of $k = \pm 1$.\\

{\boldit 1. Axisymmetric rotational perturbations}\\
Recently we solved Eqs. \eqref{5.8} and \eqref{5.8a} for all odd-parity vector perturbations, i.e. those, for example, corresponding to rotational perturbations with axial symmetry \cite{BLK1}, \cite{BLK2}. 
We decomposed perturbations in coordinates $\theta, \varphi$ on spheres only and assumed axial symmetry (spherical functions $Y_{lm}$ having $m = 0$). Since the backgrounds admit homogeneous, isotropic foliations, nonsymmetric perturbations can be found from the axisymmetric ones 
\cite{BLK2}. 
 Thus, we write in spherical coordinates of Eq. (\ref{1.2}) 
\begin{align}
\tilde{h}_{0\varphi} &= \sum^\infty_{l=1} \left[ \tilde{h}_{0\varphi} (\eta,r) \right]_l \sin \theta\; Y_{l0,\theta},\label{5.32}\\
\delta\tilde{T}^0_\varphi &= \sum^\infty_{l=1} \left[ \delta\tilde{T}^0_\varphi (\eta, r)\right]_l \sin\theta\; Y_{l0,\theta},\label{5.33}
\end{align}
where $Y_{l0,\theta} = \partial_\theta Y_{l0}$, and $\delta T^\mu_\nu$ may represent any 
 perturbation. In the case of perfect fluid, 
the fluid angular velocity [cf. Eq. \eqref{3.42}] is
$\widetilde{\Omega} = \widetilde{V}^\varphi = d\varphi/d\eta = \Omega/a$,
and we write
\begin{equation}\label{5.36}
\tilde{V}_\varphi = -a^2 r^2 \sum^\infty_{e=1} \tilde{\Omega}_l (t,r) \sin \theta \;Y_{l0,\theta}.  
\end{equation}
Putting
\begin{equation}\label{5.37}
\left[ \tilde{h}_{0\varphi}  \right]_l = a^2 r^2 \sin^2 \theta \;\tilde{\omega}_l (t, r),
\end{equation}
we have
\begin{equation}\label{5.38}
\left[ \delta\tilde{T}_\varphi^0 \right]_l = a^2 (\overline{\rho} + \overline{p}) r^2 \sin^2 \theta \;(\tilde{\omega}_l - \tilde{\Omega}_l).
\end{equation}
These perturbations are transverse: $\nabla_k\tilde{h}_0^k = 0 = \nabla _k \tilde{T}^k_0$, 
 $\mathcal{P} = 0$.  Equations (\ref{5.8}) and (\ref{5.8a}) become
\begin{equation}\label{5.39}
\nabla^2 \tilde{h}_{0\varphi} + 2 k \tilde{h}_{0\varphi} = 2 a^2 \kappa \delta \tilde{T}^0_\varphi,
\end{equation}
respectively,
\begin{equation}\label{5.40}
\nabla^2 \tilde{h}^\varphi_0 - 2[k - 2a ({\cal{H}}/a)']\tilde{h}^\varphi_0 = -2a^2 \kappa \delta \tilde{T}^\varphi_0.
\end{equation}
The relation (\ref{3.41bb}) now implies
\begin{equation}\label{5.41}
\delta\tilde{T}^0_\varphi = -r^2 \sin^2 \theta \left[ \delta \tilde{T}^\varphi_0 - \frac{2}{\kappa a^2}\Bigl[-k + a \Bigl(\frac{{\cal{H}}}{a}\Bigr)' \,\Bigr] \,\tilde{h}^\varphi_0\right],
\end{equation}
so Eq. (\ref{5.40}) immediately follows from Eq. (\ref{5.39}) and vice versa. Nevertheless, the equations differ in the sense that in Eq. (\ref{5.39})  $\delta\tilde{T}^0_\varphi$ is considered as a source, whereas in (\ref{5.40}) the source is given by $\delta\tilde{T}^\varphi_0$. 
$\delta \tilde{T}^0_\varphi$ determines (up to factor $a^4$) the density of the angular momentum -- the perturbed Bianchi identities (\ref{3.48}) 
 imply the conservation law 
\begin{equation}\label{5.42}
(a^4\delta\tilde{T}^0_\varphi)\spdot = 0.
\end{equation}
On the other hand, $\delta \tilde{T}^\varphi_0$ determines the energy current. This is most apparent in the case of perfect fluid: $\delta\tilde{T}^0_\varphi$ is given by Eqs. (\ref{5.33}) and (\ref{5.38}), while 
\begin{equation}\label{5.43}
\delta\tilde{T}^\varphi_0 = (\overline{\rho} + \overline{p})\tilde{V}^\varphi = (\overline{\rho} + \overline{p}) \tilde{\Omega} = \sum^\infty_{l=1}[\delta\tilde{T}^\varphi_0]_l (\sin \theta)^{-1} Y_{l0,\theta},
\end{equation}
where
$\left[ \delta\tilde{T}^\varphi_0 (\eta, r) \right]_l = (\overline{\rho} + \overline{p})  \tilde{\Omega}_l$. 
Substituting the expansions 
 into Eq. (\ref{5.39}), and using the orthogonality of 
 $\sin\theta\; Y_{l0,\theta}$ for different $l$'s, we obtain the ``radial'' equation for each $l$: 
\begin{equation}\label{5.45}
-\sqrt{1-kr^2} \frac{1}{r^2} \frac{\partial}{\partial r}\left[ \sqrt{1-kr^2}\frac{\partial}{\partial r}(r^2\tilde{\omega}_l) \right] + \frac{l(l+1)}{r^2}\tilde{\omega}_l - 4k \tilde{\omega}_l\\
= 2a^2\kappa (\overline{\rho} + \overline{p}) (\tilde{\Omega}_l - \tilde{\omega}_l) \equiv \lambda^2 (\tilde{\Omega}_l - \tilde{\omega}_l).
\end{equation} 
For $l=1$ the perturbations correspond to the `rigidly rotating spherical shells' in the FRW universes 
 \cite{LKB},  
 \cite{BLK1}. 
 Each sphere rotates with no shear but it expands/contracts with the background so that its angular velocity changes. 
For $l \geq 2$ the motion of the 
fluid is ``toroidal''   
\cite{BLK2}.  
   In the case of closed universes the Legendre equation which follows from Eq. (\ref{5.45}) requires a special treatment.
For example, for $k = +1$, functions $\widetilde{\omega}_l(t,r)$ determining $\widetilde{h}_{0\varphi}$ by (\ref{5.37}) and (\ref{5.32}) turn out to be $(r = \sin \chi)$
\begin{equation}\label{5.46}
\tilde{\omega}_l = 2k (\sin \chi)^{-3/2} \left\{ \tilde{\mathcal{P}} ^l_2 \int^\chi_0 \frac{\tilde{\cal{Q}}^l_2}{W_l \sin^{1/2} \chi '}(\delta T^0_\varphi)_l d\chi '
+ \tilde{\cal{Q}}^l_2 \int^\pi_\chi \frac{\tilde{\mathcal{P}}^l_2}{W_l \sin^{1/2} \chi '}(\delta T^0_\varphi)_l d\chi '\right\},
\end{equation}
where $W_l$ is the Wronskian of the functions $\tilde{\mathcal{P}}^l_2(\chi)$, $\tilde{{\cal{Q}}}^l_2(\chi)$ which are derived from the derivatives of the appropriate Legendre functions with respect to their degree \cite{BLK2}. 

The properties of the solutions of Eq. \eqref{5.45} differ significantly according to whether we consider the right-hand side of \eqref{5.45}, i.e., the angular momentum density $\delta \tilde{T}^0_\varphi$ as the source of $\tilde{\omega}_l$, or we solve \eqref{5.45} for $\tilde{\omega}_l$ with $\tilde{\Omega}_l$ given, i.e., with the angular velocity as the source. 

The rotation 
 of inertial frames \eqref{2.67} is given by  the angular velocity 

\begin{equation}\label{5.47}
-\omega^j = \frac{1}{2a} \left[ \sum^\infty_{l=1}l(l+1)\omega_l Y_{l0}, \;\;\sum^\infty_{l=1}\frac{1}{r^2}\frac{d}{dr}(r^2\omega_l)Y_{l0,\theta},{}{}0 \right].
\end{equation}
The complete solutions for $\omega_l$ 
 for both $\delta \tilde{T}^0_\varphi$ and $\delta \tilde{T}_0^\varphi$ 
  given are determined in \cite{BLK2}. Roughly speaking, in 
   flat and open universes the effects of torodial motions beyond the cosmological horizons are exponentially damped when the angular velocity of matter is given. For flat universes, this was first noticed by Schmid \cite{Schm}. We shall see it occurs also for accelerations. However, these dragging effects are not damped when angular momenta are given as 
 sources. In \cite{BLK1} we give the physical explanation. 
 Since 
 Eqs. \eqref{5.8} and \eqref{5.8a} 
  are 
   elliptic equations, in both cases the inertial influences of ``distant matter'' are 
    expressed instantaneously.

We found toroidal perturbations to cause the rotation of local inertial frames by the angular velocity (\ref{5.46}). Do they 
 cause their acceleration? Since now 
  only $\tilde{h}_{0\varphi}\neq 0$ among all $\tilde{h}_{\mu\nu}$, the only nonvanishing component of the acceleration (\ref{2.50}) is
\begin{equation}\label{5.48}
\alpha^\varphi = \frac{1}{r^2\sin^2\theta}(a \tilde{h}_{0\varphi})\spdot.
\end{equation}
Substituting for $\widetilde{h}_{0\varphi}$ from Eqs. (\ref{5.32}) and (\ref{5.37}), and expressing the acceleration in the `background' frame, 
we find
\begin{equation}\label{5.49}
\alpha_{(\varphi)} = -r \sum^\infty_{l=1}\frac{1}{a}(a^2\omega_l)\spdot Y_{l0,\theta.}
\end{equation}
The acceleration vanishes at the axis of rotation. 
 More interestingly, it vanishes everywhere in a static (Einstein) universe if the matter rotates uniformly.  
 Indeed, the angular momentum density conservation law 
  requires $\bigl[a^5 (\rho + p)(\omega - \Omega)\bigr]\spdot = 0$, which implies $\dot{\omega}=0$ and hence $\alpha_{(\varphi)} = 0$ for time-independent $\Omega$ and $a =$ constant. In the 
 FRW universes the acceleration (\ref{5.49}) is nonvanishing, thus resembling the acceleration of the local inertial frames with respect to the static frames inside a collapsing, slowly rotating shell where particles at rest with respect to infinity experience the Euler acceleration, although the spacetime inside the shell is flat  \cite{KLB}. 
 For all 
  types of the FRW universes the solutions 
  are of the form $\omega_l = g_l(r) a^{-3}(t)$, where $g_l(r)$  are explicitly given in terms of the integrals of the special functions mentioned above and the sources $[\delta T^0_\varphi ]_l$. 
 Hence, the accelerations (\ref{5.48}) are of the form
\begin{equation}\label{5.50}
\alpha_{(\varphi)} = -(H/a^2) r \sum^\infty_{l=1}g_l(r) Y_{l0,\theta}.
\end{equation}
As an illustration, for $k = 0$ and $l=1$ perturbation we get
\begin{equation}\label{5.51}
\alpha_{(\varphi)} = 2(H/a^2)r \sin\theta {}\Bigl( \frac{J(<r)}{r^3} + \int^\infty_r \frac{dJ}{dr'}r'^{-3}dr'\Bigr),
\end{equation}
where $J(<r)$ is the angular momentum inside $r$.
With angular velocity 
 $\Omega$ considered as a source, $\omega$ shows the exponential decline near the origin when the source 
  is beyond the horizon 
   \cite{BLK2}. 
   As a consequence of Eq. (\ref{5.49}) the acceleration behaves similarly.\\

{\boldit 2. Perturbations of potential type}\\
In the example of toroidal perturbations, we had $\delta\tilde{T}^0_0= \nabla_k\tilde{V}^k = \nabla_k\delta\tilde{T}_0^k =  \nabla_k\tilde{h}^k_0 = \tilde{h}_{00}= 0$. As a second example, consider 
 briefly the case in which these quantities may be nonvanishing but $\tilde{V}_k$ and $\tilde{h}_{k0}$ have a vanishing transverse part so that $\tilde{h}_{k0} = \nabla_k h$ for some scalar $h$, and similarly for $\tilde{V}_k = \nabla_{k}w$. Physically, such perturbations describe a change in the matter density and a curl-free velocity field. No rotation of local inertial frames arises for such perturbations -- the vorticity vector (\ref{2.66}) vanishes for these ``scalar perturbations''. 

In order to determine the acceleration, we can use the gauge condition (\ref{5.1}). It enables us to find $\tilde{h}_{00}$ in terms of $\dot{\tilde{h}}{}^n_n$ and $\mathcal{P}$. Both $\tilde{h}^n_n$ and $\dot{\tilde{h}}{}^n_n$ can be determined from (\ref{5.2}) and its time derivative. With $\delta\tilde{T}_0^0$ given, the Green's functions (\ref{5.19})- (\ref{5.21}) yield $\dot{\tilde{h}}{}_n^n$. In the simplest case of a flat universe,
\begin{equation}\label{5.52}
\dot{\tilde{h}}{}^n_n = \kappa \int \frac{\bigl(a^2\delta\tilde{T}^0_0 \bigr)\spdot}{|x - x'|}d^3x'.
\end{equation}
The scalar $\mathcal{P}$ can 
 be obtained by solving either Eq. (\ref{5.7}) or Eq. (\ref{5.7a}). As with toroidal perturbations, 
 when the angular momentum $\delta\widetilde{T}^0_k$ is 
  prescribed, 
   $\mathcal{P} = \nabla_k\tilde{h}^k_0$ will not be suppressed at the origin if $\delta\tilde{T}^0_k$ occurs beyond a 
 horizon. 
  A suppression 
   takes place if the velocity, or the energy current $\delta\widetilde{T}^k_0$, is prescribed, as it corresponds to solving Eq. (\ref{5.7a}). In the flat universe, for example, Eq. (\ref{5.7a}) reads
\begin{equation}\label{5.53}
\nabla^2\mathcal{P} + 3a^2 \Bigl( \frac{\dot{a}}{a}\Bigr)\spdot\mathcal{P} = -\frac{3}{2}a^2\kappa\nabla_l \delta\tilde{T}^l_0,
\end{equation}
which is a Yukawa-type equation with the solution [$\lambda^2 (t) = -3a^2(\dot{a}/a)\spdot$]
\begin{equation}\label{5.54}
\mathcal{P} = -\frac{3a^2\kappa}{8\pi} \int (\nabla_l\delta\tilde{T}^l_0)(x') \frac{e^{-\lambda|x-x'|}}{|x-x'|}.
\end{equation}

If we start from Eq. (\ref{5.7}) with $\delta\tilde{T}^0_k$ given as a source, the solutions can be written in terms of the Green's functions (\ref{5.19}) - (\ref{5.21}). Taking $\partial/\partial t$ of \eqref{5.7} and assuming $\delta\dot{\tilde{T}}{}^0_k$ given, the same Green's functions will yield $\mathcal{\dot{P}}$. Since 
$\tilde{h}_{0k}= \nabla_k h$, $\mathcal{P} = \nabla^2 h$, $\mathcal{\dot{P}}= \nabla^2\dot{h}$, we can find $\tilde{h}_{k0}$ and $\dot{\tilde{h}}{}_{k0}$ by solving Laplace equations for $h$ and $\dot{h}$ with $\mathcal{P}$ and $\dot{\mathcal{P}}$ given. The solutions are unique up to an additive function of time which does not contribute to $\tilde{h}_{0k}$. 
 Knowing $\dot{\tilde{h}}{}_{k0}$ and $\tilde{h}_{00} = \frac{2}{3}\dot{a}^{-1}(\mathcal{P} - \frac{1}{2}a \dot{\tilde{h}}{}^n_n)$ from the gauge condition, the acceleration (\ref{2.50}) of the local inertial frames can be determined in terms of the sources.

Alternatively, we can 
 start from Eq. (\ref{5.4}) to determine directly $\tilde{h}_{00}$ in terms of the matter perturbations $\delta\tilde{T}^0_0$ and $\delta\tilde{T}^n_n$. 
 Equation (\ref{5.4}) for $k = 0$ 
  becomes just a Yukawa-type equation exactly in the form of Eq. (\ref{5.54}). 
   $\tilde{h}_{00}$
exhibits an exponential suppression 
 near the origin if the source $\delta\tilde{T}_0^0$, $\delta\tilde{T}_n^n$ occurs beyond the cosmological horizon. 
 The suppression 
  enters the formula (\ref{2.50}) for the acceleration. However, the term $\dot{\tilde{h}}{}_{0k}$ 
 will not be suppressed if the angular momentum $\delta\tilde{T}{}^0_k$ is prescribed. 
  If 
   the energy current $\delta\tilde{T}^k_0$ is considered as a source of $\dot{\tilde{h}}{}_{0k}$, the total acceleration will be exponentially suppressed. In both cases, however, the acceleration is determined instantaneously. 

\subsection{The determination of local inertial frames}

We described how the accelerations and rotations of the local inertial frames can be determined explicitly in these specific examples in order to illustrate the general framework. Finally, let us discuss, within the Mach 1 gauge, a question of \textit{the uniqueness of the solutions} of the field equations for general perturbations and of the resulting expressions for the acceleration and rotation of local inertial frames. The homogeneous equation corresponding to Eq. (\ref{5.3}) for $\tilde{h}_{k0}$ is identical to Eq.(\ref{4.39}), the well-behaved solutions of which were analyzed in detail between Eqs. (\ref{4.39}) and (\ref{4.50}). They 
 do not exist in $H^3$;  
  in $E^3$ 
   they describe the time-dependent linear combination of translations, 
    which can be eliminated by requiring $\widetilde{h}_{k0}$ to decay at infinity. In $S^3$, 
     they correspond to the time-dependent linear combinations of 10 conformal Killing vectors. 
      However, by imposing our integral gauge conditions (\ref{4.96}), we dispose of the four conformal Killing vectors which are not Killing. 

Therefore, the complete general solution of Eq. (\ref{5.3}) has for $k = 1$ the form
\begin{equation}\label{5.55}
\tilde{h}_{0k} = \tilde{h}_{0k\;(inh)} + \sum^6_{A=1} f_A(t)\xi^{(A)}_k,
\end{equation}
where $\tilde{h}_{0k\;(inh)}$ is a solution of the inhomogeneous equation (\ref{5.3}) and $f_A(t)$ are arbitrary functions of time; $\xi^{(A)}_k $ are 6 Killing vectors of $S^3$ describing rotations and quasitranslations (see Appendix \ref{App.C}). 
 Owing to our integral gauge conditions, Eq. (\ref{5.2}) admits a unique solution; 
 ${\mathcal{P}} = \nabla^k\tilde{h}_{0k}$ is not affected by the Killing vectors in (\ref{5.55}), and hence, the Mach 1 gauge condition (\ref{5.1}) determines a unique $\tilde{h}_{00}$. If we start from Eq. (\ref{5.4}) to determine $\tilde{h}_{00}$ directly in terms of $\delta\tilde{T}^0_0$ and $\delta\widetilde{T}^n_n$, we also arrive at a unique solution because the homogeneous equation corresponding to Eq. \eqref{5.4} coincides precisely with Eq. \eqref{4.56}.  
 Therefore, with $\delta\tilde{T}_0^0$, $\delta\tilde{T}_n^n$, and angular momenta $\delta\tilde{T}_k^0$  given, the accelerations and rotations of local inertial frames in closed universes are determined by formulas (\ref{2.50}) and (\ref{2.68}) only up to the freedom exhibited in Eq. (\ref{5.55}). This freedom corresponds precisely to changing the coordinate system by the infinitesimal transformation 
 in which $\zeta^0 = 0$, $\zeta^i = \sum^6_{A=1}F_A (t) \xi^{(A)i}$.
 Then $\triangle \tilde{h}_{00} = \triangle \tilde{h}_{n}^n = \triangle\tilde{h}_{kl} = 0$ because $\zeta^i$ is a linear combination of the Killing vectors. 
 The six spacelike Killing vectors generate motions which preserve the symmetries of the space. However,
\begin{equation}\label{5.56}
\triangle\tilde{h}_{0k} = -a\dot{\zeta}_k = -a \sum^6_{A=1}\dot{F}_A(t) \xi^{(A)}_k ,
\end{equation}
which is equal to the additional term in Eq. (\ref{5.55}) provided that $f_A (t) = - a(t) \dot{F}_A(t)$. The transformations
\begin{equation}\label{5.57}
x'^i = x^i + \sum^6_{A=1} F_A(t) \xi^{(A)i}
\end{equation}
with arbitrary coefficients $F_A(t)$ lead to mutually accelerated frames. The accelerations have special forms when regarded as functions in space. Putting $x^i =$ constant in (\ref{5.57}) we get $d^2x'^i/dt^2 = \sum^6_{A=1} \ddot{F}_A(t) \xi^{(A)i}$ where $\xi^{(A)i}$ are specific functions of $x^i$ (cf. Appendix \ref{App.C}). An example of accelerations and rotations generated by this type of transformations with one translational and one rotational Killing vector is given in Eqs. \eqref{4.90} and \eqref{4.91} above.


Consider now for simplicity just one rotational Killing vector and again the case of toroidal perturbations. The arbitrariness exhibited by the additional terms in Eq. (\ref{5.55}) can then be seen distinctly. 
 The homogeneous equation corresponding to Eq. (\ref{5.45}) for $l = 1$ with 
 the angular momentum given as a source is solved by $\widetilde{\omega}_{(0)} = \widetilde{\omega}_{(0)}(t) $, where $\widetilde{\omega}_{(0)}$ is an arbitrary function of $t$. This implies [cf. Eq. \ref{5.37}]
\begin{equation}\label{5.58}
\widetilde{h}_{0\varphi} = \widetilde{\omega}_{(0)} a^2 \sin^2 \chi \sin^2 \theta = \widetilde{\omega}_{(0)} \xi^{(\varphi)}_\varphi ,
\end{equation}
where $\xi^{(\varphi)i} = (0, 0, 0, 1)$ is the rotational Killing vector, 
which is  a special case of Eq. (\ref{5.56}). The transformation $\varphi ' = \varphi - \int \tilde{\omega}_{(0)} (t) dt$ [a special case of \eqref{5.57}] would make the term (\ref{5.58}) vanish. Since, however, in closed universes such an arbitrary `integration constant' $\widetilde{\omega}_{(0)} (t)$ cannot be  eliminated by boundary conditions, all frames with different $\widetilde{\omega}_{(0)} (t)$ are admitted. In this sense, only \textit{relative} rotations of the local inertial frames can be determined if the angular momentum is considered as a source of their dragging.

In the case of \textit{general} perturbations of the spherical universes with the distributions of energy $\delta\widetilde{T}^0_0$, angular momenta 
 $\delta\tilde{T}^0_k$ given, the freedom 
  described by 
   $F_A(t)$ in Eq. \eqref{5.57} which preserves the symmetries of the space cannot be eliminated by boundary conditions. In this sense, only \textit{relative rotations and accelerations of the local inertial frames can be determined}. The perturbations $\widetilde{h}_{00}$ and $\widetilde{h}_{0i}$, which imply these rotations and accelerations, are determined by the field equations and the Mach 1 gauge conditions \textit{instantaneously} from appropriate averages over the distributions of $\delta\widetilde{T}^0_0$ and $\delta\widetilde{T}^0_k$. An explicit example 
 is the expression \eqref{5.22} with the source term $P\sim \delta\widetilde{T}_0^0$ and the Green's function given in $S^3$ by Eq. \eqref{5.19} determining $\tilde{h}_{n}^{n}$ and thus $\tilde{h}_{00}$. Other 
 example 
  are the functions $\widetilde{\omega}_l (t, r)$ 
   in Eq. \eqref{5.46}. 

\textit{This instantaneous determination of the local inertial frames by such averages, up to global rotations and accelerations given by the symmetries of the space, is the crucial feature exhibiting the validity of Mach's principle in relativistic cosmology}, at least in the first-order perturbation theory. The ability to describe the same physical situation using these differently rotating and accelerating frames is a consequence of the dynamics having a higher degree of symmetry than the realisation of the world in terms of the positions of actual bodies. The frame in which we choose to describe the motions is not of importance; what matters is the relative motions of the bodies, not that of the frame relative to the bodies.

Finally, consider now energy currents 
 $\delta\tilde{T}^k_0$ 
 together with $\delta\tilde{T}^0_0$ as the sources. In case of the perfect-fluid perturbations we thus take the fluid velocity $\tilde{V}^k$ and $\delta\rho$ as the sources. 
 When velocities and accelerations of ``heavenly bodies'' are given, the rotations and accelerations of the local inertial frames are determined uniquely in spherical universes. As we have shown below Eq. \eqref{4.56}, the homogeneous equation corresponding to Eq. \eqref{5.7a} for $\mathcal{P} = \nabla_k\tilde{h}^k_0$ admits only the trivial solution $\mathcal{P} = 0$. Hence, the inhomogeneous Eq. \eqref{5.7a} determines $\mathcal{P}$ uniquely when 
 $\nabla_k\delta\tilde{T}^k_0$ is given. The same is true for 
  \eqref{5.4} for $\tilde{h}_{00}$ because the 
   homogeneous equation is the same. To determine perturbation $\tilde{h}_{00}$ 
    we need to know both $\delta\rho$ and $\delta p$ (resp. $\delta\tilde{T}^n_n$). The gauge condition \eqref{5.1} then implies $\dot{\tilde{h}}{}_n^n$. Alternatively, we can solve \eqref{5.2} for $\tilde{h}^n_n$ and $\dot{\tilde{h}}{}^n_n$ by giving $\delta\rho$ and $\delta\dot{\rho}$, and extract $\tilde{h}_{00}$ from the gauge condition.  
Solutions for $\tilde{h}^n_n$, $\dot{\tilde{h}}{}^n_n$ are unique due to our integral gauge conditions, so 
 a unique $\tilde{h}_{00}$ can also be found. Finally, the homogeneous part of 
  \eqref{5.8a} for $k=1$ with sources $\delta\tilde{T}^k_0$ and $\mathcal{P}$ given reads
\begin{equation}\label{5.59}
\nabla^2\tilde{h}^k_0 - 2 \left[1 - 2a^2 \Bigl( \frac{\dot{a}}{a}\Bigr)\spdot\right] \tilde{h}^k_0 = 0.
\end{equation}
This 
 admits only 
  $\tilde{h}^k_0 = 0$: multiply  by $\tilde{h}_{k0}$ and integrate by parts over 
   closed space, 
\begin{equation}\label{5.60}
- \int_D f^{ij} f^{kl}\nabla_i\tilde{h}_{k0} \nabla_j\tilde{h}_{l0}d^{(3)} V = 2 \mathcal{A} (t) \int_D f^{kl}\tilde{h}_{k0}\tilde{h}_{l0}d^{(3)} V,
\end{equation}
where $\mathcal{A} (t) = 1 - 2a^2(\dot{a}/a)\spdot = a^2[\frac{1}{2}\kappa(\overline{\rho} + \overline{p}) - \dot{H}]$. 
 Since the integrands on both sides are spatial scalars, we can calculate them at each point by using $f^{ij} =$ diag $(1, 1, 1)$. In this way we find that both are non-negative so that the only way how to satisfy \eqref{5.60} for $\mathcal{A}(t)>0$ is by putting $\tilde{h}_{k0} = 0.$ For standard models $\dot{H} < 0$ and indeed $\mathcal{A}(t)>0$. 
In the example of toroidal perturbations, the uniqueness of the solutions is reflected by the fact that Eq. \eqref{5.45} has unique solutions for given angular velocity $\widetilde{\Omega}_l$ of matter. A purely time-dependent $\widetilde{\omega}_{(0)l}(t)$ does not solve \eqref{5.45} when 
 $\widetilde{\Omega}_l$ is prescribed and not the whole r.h.s. $\tilde{\Omega}_l - \tilde{\omega}_l$.

We thus arrive at another important \textit{aspect of Mach's principle} in relativistic cosmology: \textit{if the velocities, density and pressure perturbations of cosmic fluid are given, the (linearized) field equations in a closed universe provide a unique determination of the rotations and accelerations of the local inertial frames}.

\section{Concluding remarks}

Although this paper includes also items which have a review character, primarily it contains new developments: the analysis of accelerations and rotations of local inertial frames and of gyroscopes in perturbed FRW universes; the general forms of the perturbed Einstein field equations and Bianchi identities 
are formulated without gauge conditions, harmonics, or splittings, the motivation for and the analysis of the instantaneous, Machian-based gauges, including the integral gauge conditions and their relation to Traschen's integral constraints; and the manifestation of Mach's ideas in the framework of general linear cosmological perturbations of FRW universes. In particular, those who wish to study cosmological perturbation problems in position space, as advocated recently in Ref. \cite{Bash}, may find here useful relations not given before. Various specific perturbation problems can be attacked by applying the results presented here. We already used the formalism to investigate rotational and  toroidal vector perturbations of FRW universes \cite{BLK1}, \cite{BLK2}, as mentioned and applied in Section V; there we also discussed vector perturbations of potential type.

 For given distributions of energy-momentum and angular momentum of matter sources, the rotations and accelerations of local inertial frames are uniquely given in the Machain gauges in open universes under suitable boundary conditions, whereas in closed universes they are determined up to motions generated by the Killing vectors, i.e., by symmetries of the background. They are determined uniquely also in closed universes if velocities, density, and pressure perturbations of cosmic fluid are given. As a consequence of the constraint equations and the choice of gauges which imply suitable slicing of perturbed universes, these inertial properties are determined instantaneously. In this sense Mach's principle is embodied in the cosmological linear perturbation theory.

The dragging of inertial frames is an essentially \textit{global effect} which, at least in linear perturbation theory, has to be seen as an instantaneous phenomenon. This was first 
 demonstrated by Lindblom and Brill \cite{LiBr}, who investigated rotational dragging by a slowly rotating, massive spherical shell freely falling under its own gravity. We reconsidered
 the problem  and explored its electromagnetic analogue 
 \cite{KLB}. 
 The need to introduce a suitable coordinate frame  (the `gauge') to describe the dragging is well illustrated inside the 
 shell. Spacetime is flat there; no local geometrical (gauge invariant) perturbations occur. 
  The time-dependent rotation of inertial frames is exhibited by considering the congruence of static observers, i.e. those who are at rest with respect to static observers at
 infinity. 
 They play the role analogous to that of the cosmological observers in the present paper. They experience acceleration, and the congruence of their worldlines  twists. Both quantities, characterizing  their congruence, can be expressed in a covariant manner as in formulae \eqref{2.17} and \eqref{2.18} for cosmological observers.
Also, massive, slowly rotating shells immersed in FRW universes were analyzed \cite{Klein}, \cite{DoBi}, including their observational consequences on the appearance  of sources behind the shells \cite{DoBi}, \cite{Dol}. 
 In \cite{LBK} we considered {\it strong} rectilinear dragging using exact conform static solutions of the Einstein-Maxwell equations with charged dust. 

A thorough nonlinear study of Mach's ideas within the framework of general relativity lies in the future. Quoting from the same source by which we started (see \cite{MTW}, p. 546), ``Much must still be done to spell out the physics behind these equations [the initial-value equations] and to see this physics in action''.\\

\textit{\textbf{Acknowledgements}}

We thank E. Bertschinger for the correspondence regarding Ref. \cite{LKB}, and J. Barbour, J. Bardeen, H. Bondi (since deceased), G. Efstathiou, J. Ehlers, A. Higuchi, V. Moncrief, H. Pfister, W. Rindler, B. Schmidt, and C. Schmid, for discussions. J. B. and J. K. are grateful to the Institute of Astronomy, Cambridge and the Royal Society for hospitality and support. J. B. acknowledges also the hospitality of the Albert Einstein Institute in Golm, the support of the Alexander von Humbolt Foundation, and a partial support from the grant GA\v{C}R 202/06/0041 of the Czech Republic and of the grants No. LC06014 and MSM0021620860 of the Ministry of Education.


\appendix
\renewcommand{\thesubsection}{\arabic{subsection}}  
\section{Perturbed field equations with cosmic time $t$}\label{App.A}

We write the perturbed FRW metric in the form \eqref{1.1}
\begin{equation}\label{3.1}
ds^2 = \left( \bar g_{\mu \nu} + h_{\mu \nu} \right) dx^\mu
dx^\nu = dt^2 - a^2(t) f_{kl}dx^k dx^l + h_{\mu \nu}
dx^\mu dx^\nu.
\end{equation}
The background Christoffel symbols are
\begin{equation}\label{3.10}
{\bar{\Gamma}}_{0l}^m=H\delta^m_l\ ,\ \
{\bar{\Gamma}}_{kl}^0=-H\bar{g}_{kl}\ ,\ \
{\bar{\Gamma}}_{kl}^m=f^{mn}\left(\partial_{(k}f_{l)n}-\pul\partial_n f_{kl}\right);
\end{equation}
hereafter the symmetrization brackets (\;) include the factor $\tfrac{1}{2}$, as do antisymmetrization ones [\;];
 $H=\dot{a}/a$ is the Hubble ``constant''.
The non-vanishing components of the background Einstein equations,
$\bar{G}^\nu_\mu =\bar{R}^\nu_\mu
-\pul\delta^\nu_\mu\bar{R}
= \kappa\bar T_\mu^\nu \;+ \;\Lambda\delta_\mu^\nu$, read
\begin{equation}\label{3.11}
\bar{G}^0_0 =3\left( \frac{k}{a^2}+H^2\right)=\kappa\bar{\rho} + \Lambda\ ,\ \
\bar{G}^l_k =
   {\delta}^l_k \left( \frac{k}{a^2}+3H^2+2\dot{H}\right)
        =-(\kappa\bar{p} - \Lambda) \ \delta^l_k,
\end{equation}
with $\kappa =8\pi G/c^4$, $k=0,\ \pm 1$ denoting the curvature
index, and $\Lambda$ cosmological constant; the background energy-momentum
tensor $\bar{T}^{\nu}_{\mu}$ of perfect fluid  is given by \eqref{2.75}.
The~indices of $h_{\mu\nu}$ 
are raised or lowered with $\bar{g}^{\mu\nu}$
and $\bar{g}_{\mu\nu}$; thus $h^0_0 =h_{00},\,
h^k_0=\bar{g}^{kl}h_{0l}=-\frac{1}{a^2}f^{kl}h_{0l}, \,\mbox{etc}.$
No spatial index is ever  displaced with $f_{kl}$ alone.
We 
 introduce the covariant derivative,
\begin{equation}\label{3.13}
\nabla_k h^m_0=\partial_k h^m_0 +\bar{\Gamma}^m_{kl}h^l_0.
\end{equation}
The background curvature tensor of spatial sections $t = {\rm constant}$ is 
${\bar{\cal{R}}}^r{}_{ksl}=k\left(\delta^r_s f_{kl}
-\delta^r_l f_{ks} \right)$
and the 
 Ricci 3-tensor ${\bar{\cal{R}}}_{kl}=2kf_{kl}$.
Useful identities are ($\nabla_{kl}\equiv\nabla_k\nabla_l$, $\nabla^k =
f^{kl}\nabla_{l}$, $\nabla^2 = f^{kl} \nabla_k \nabla_l$):
\begin{equation}\label{3.2}
(\nabla_{kl} - \nabla_{lk})V^l = -2 k f_{kl} V^l = \frac{2k}{a^2} \bar{g}_{kl} V^l
= \frac{2k}{a^2} V_k\,,
\end{equation}
\begin{equation}\label{3.3}
\nabla_k \nabla^2 V^k = \nabla^2 (\nabla_k V^k)
+ 2k (\nabla_k V^k)\,,
\end{equation}
\begin{equation}\label{3.16}
(\nabla_{kl} -\nabla_{lk})h^l_0 =2\frac{k}{a^2}h_{k0}\,,
\end{equation}
\begin{equation}\label{3.17}
(\nabla_{kl}-\nabla_{lk})h^n_m=h^r_m{\bar{\cal{R}}}^n{}_{rkl}
         -h^n_r{\bar{\cal{R}}}^r{}_{mkl}=
2k\left(  \delta^n_{[k}h^r_{l]}-h^n_{[k}\delta^r_{l]}\right)
    f_{rm}.
\end{equation}
The~perturbed Einstein equations,
$\delta G^\nu_\mu=\kappa\delta T^\nu_\mu$, are expressed in terms of
$h^0_0$, $h^0_k$, and $h^l_k$. In this `mixed' form 
 $\Lambda$ does not appear.
The~left-hand sides, $\delta G^\nu_\mu$ , read as follows:
\begin{align}
\delta G^0_0 &= -\pul
              \nabla_{rs}\left(\bar{g}^{rn}h^s_n-\bar{g}^{rs}h^n_n\right)
              -\frac{k}{a^2}h^n_n
            -2H\left(\tripul Hh^0_0
              -\pul\dot{h}^n_n+\nabla_n h^n_0\right)\ ,
         \label{3.18}\\
\delta G^0_k &= \pul
            \nabla_l \left(\dot{h}^l_k-\delta^l_k \dot{h}^n_n\right)
            +\bar{g}^{rs}\left(\nabla_{kr}h^0_s-\nabla_{r(k}h^0_{s)}\right)
           +H\nabla_k h^0_0\ ,\label{3.19}\\
\delta G^k_0 &= \bar g^{kl}\left[\delta G_l^0 - 2\left(\frac{k}{a^2} - \dot H\right) h_l^0\right]\label{3.19a}\ , \\
\delta G^m_k &= -\pul
               \left(\ddot{h}^m_k-\delta^m_k \ddot{h}^n_n\right)
                -\tripul  H
                \left(\dot{h}^m_k-\delta^m_k\dot{h}^n_n\right)
                -\frac{k}{a^2}h^m_k\nonumber\\
             &\ +\pul
                \nabla_{kl}\left(\bar{g}^{ln}h^m_n
                               -\bar{g}^{lm}h^n_n\right)
               -\pul\bar{g}^{rs}
                \nabla_{rs}\left(h^m_k-\delta^m_k h^n_n\right)
                    +\pul
                  \nabla_{rs}\left(\bar{g}^{mr}h^s_k-\delta^m_k h^{rs}\right)
                   \nonumber\\
              &\ +\bar{g}^{ml}\left(\nabla_{(k}\dot{h}^0_{l)}
                    +H\nabla_{(k}h^0_{l)}\right)
                  -\delta^m_k\bar{g}^{rs}
                 \left(\nabla_r \dot{h}^0_s+H\nabla_r h^0_s\right)\nonumber\\
              &\ -\pul
                 \left(\bar{g}^{ml}\nabla_{lk}h^0_0
                  -\delta^m_k\bar{g}^{rs}\nabla_{rs}h^0_0\right)
                 -\delta^m_k\left[ H\dot{h}^0_0
                       +\left(2\dot{H}+3H^2\right) h^0_0\right]\ .\label{3.20}
\end{align}
If $\delta T^\nu_\mu$ is a perfect-fluid perturbation 
then the right-hand side 
 is given by [see Eqs.~(\ref{2.85})]
\begin{eqnarray}\label{3.22}
&\;&\delta T^0_0=\delta\rho\ ,\ \ \delta T^0_k = (\overline{\rho} + \overline{p}) (V_k + h_k^0) =
   \frac{2}{\kappa}\left( \frac{k}{a^2}-\dot{H} \right)\left( V_k+h^0_k \right)\
   ,\nonumber\\
&\;&\delta T^k_0 = (\overline{\rho} + \overline{p}) V^k = \frac{2}{\kappa}(\frac{k}{a^2} - \dot{H}
) V^k ,\;\;\; \delta T^l_k=-\delta^l_k\delta p\ .
\end{eqnarray}
In the last equations we used the relation
\begin{equation}\label{3.22a}
(\bar\rho + \bar p) = \frac{2}{\kappa}\left(\frac{k}{a^2} - \dot H\right)\,,
\end{equation}
which follows from the background Einstein equations \eqref{3.11} for all
$\bar\rho$, $\bar p$, $k$, $\Lambda$.


The relations between various $h_{\mu\nu}$'s and $\tilde{h}_{\mu\nu}$'s used in the main text are
\begin{equation}\label{3.32}
\begin{array}{lll}
h_{00}={\tilde{h}}_{00}\ ,&h_{0l}=a{\tilde{h}}_{0l}\ ,\ \ \ \ \
        &h_{kl}=a^2{\tilde{h}}_{kl}\ ,\ \ \ \ \ \ \\[1mm]
h^0_0={\tilde{h}}_{00}\ ,&h^l_0=-a^{-1}{\tilde{h}}^l_0\ ,
        &h^0_l=a{\tilde{h}}_{0l}\ ,\ \ h^l_k=-{\tilde{h}}^l_k\ , \\[1mm]
h^{00}={\tilde{h}}_{00}\ ,&h^{0l}=-a^{-1}{\tilde{h}}^l_0\ ,
        &h^{kl}=a^{-2}{\tilde{h}}^{kl}\ .
\end{array}
\end{equation}
Eqs.~\eqref{3.13} and \eqref{3.3} hold also for ${\tilde{h}}^m_{0}$ and $\tilde V^k$,
but Eqs.~\eqref{3.2} and \eqref{3.16} take the form
\begin{equation}\label{3.33a}
(\nabla_{kl} - \nabla_{lk}) \tilde V^l = -2kf_{kl} \tilde V^l = -2k\tilde V_k\ ,
\end{equation}
\begin{equation}\label{3.33}
(\nabla_{kl} -\nabla_{lk}){\tilde{h}}^l_0
      =-2kf_{kl}{\tilde{h}}^l_0=-2k{\tilde{h}}_{0k}\ ,
\end{equation}
and Eq. \eqref{3.17} becomes
\begin{equation}\label{3.34}
(\nabla_{kl}-\nabla_{lk}){\tilde h}^n_m=
2k\left( \delta^n_{[k}{\tilde h}_{l]m}-{{\tilde h}^n}_{[k}{f}_{l]m}\right)\,.
\end{equation}

\section{Gauge transformations of perturbations}\label{App.B}

As a consequence of infinitesimal transformations 
\begin{equation}\label{4.1}
x^0\rightarrow x'^{0} = x^0 + \zeta^0(x),
\end{equation}
\begin{equation}\label{4.2}
x^i\rightarrow x'^{i} = x^i + \zeta^i(x),
\end{equation}
 we find the following changes of various
metric components under the change of gauge (notice that
$\Delta Q \equiv Q - Q'$ for any $Q$):
\begin{equation}\label{4.4}
\Delta h_{00} = 2\dot{\zeta}^0 = \Delta \tilde{h}_{00},
\end{equation}
\begin{equation}\label{4.5}
\Delta h_{0l} = \partial_l \zeta^0 - a^2 \dot{\zeta}_l = a
\Delta\tilde{h}_{0l}, \;\;\;\; (\zeta_l = f_{lk} \zeta^k )
\end{equation}
\begin{equation}\label{4.6}
\Delta h_{kl} = -2a^2 \left[ \nabla_{(k} \zeta_{l)} +
\frac{\dot{a}}a{}f_{kl}\zeta^0 \right] = a^2
\Delta\tilde{h}_{kl},
\end{equation}
\begin{equation}\label{4.7}
\Delta h^n_n = 2a^2 \left[ \nabla_n \zeta^n + 3 \frac{\dot{a}}{a}
\zeta^0 \right] = -\Delta \tilde{h}^n_n.
\end{equation}
Similarly, the perturbations of the energy-momentum
tensor components change under the transformations \eqref{4.1} and \eqref{4.2}
as follows:
\begin{eqnarray}
\Delta\,\delta T_0^0 &=& \dot{\bar\rho}\zeta^0 = \Delta\,\delta\tilde T_0^0\,,\\
\Delta\,\delta T_k^0 &=& (\bar\rho + \bar p)\partial_k\zeta^0 = a \Delta\,\delta\tilde T_k^0\,,\\
\Delta\,\delta T_0^k &=& -(\bar\rho + \bar p)\dot\zeta^k = a^{-1} \Delta\,\delta \tilde{T}_0^k\,,\\
\Delta\,\delta T_k^l &=& -\dot{\bar p}\,\zeta^0\delta_k^l = \Delta\,\delta\tilde{T} _k^l\,,
\end{eqnarray}
where we substituted from Eq.~\eqref{2.75} for the background values of $\bar T_\mu^\nu$. In particular, in the fluid case
\begin{gather}
\Delta\,\delta \rho = \dot{\bar\rho}\;\zeta^0\;, \qquad \Delta\,\delta p = \dot{\bar p}\,\zeta^0\,,\\
\Delta\,\delta U^0 = -\dot{\zeta}^0 = -\Delta\,\delta U_0\;, \qquad \Delta\,\delta U^m = - \dot{\zeta}^m
\;,\qquad\Delta\,\delta U_m = \zeta^0_{,m} \;,\\
\Delta\,V^m = -\dot{\zeta}^m = a^{-1}\Delta\,\tilde{V}^m\;,\qquad \Delta\,V_m = a^2 \dot{\zeta}_m = -a\Delta\,\tilde{V}_m\;.
\end{gather}
Let us emphasize that the above results for the changes of both
$h_{\mu\nu}$'s and $\widetilde{h}_{\mu\nu}$'s are expressed in $x^\mu = (t, x^i)$
coordinates. In $\widetilde{x}^\mu =
(\eta,x^i)$ coordinates we find, for example,
\begin{equation}\label{4.8}
\tilde{h}'_{00} = \tilde{h}_{00} -
\Delta\tilde{h}_{00} = \tilde{h}_{00} - 2 \dot{\zeta}^0 =
\tilde{h}_{00} - 2\frac{d(a \tilde{\zeta}^0)}{d\eta} a^{-1} =
\widetilde{h}_{00} - 2 \mathcal{H}\tilde{\zeta}^0 - 2
\frac{d\tilde{\zeta}^{0}}{d\eta}.
\end{equation}

\section{Killing and conformal Killing vectors on the FRW backgrounds}\label{App.C}

Killing vectors are 
 also conformal Killing vectors but
here we call ``conformal Killing vectors'' -- sometimes more
explicitly ``proper'' conformal Killing vectors -- those which are
not Killing vectors. All these vectors on 
 the FRW backgrounds
are well-known. Since, however, we did not find listed all of them
in a transparent manner in one place, 
 we give them
here. Their relation to the scalar and vector harmonics will also be
elucidated. The 3-dimensional spatial vectors are 
 frequently
used in the main text. There exists an extensive literature on the harmonics in $S^3$ and $H^3$, 
 see., e.g., \cite{To} and \cite{HI}.

\subsection{Killing and conformal Killing 3-vectors in $\boldsymbol{E^3 (k = 0), S^3 (k = 1)}$ and $\boldsymbol{H^3 (k = -1)}$}

The standard Killing equation
\begin{equation}\label{6.1}
\nabla_k\xi_i + \nabla_i\xi_k = 0
\end{equation}
in the FRW 3-backgrounds with curvature tensor \eqref{3.17} can be
written in an equivalent form
\begin{equation}\label{6.2}
\nabla^2\xi_i + 2k\xi_i = 0.
\end{equation}

The Killing vectors have their simplest form in the coordinates $x^m$
in which the metric is (see e.g. \cite{weinb}, Ch.13),
\begin{equation}\label{6.3}
ds^2 = dt^2 - a^2(t) \left[ \delta_{kl} + \frac{kx^k x^l}{1 - kr^2} \right] dx^k\;dx^l ,
\end{equation}
where
\begin{equation}\label{6.4}
r^2 = (x^1)^2 + (x^2)^2 + (x^3)^2.
\end{equation}
The 3 (quasi)translational Killing vectors are given by
\begin{equation}\label{6.8}
\xi^{(J)i} = \sqrt{1-kr^2}\delta^i_J,\;\;\;\; J=1,2,3,
\end{equation}
and the 3 rotational Killing vectors by
\begin{equation}\label{6.9}
\xi^{(J)i} = \varepsilon^{iJ}{}_{l}{}\; x^l,
\end{equation}
Here $\varepsilon$ is the usual permutation symbol, $\varepsilon_{123} =
+1$, with indices moved by $\delta_{ik}$, resp.~$\delta^{ik}$.

The 4 conformal Killing vectors are also simple in the $x^i$
coordinates:
\begin{equation}\label{6.10}
\psi^i = \sqrt{1-kr^2}x^i,
\end{equation}
\begin{equation}\label{6.11}
\psi^{(J)i} = \delta^{iJ} - k x^i x^J,\qquad k = \pm 1,
\end{equation}
\begin{equation}\label{6.12}
\psi^{(J)i} = \frac{1}{2} \delta^{iJ} r^2 - x^i x^J,\qquad
k = 0.
\end{equation}

The same Killing and conformal Killing vectors in hyperspherical
coordinates are more complicated but are directly connected with
better known forms of (hyper)spherical harmonics. Here we denote $r =
\sin \chi\ \;(k = +1)$, $r = \chi\; (k = 0)$, $r = \sinh\chi\ \;(k = -1)$ and,
correspondingly, $r' = \cos \chi, 1, \cosh \chi$. 
 As in the main text, we denote the six Killing vectors by $\xi^{(A)i}, A = 1, ..., 6$, and the four conformal Killing vectors by $\psi^{(A)i}, A = 1, ..., 4$. 
The 3 (quasi)translational Killing vectors read
\begin{equation}\label{6.7b}
\begin{aligned}
\xi^{(1)i} &= (\sin \theta \cos \varphi, r'r^{-1} \cos\theta
\cos\varphi, -r'r^{-1} \sin \varphi/\sin\theta), \\
\xi^{(2)i} &= (\sin \theta \sin \varphi , r' r^{-1} \cos \theta \sin \varphi , r' r^{-1} \cos \varphi / \sin \theta ), \\
\xi^{(3)i} &= (\cos \theta , -r'{}r^{-1}\sin \theta , 0 ).
\end{aligned}
\end{equation}
The 3 rotational Killing vectors \eqref{6.9} turn into the same forms independent of $k$:
\begin{equation}\label{6.7e}
\xi^{(4)i} = (0, -\sin \varphi , -\cot \theta \cos \varphi), \;\;
\xi^{(5)i} = (0, \cos \varphi , -\cot \theta \sin \varphi), \;\;
\xi^{(6)i} = (0, 0, 1).
\end{equation}
The `dilatation' conformal Killing vector \eqref{6.10} is for all $k$ simply given by
\begin{equation}\label{6.7f}
\psi^{(1)i} = (r, 0, 0),
\end{equation}
whereas the other three conformal Killing vectors read for $k = \pm1$,
\begin{equation}\label{6.7i}
\begin{aligned}
\psi^{(2)i} &=( r'\sin \theta \cos \varphi , r^{-1} \cos \theta \cos \varphi, - r^{-1} \sin \varphi/\sin \theta),\\
\psi^{(3)i} &= ( r'\sin \theta \sin \varphi, r^{-1} \cos \theta \sin \varphi , r^{-1} \cos \varphi/\sin \theta),\\
\psi^{(4)i} &=( r'\sin \theta , -r^{-1}\sin \theta, 0),
\end{aligned}
\end{equation}
and for $k = 0$,
\begin{equation}\label{6.7l}
\begin{aligned}
\psi^{(2)i} &= \begin{matrix} {\frac{1}{2}}\end{matrix}r^2(-\sin \theta \cos \varphi, r^{-1}\cos \theta \cos \varphi, -r^{-1}\sin \varphi/\sin \theta),\\
\psi^{(3)i} &= \begin{matrix} {\frac{1}{2}}\end{matrix}r^2(-\sin \theta \sin \varphi , r^{-1}\cos \theta\sin \varphi , r^{-1}\cos \varphi/\sin \theta),\\
\psi^{(4)i} &= \begin{matrix} {\frac{1}{2}}\end{matrix}r^2(-\cos \theta , -r^{-1}\sin \theta , 0).
\end{aligned}
\end{equation}

\subsection {Scalar harmonics in $\boldsymbol S^3$ and $\boldsymbol H^3$}

In $S^3$, the scalar harmonics $Q_{Llm}$ with $L\geq l \geq 0$, $L, l$ integers, $m = -l, ..., +l$, satisfy
\begin{equation}\label{6.7m}
\nabla^2 Q_{Llm} + L(L+2) Q_{Llm} = 0.
\end{equation}
In the normalized form they read
\begin{equation}\label{6.7n}
Q_{Llm} = \sqrt{N_{Ll}} \frac{1}{\sqrt{\sin\chi}} P^{-(l+\frac{1}{2})}_{L+\frac{1}{2}}(\cos \chi)Y_{lm}(\theta ,\varphi),
\end{equation}
where $N_{Ll}= \frac{(L+1)(L+l+1)!}{(L-l)!}, P ^{-\mu}_\nu$ are Legendre functions of the first kind, and $Y_{lm}$ are the usual spherical harmonics.
In $H^3$, the harmonics are $Q_{\lambda lm}, \lambda \geq 0$ is continuous and must be real for square integrability; they satisfy
\begin{equation}\label{6.7o}
\nabla ^2 Q_{\lambda lm} + (\lambda^2 + 1) Q_{\lambda lm} = 0.
\end{equation}
The normalized form is
\begin{equation}\label{6.7p}
Q_{\lambda lm} = \sqrt{N_{\lambda l}} \frac{1}{\sqrt{\sinh\chi}} {}P^{-(l+\frac{1}{2})}_{-\frac{1}{2}+i\lambda} (\cosh \chi) Y_{lm}(\theta ,\varphi),
\end{equation}
\begin{equation}\label{6.7r}
N_{\lambda l} = \lambda^2(\lambda^2 + 1)(\lambda^2 + 2^2)...(\lambda^2 + l^2).\\ \nonumber
\end{equation}
It can be easily seen that, for $L = 1$, $l = 0$, $m = 0$ and $L = 1$, $l = 1$, $m = -1, 0, +1$, the expression \eqref{6.7n} leads, up to multiplicative constants, to the four functions given in Eq. \eqref{4.42}, whereas nonintegrable harmonics \eqref{6.7p} for $\lambda = 2i$, $l = 0$, $m = 0$ and $\lambda = 2i$, $l = 1$, $m = -1, 0, +1$ imply Eq. \eqref{4.43}. Their gradients \eqref{4.41} yield the covariant components of the conformal Killing vectors \eqref{6.7f} and \eqref{6.7i}.

Let us remark that for $k = \pm 1$ the translational Killing vectors are not gradients of scalars (as they are for $k = 0$). They are proportional to the vector spherical harmonics with even parity and $L = 1$, $l = 1$ (for $k = \pm 1$) and $\lambda = 2i, l = 1$ (for $k = -1$). The rotational Killing vectors are proportional to vector harmonics with odd parity and $L = 1$, $l = 1$, respectively,  $\lambda = 2i$, $l = 1$.

\section{Field equations and solutions in other gauges}\label{App.D}

\subsection{Mach 2 gauge}

The gauge conditions ${\cal {T}}_k  = \nabla_l{\tilde{h}}{}^l_{T k} = 0$ and $\nabla^2 \tilde{h}^n_n + 3 k\tilde{h}_n^n = 0$ [cf. \eqref{4.17} and \eqref{4.25}] simplify the field equation \eqref{3.38} into the relation
\begin{equation}\label{B1}
-2{\cal{H}}\mathcal{K} = a^2\kappa\delta {\tilde T}^0_0 , 
\end{equation}
from which ${\cal{K}}$ can be expressed and $\nabla^2 {\cal{K}}$ needed in the following step can easily be calculated. Applying $\nabla^k$ to Eq. (\ref{3.39}) we obtain the equation for ${\cal{P}} = \nabla_l \tilde{h}^l_0$,\\
\begin{equation}\label{B2}
\nabla ^2{\mathcal{P}} + 3k\mathcal{P} = \frac{3}{2}a^2 \kappa (\nabla 
^k \delta \tilde{T}^0_k + \frac{1}{3{\cal{H}}}\nabla ^2 \delta\tilde{T}^0_0).
\end{equation}
Solving for ${\cal{P}}$ and substituting back into Eq. (\ref{3.38}) we get the elliptic equation for $\tilde{h}_{0k}$:
\begin{equation}\label{B3}
\nabla^2 \tilde{h}_{0k} + 2 k \tilde{h}_{0k} = 2a^2\kappa (\delta\tilde{T}_k^0 + \frac{1}{3{\cal{H}}}\nabla_k \delta\tilde{T}^0_0) - \frac{1}{3}\nabla_k {\cal{P}}.
\end{equation}
From Eq. (\ref{3.40}) $\tilde{h}_{00}$ can be determined. ${\cal{K}}$ can be expressed in terms of $\delta \tilde{T}^0_0$ from relation (\ref{B1}), and the last the term on the r.h.s. involving the time derivative $(a^2 {\cal{H}}^{-1} \delta \widetilde{T}^0_0)'$ can be calculated by employing the perturbed Bianchi identities, Eq. (\ref{3.47}). 
 Eq. (\ref{3.40}) becomes 
\begin{equation}\label{B4}
\nabla ^2\widetilde{h}_{00}+3 k \widetilde{h}_{00} =
\frac{a^2\kappa}{\cal{H}}(\nabla^k \delta\widetilde{T}^0_k - \frac{k}{{\cal{H}}}\delta\widetilde{T}^0_0). 
\end{equation}
From the elliptic equations (\ref{B2})--(\ref{B4}) the metric perturbations $\tilde{h}_{00}$, $\tilde{h}_{0k}$ follow instantaneously if the sources $\delta \tilde{T}^0_0$ and $\delta\tilde{T}^0_k$ (resp. $\delta\tilde{T}^k_0$) are given.


\subsection{Mach 3 gauge}

Together with ${\cal{T}}_k = \nabla_l{\tilde{h}}^l_{T k} = 0$ it is now assumed $\nabla ^2{\mathcal{P}} + 3k\mathcal{P} = 0$, or simply ${\cal{P}}=\nabla_l \widetilde{h}^l_0 = 0$. Applying $\nabla^k$ to Eq. (\ref{3.39}), one gets
\begin{equation}\label{B5}
\nabla^2 {\cal{K}} = \frac{3}{2} a^2 \kappa \nabla^k \delta \widetilde{T}^0_k.
\end{equation}
With ${\cal{K}}$ known we obtain the elliptic equation for $\widetilde{h}^n_n$ from Eq. (\ref{3.38}):
\begin{equation}\label{B6}
\nabla^2 \widetilde{h}^n_n + 3 k \widetilde{h}^n_n = 3 a^2 \kappa {} \delta \widetilde{T}^0_0 + 6 {\cal{H K}}.
\end{equation}
Next, we make the time-derivative of (\ref{B5}) and substitute for $\nabla^k \delta \widetilde{T'}^0_{k}$ from the perturbed Bianchi identities. Applying then $\nabla^2$ on Eq. (\ref{3.40}) we arrive at the elliptic equation for $\widetilde{h}_{00}$: 
\begin{equation}\label{B7}
\nabla^2 (\nabla^2 \tilde{h}_{00} + 3 k \tilde{h}_{00}) = a^2 \kappa \left[ \nabla^2 (\delta\tilde{T}_0^0 - \delta\tilde{T}_n^n) + 3{\cal{H}} \nabla^k \delta\tilde{T}_k^0
 + 3 \nabla^k \nabla_m \delta\tilde{T}^m_k \right].
 \end{equation}
There is another simple elliptic equation satisfied by the quantity $\chi = \widetilde{h}_{00} - \frac{1}{3} \widetilde{h}^n_n$. Taking $\nabla^2$ of Eq. (\ref{B6}) and regarding Eq. (\ref{B5}), we combine it with Eq. (\ref{B7}) to obtain
\begin{equation}\label{B8}
\nabla^2 (\nabla^2 \chi + 3 k \chi) = 3 a^2 \kappa (\nabla^k \nabla_l \delta \tilde{T} ^l_k - \frac{1}{3} \nabla^2\delta\tilde{T}^n_n).
\end{equation}
The quantity $\chi$ appears directly also in Eq. (\ref{3.41}) for the spatial components $\delta \underset{T}{\tilde{G}}{}^l_k$. Applying $\nabla^k\nabla_l$ on this equation, one arrives again at Eq. (\ref{B8}) above.

With ${\cal{P}}$ known (${\cal{P}} = 0$ in the simplest choice of the Mach $3^*$ gauge) and ${\cal{K}}$ determined from Eq. (\ref{B5}), the constraint equation (\ref{3.39}) becomes a simple elliptic equation for $\widetilde{h}_{k0}$:
\begin{equation}\label{B9}
\nabla^2 \widetilde{h}_{k0} + 2k\widetilde{h}_{k0} = 2 a^2 \kappa\delta \widetilde{T}^0_k - \frac{1}{3}\nabla_k{\cal{P}} - \frac{4}{3}\nabla_k{\cal{K}}.
\end{equation}


The equations for $\widetilde{h}_{00}$, $\widetilde{h}_{0k}$ in both Mach 2 and 3 gauge are elliptic. 
Their form 
 is very similar to the 
  equations in the Mach 1 gauge. We can solve them by the same methods.


\subsection{Generalized Lorenz-de Donder gauge}

We start from the gauge conditions (\ref{4.31}), expressed explicitly 
 in Eqs. (\ref{4.32}) and (\ref{4.33}). Now in general ${\cal{T}}_k = \nabla_l{\tilde{h}}{}^l_{T k} \neq 0$, and also ${\cal{P}} = \nabla_l\tilde{h}^l_0$ and ${\cal{K}}$ 
 are nonvanishing. Nevertheless, the field equations \eqref{3.38}-\eqref{3.41} can be rewritten into a quite telling form. Denoting
$\widetilde{h}_{00} = \varphi, \; \tfrac{1}{3}\widetilde{h}^n_n = \psi ,\; {\cal{P}} = \nabla_l\widetilde{h}^l_0$,
we arrive at the following system: 




\begin{align}
\nabla^2\varphi - \varphi\;'' - 2 {\cal{H}} a^3 \left( \frac{\varphi}{a^3}\right) ' - 6a \left( \frac{\cal{H}}{a}\right) ' {}\psi - 4 {\cal{HP}} &=  a^2 \kappa (\delta\widetilde{T}^0_0 - \delta\widetilde{T}^n_n),\label{S4a}\\
\nabla ^2 \psi - \psi{}'' + 4 k \psi - 2 {\cal{H}}{}a^3 \left( \frac{\psi}{a^3}\right) ' - 2a \left( \frac{\cal{H}}{a}\right) '\varphi - \frac{4}{3} {\cal{HP}} &=  a^2\kappa (\delta \widetilde{T}^0_0 + \frac{1}{3}\delta \widetilde{T}^n_n), \label{S4}\\
\nabla^2\widetilde{h}_{0k} - \widetilde{h}{}''_{0k} + 2 k\widetilde{h}_{0k} - 4 ({\cal{H}}\widetilde{h}_{0k})' - {\cal{H}} \nabla _k (\varphi + 3\psi) &= 2a^2 \kappa \delta \widetilde{T}^0_k, \label{S5}\\
\nabla^2 {\cal{P}}- {\cal{P}}{}'' + 4 k {\cal{P}}-4({\cal{HP}})' - {\cal{H}}\nabla^2 (\varphi+3\psi) &= 2a^2 \kappa\nabla^k \delta\widetilde{T}^0_k, \label{S6}
\end{align}
\begin{equation}
-\nabla^2{\tilde{h}}{}^l_{T k} + {\tilde{h}}^{l}{}''_{T k} + 2k{\tilde{h}}{}^l_{T k} + 2{\cal{H}}{\tilde{h}}^l{}'_{T k} + 4{\cal{H}} (f^{lm}\nabla_{(m} \tilde{h}_{k)0} - \frac{1}{3}\delta ^l_k \nabla _n \tilde{h}^n_0) = 2a^2 \kappa(\delta \tilde{T}^l_k - \frac{1}{3}\delta^l_k\delta\tilde{T}^n_n). \label{S7}
\end{equation}
All equations 
 now have the character of hyperbolic generalized wave equations. The metric perturbations are {\it not} determined instantaneously in terms of the sources $\delta \widetilde{T}^\mu_\nu$. Although the main parts of the equations are given by the standard wave operators $\nabla^2 - {\mathrm{d}}^2/{\mathrm{d}}\eta^2$, there are terms involving lower derivatives of the metric perturbations which make the system coupled. 
These equations may turn out to be useful in cosmology as the standard harmonic gauge is in the post-Minkowskian approximations to general relativity.


\end{document}